\documentclass[a4paper,11pt]{article}
\usepackage[utf8]{inputenc}
\usepackage[english]{babel}
\usepackage[T1]{fontenc}
\usepackage[table,dvipsnames]{xcolor}
\usepackage{authblk}
\usepackage{url}
\usepackage{amsmath}
\usepackage{amssymb}
\usepackage{csquotes}
\usepackage{graphicx}
\usepackage{subfig}
\usepackage{tabularx}
\usepackage{booktabs}
\usepackage{footnote}
\usepackage{tablefootnote}
\usepackage{longtable}
\usepackage{multirow}
\usepackage{tabulary}
\usepackage{physics}
\usepackage[colorlinks=true,linkcolor=black,citecolor=blue,urlcolor=blue]{hyperref}
\usepackage{algorithm, algorithmic}

\usepackage[noabbrev,capitalize]{cleveref}
\crefname{equation}{Eq.}{Eqs.}
\crefname{figure}{Fig.}{Figs.}
\crefname{table}{Tab.}{Tabs.}
\crefname{section}{Sec.}{Secs.}
\crefname{subsection}{Sec.}{Secs.}
\crefname{subsubsection}{Sec.}{Secs.}

\usepackage[
    style=alphabetic,
    giveninits=true,
    backend=biber,
    citestyle=alphabetic
]{biblatex}

\DeclareFieldFormat{pages}{\mkfirstpage{#1}}
\renewbibmacro{in:}{}

\appto{\bibsetup}{\sloppy}
\usepackage{geometry}
\usepackage[acronym]{glossaries}
\makeglossaries
\newacronym{dl}{DL}{deep learning}
\newacronym{drl}{DRL}{deep reinforcement learning}
\newacronym{dnn}{DNN}{deep neural network}
\newacronym{cnn}{CNN}{convolutional neural network}
\newacronym{qcnn}{QCNN}{quantum convolutional neural network}
\newacronym{distrl}{DistRL}{distributional reinforcement learning}
\newacronym{nn}{NN}{neural network}
\newacronym{qnn}{QNN}{quantum neural network}
\newacronym{qrnn}{QRNN}{quantum recurrent neural network}
\newacronym{qlstm}{QLSTM}{quantum long short-term memory}
\newacronym{dql}{DQL}{deep $Q$-learning}
\newacronym{ddql}{DDQL}{double deep $Q$-learning}
\newacronym{ml}{ML}{machine learning}
\newacronym{fim}{FIM}{Fisher information matrix}
\newacronym{nisq}{NISQ}{noisy intermediate-scale quantum}
\newacronym{ps}{PS}{projective simulation}
\newacronym{qirl}{QiRL}{quantum-inspired reinforcement learning}
\newacronym{qml}{QML}{quantum machine learning}
\newacronym{qpu}{QPU}{quantum processing unit}
\newacronym{qrl}{QRL}{quantum reinforcement learning}
\newacronym{rl}{RL}{reinforcement learning}
\newacronym{qpg}{QPG}{quantum policy gradient}
\newacronym{qnpg}{QNPG}{quantum natural policy gradient}
\newacronym{qa3c}{QA3C}{quantum asynchronous advantage actor critic}
\newacronym{qddpg}{QDDPG}{quantum deep deterministic policy gradient}
\newacronym{dqas}{DQAS}{differential quantum architecture search}
\newacronym{td}{TD}{temporal difference}
\newacronym{vqc}{VQC}{variational quantum circuit}
\newacronym{vqa}{VQA}{variational quantum algorithm}
\newacronym{qc}{QC}{quantum computing}
\newacronym{pdf}{PDF}{probability density function}
\newacronym[\glslongpluralkey={Markov decision processes}]{mdp}{MDP}{Markov decision process}
\newacronym{mse}{MSE}{mean square error}
\newacronym{dlp}{DLP}{discrete logarithm problem}
\newacronym{vqe}{VQE}{variational quantum eigensolver}
\newacronym{sac}{SAC}{soft actor-critic}
\newacronym{marl}{MARL}{multi-agent reinforcement learning}
\newacronym{qmarl}{QMARL}{quantum multi-agent reinforcement learning}
\newacronym{ctde}{CTDE}{centralized training with decentralized execution}
\newacronym{pomdp}{POMDP}{partially observable Markov decision process}
\newacronym{tn}{TN}{tensor network}
\newacronym{mps}{MPS}{matrix product state}
\newacronym{ppo}{PPO}{proximal policy optimization}
\newacronym{vrp}{VRP}{vehicle routing problem}
\newacronym{cvrp}{CVRP}{capacitated vehicle routing problem}
\newacronym{tsp}{TSP}{traveling salesman problem}
\newacronym{cv}{CV}{continuous-variable}
\newacronym{qr}{QR}{quantile regression}
\newacronym{cptp}{CPTP}{completely positive trace preserving}
\newacronym{bcq}{BCQ}{batch-constrained deep $Q$-learning}
\newacronym{bcqq}{BCQQ}{batch-constrained quantum $Q$-learning}
\newacronym{dru}{DRU}{data re-uploading}
\newacronym{cql}{CQL}{conservative $Q$-learning}
\newacronym{cq2l}{CQ2L}{conservative quantum $Q$-learning}
\newacronym{dqn}{DQN}{deep $Q$-network}
\newacronym{vqdqn}{VQ-DQN}{variational quantum deep $Q$-networks}

\title{\textbf{A Survey on Quantum Reinforcement Learning}\vspace{1.5cm}}
\author{Nico Meyer, Christian Ufrecht, Maniraman Periyasamy, Daniel D.\ Scherer, Axel Plinge, and Christopher Mutschler\\Fraunhofer IIS, Fraunhofer Institute for Integrated Circuits IIS, Nuremberg, Germany\\\texttt{\{firstname.\@lastname|daniel.scherer2\}@iis.\@fraunhofer.\@de}}

\date{January 1, 2024}
\begin{document}

\maketitle

\vfill

%------------------------------------------------------------
\begin{abstract}
    \Glsentrylong{qrl} is an emerging field at the intersection of \glsentrylong{qc} and \glsentrylong{ml}. While we intend to provide a broad overview of the literature on \glsentrylong{qrl} -- our interpretation of this term will be clarified below -- we put particular emphasis on recent developments. With a focus on already available \glsentrylong{nisq} devices, these include \glsentrylongpl{vqc} acting as function approximators in an otherwise classical \glsentrylong{rl} setting. In addition, we survey \glsentrylong{qrl} algorithms based on future fault-tolerant hardware, some of which come with a provable quantum advantage. We provide both a birds-eye-view of the field, as well as summaries and reviews for selected parts of the literature.
\end{abstract}
%------------------------------------------------------------

\vfill

\tableofcontents

\vfill

\glsresetall

\newpage

%------------------------------------------------------------
\section{Introduction and Overview}
\label{sec:IntroAndOverview}
%------------------------------------------------------------

With recent advances in the fabrication and control of hardware for quantum information processing, the possibilities of merging \gls{qc} with \gls{ml} have received a huge amount of attention within the growing research community. Hereby, \gls{rl} is the third paradigm besides supervised and unsupervised learning. In this survey article, we provide an overview over so-called \gls{qrl} algorithms. We understand these as quantum-assisted approaches, that solve a particular task (be they classical or quantum in nature) by employing quantum resources (either in simulation and/or in experiment).

In order to keep this contribution as self-contained as possible, we provide the necessary backgrounds before venturing into the \gls{qrl} literature. We start out with a brief recap of the essentials of the \gls{rl} paradigm in the fully classical setting in \cref{sec:IntroRL}. Further, in \cref{sec:IntroQC} we provide a quick introduction to \gls{qc} and \glspl{vqc}. Readers familiar with either of the topics may safely skip these sections. 

%
%------------------------------------------------------------
\begin{figure}[b!]
    \centering
    \includegraphics[width=1.0\textwidth]{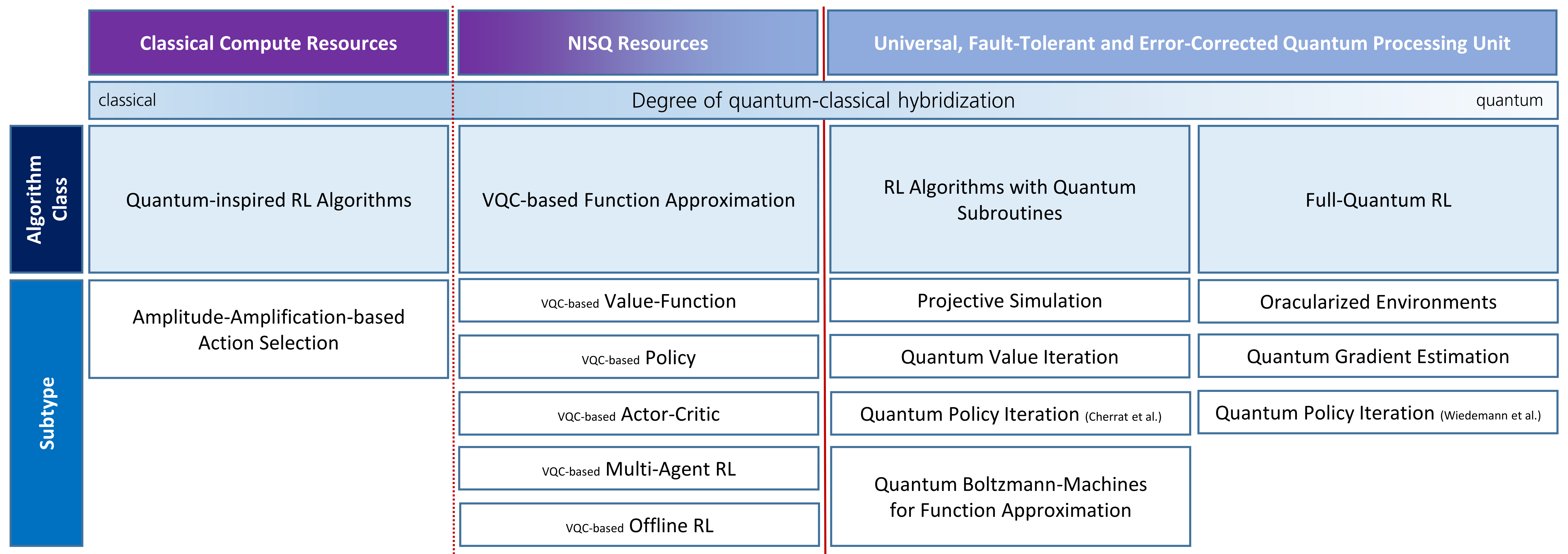}
    \caption{A possible classification matrix for \gls{qrl} algorithms, where we took into account only those variants of \gls{qrl} which we focus on in \cref{sec:QRLAlgs}. The algorithm classes are ordered according to their degree of quantum-classical hybridization, ranging from purely classical to purely quantum. A more detailed review of the $22$ selected works on \gls{qirl}-algorithms can be found in \cref{subsec:quantum_inspired}. \gls{vqc}-based approaches are summarized in quite some detail in \cref{subsec:VQC_based} -- comprising of $68$ papers. \gls{qrl}-algorithms employing post-\gls{nisq} quantum algorithms as subroutines or even fully quantum approaches to \gls{qrl} are described in \cref{subsec:projective_simulaton}, \cref{subsec:boltzman_machines}, \cref{subsec:QPI} and \cref{subsec:Oracles}, based on $30$ selected manuscripts. The dashed vertical line between classical and \gls{nisq} compute resources indicates that presently it is unclear whether \gls{qrl} with \gls{nisq}-compatible algorithms offers robust quantum advantage on a broad range of learning problems. The solid vertical line distinguishes post-\gls{nisq} algorithms from both classical and \gls{nisq}-compatible algorithms, as they typically come with guaranteed quantum advantage (at least relative to their classical counterparts).}
    \label{fig:QRLAlgs}
\end{figure}
%------------------------------------------------------------

In \cref{sec:QRLAlgs} we turn our attention to the emerging field of \gls{qrl}, starting out with a quick overview of the literature. Then we delve into summaries of the most prominent contributions. This selection is necessarily subjective and reflects our own research interests -- overall we identified $177$ relevant manuscripts, of which we reviewed $120$ explicitly. For a detailed overview on paper counts see~\cref{tab:number_publications}. We organized our summaries into several blocks, that are ordered by what one could call an increasing degree of `quantiziation'. The first of these blocks in \cref{subsec:quantum_inspired} covers what we refer to as `quantum-inspired' \gls{rl} algorithms. The second block in \cref{subsec:VQC_based} takes a rather detailed look at \gls{qrl} algorithms that employ so-called \glspl{vqc} as function approximators. In many cases, the corresponding algorithms are obtained by simply replacing a standard \glsentrylong{nn} function approximator (or any other sort) by an appropriate \gls{vqc}. We provide detailed summaries for most papers in this category, as \glsentrylongpl{vqa} are believed to offer the potential to obtain quantum advantage despite the limitations of present day \gls{nisq} hardware. In \cref{subsec:projective_simulaton,subsec:boltzman_machines}, we take a look at realizations of \gls{qrl} based on so-called \glsentrylong{ps} and the use of Boltzmann machines as function approximators, respectively. In \cref{subsec:QPI} we move to a class of approaches that employ quantum algorithms as subroutines. The corresponding hardware requirements will likely be compatible only with universal, fault-tolerant and error-corrected \glspl{qpu}. Finally, \cref{subsec:Oracles} provides a summary for a formal approach to \gls{qrl}, which treats all components of \gls{rl} `quantumly'. From our point of view, the highest degree of quantization can thus be found in these approaches. \cref{fig:QRLAlgs} gives an overview of the \gls{qrl} literature as understood in this survey.

\begin{table}[tb]
\centering
    \begin{tabular}{@{}l|ccccccc@{}}
    \toprule
     & $\leq 2018$  & $2019$ & $2020$ & $2021$ & $2022$ & $2023$ & 
     $\Sigma$~ \\
    \midrule
    ~Quantum-inspired QRL & $12$  & $1$  & $2$ & $5$ & $2$ & $0$ & $22$~ \\
    ~VQC-based QRL & $0$ & $2$ & $2$ & $9$ & $21$ & $34$ & $68$~ \\
    ~QRL application$^{\boldsymbol{\mathrm{a}}}$ & $0$ & $0$ & $0$ & $1$ & $7$ & $18$ & $26$~ \\
    ~Post-NISQ QRL & $12$ & $2$ & $2$ & $6$ & $4$ & $4$ & $30$~ \\
    \bottomrule
    \end{tabular}
    \flushleft
    \centering
    \footnotesize{$^{\boldsymbol{\mathrm{a}}}$The QRL application papers are VQC-based and also counted towards that category.}
    \caption{The year-wise paper count for the different classes of \gls{qrl}. The data not necessarily reflects publication, but rather the first public availability, e.g. via preprint servers.}\label{tab:number_publications}%
\end{table}

Finally, in \cref{sec:Outlook} we state our concluding thoughts on the current state-of-the-art of \gls{qrl}. Before moving to more technical content, we would like to express our hope that this literature survey on \gls{qrl} will be of use to colleagues and collaborators and the wider \gls{qc} research community. It represents our effort to familiarize ourselves with \gls{qrl} and its main research directions.  

%------------------------------------------------------------
\section{Classical Reinforcement Learning}
\label{sec:IntroRL}
%------------------------------------------------------------

Compared to the methods of supervised and unsupervised learning, which are typically implemented as passive learning, \gls{rl} falls into the class of interaction-based learning~\cite{Sutton_2019}. On an abstract level, the learner interacts with its environment, the state of which it can either fully or only partially observe through a corresponding observation obtained after executing an action according to an underlying policy. In the \gls{rl} paradigm, the learner is therefore appropriately referred to as an agent: it can - be it in simulation or in the real world - interact with its environment according to its abilities. The aim of \gls{rl} is to learn a policy through the interaction of the agent with the environment, which is optimal with regard to a reward adapted to the problem. In other words, the agent should find an optimal policy during the learning process in the abstract space of all policies, which maximizes the expected cumulative reward. The theoretical basis for \gls{rl} is formed by so-called \glspl{mdp} and the associated Bellman equation, which represents a consistency equation for the so-called value function. In turn, an optimal policy can be extracted from the optimal value function. Alternatively, the optimal policy can also be learned directly. Under certain conditions, the elements of \gls{rl} can be mapped to their respective equivalents in control theory, where typically a dynamic optimization problem is solved by gradient-based methods with simulation of the corresponding model dynamics. On the \gls{rl} side, there are both model-based and model-free approaches. The model-free approach in particular is one of the strengths of the \gls{rl} method, since in many cases state and action spaces are too high-dimensional to design realistic dynamical models and simulate them to efficiently find optimal control strategies. The large dimensions of the spaces that occur in realistic problems make the use of approximation methods for the value function necessary. Driven by the breakthroughs in \gls{dl}, artificial \glspl{nn} have established themselves as function approximators for both value function and policy (understood as a deterministic or probabilistic mapping of states to actions), thus establishing the field of \gls{drl}.

\begin{figure}[ht]
    \centering
    \includegraphics[width=0.4\textwidth]{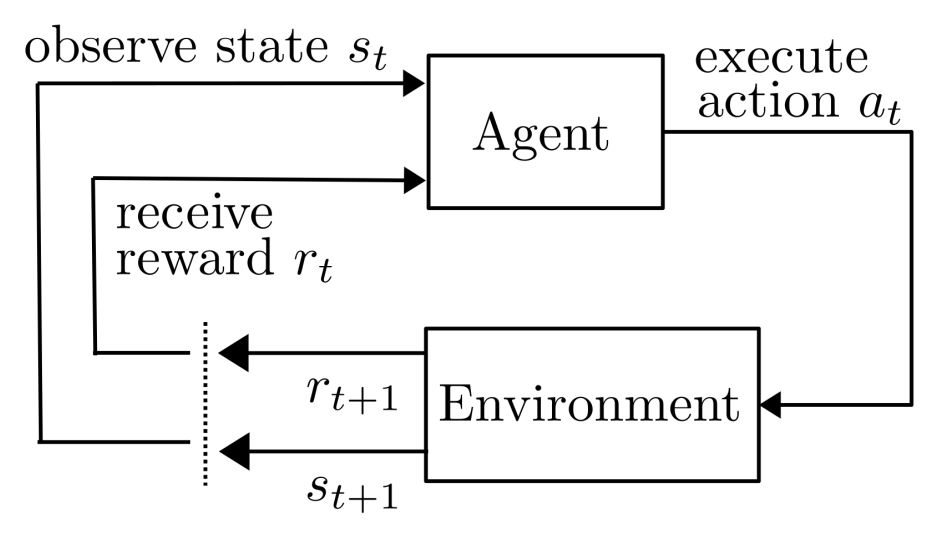}
    \caption{Interaction between agent and environment for one timestep of a \gls{rl} task.}
    \label{fig:IntroRL_pipeline}
\end{figure}

In the following, we will introduce the various notions pertaining to \gls{rl} in a more formal way and provide the background necessary to understand the basic \gls{rl} terminology. In an \gls{rl} scenario, the algorithm, also referred to as agent, generates its own data by interacting with an environment. This interaction happens over some discrete timesteps $t$, which are accumulated to episodes with either finite or infinite horizon. In each timestep, the agent is able to make an observation $s_t \in \mathcal{S}$ of the environment. Based on this state information, an action $a_t \in \mathcal{A}$ acting on the environment is selected according to a policy. Based on the (usually unknown) environment dynamics, the next state $s_{t+1} \in \mathcal{S}$ is observed from the environment and the agent receives a reward $r_t \in \mathcal{R}$ for its choice. The agent should select the actions in such a way that some objective is optimized, usually related to the long term reward. A sketch of this pipeline can be found in \cref{fig:IntroRL_pipeline}. In this survey article, we follow the formalism and notation of Sutton et al.~\cite{Sutton_2019}, with small adaptions wherever we feel that it eases comprehension.

\paragraph{Reinforcement Learning as a Markov Decision Process} More formally, this setup is usually described as an \gls{mdp}. A finite \gls{mdp} is a 5-tuple $(\mathcal{S}, \mathcal{A}, \mathcal{R}, p, \gamma)$, where the sets $\mathcal{S}$, $\mathcal{A}$ and $\mathcal{R}$ are finite. It is defined by the following components:
\begin{itemize}
    \item A set of states $\mathcal{S}$ the agent can observe from the environment
    \item A set of actions $\mathcal{A}$ the agent can execute in the environment
    \item A set of rewards $\mathcal{R} \subset \mathbb{R}$ the agent can receive from the environment
    \item The environment dynamics $p: \mathcal{S} \times \mathcal{R} \times \mathcal{A} \times \mathcal{S} \to [0,1]$; The value $p(s', r | s, a) := \text{Pr} \lbrace s_{t+1}=s', r_{t}=r | s_{t}=s, a_{t}=a \rbrace$ gives the probability that the environment transitions to state $s_{t+1}$ and the agent receives reward $r_{t}$, if the agents executes action $a_t$ in state $s_t$ at time $t$.
    %\item The reward function $R: \mathcal{S} \times \mathcal{A} \to \mathbb{R}$; The value $R(s, a) := \mathbb{E} \left[ r_t | s_t=s, a_t=a \right]$ gives the reward for the case, that action $a_t$ is executed in state $s_t$.
    %\item The environment dynamics $T: \mathcal{S} \times \mathcal{A} \times \mathcal{S} \to [0,1]$; The value $T(s' | s, a) := \text{Pr} \lbrace s_{t+1}=s' | s_{t}=s, a_{t}=a \rbrace$ gives the probability that the environment transitions to state $s_{t+1}$, if the agents executes action $a_t$ in state $s_t$ at time $t$.
    \item The discount factor $0 \leq \gamma \leq 1$, more on this below;
\end{itemize}
The dynamics of the environment are often not accessible to the agent, otherwise the task collapses to (not necessarily trivial) dynamic programming. The function $p$ satisfies the properties of a \gls{pdf}, i.e., it holds $\sum_{s' \in \mathcal{S}, r \in \mathcal{R}} p(s', r | s, a) = 1$, for all choices of $s \in \mathcal{S}$ and $a \in \mathcal{A}$. According to the \emph{Markov property}, the dynamics are completely described by $p$, i.e., the consecutive state $s_{t+1}$ and reward $r_t$ depend solely on the directly preceding state $s_t$ and action $a_t$.

With this framework in mind, the interaction between agent and environment can be described as a \emph{trajectory} $\tau$. For a finite or infinite horizon $H$, one \emph{episode} is therefore given by the sequence
\begin{equation}
        \tau = \left[ s_0, a_0, r_0, s_1, a_1, r_1, s_2, \cdots, s_{H-1}, a_{H-1}, r_{H-1} \right],
\end{equation}
with $s_t \in \mathcal{S}$, $a_t \in \mathcal{A}$, and $r_t$ sampled following the environment dynamics for each timestep $t$.

\paragraph{Long Term Reward as Objective} The agent gets feedback from the environment through the immediate rewards $r_t$. However, instead of maximizing these short-term rewards, it is much more appropriate to use some long term measure as objective. A natural choice is to go for the cumulative reward, also referred to as the \emph{expected return}
\begin{equation}
    G_t := r_t + r_{t+1} + r_{t+2} + \cdots + r_{H-1}.
\end{equation}
For \emph{episodic} tasks ($H<\infty$) it is often desirable and for \emph{continuous} tasks ($H=\infty$) it is necessary to use a \emph{discount factor} $\gamma$. This leads to the \emph{discounted (expected) return}
\begin{equation}
    G_t := \sum\limits_{t'=t}^{H-1} \gamma^{t'-t} \cdot r_{t'},
\end{equation}
where each choice of $\gamma$ defines a different \gls{mdp}. For $\gamma < 1$ the value of $G_t$ is guaranteed to be finite and emphasis on individual rewards decreases with distance from the current time-step. For $\gamma = 0$ the sum reduces to just the immediate reward, so an appropriate choice of this hyperparameter is crucial for the potential success of the \gls{rl} agent.

\paragraph{Policy, Value Functions and Optimality} In order to describe a meaningful \gls{rl} setup, there are still some concepts missing. As described above, the agent needs to decide for an action in every timestep, depending on the state information that is observed. This decision making process can be understood as a (stochastic) \emph{policy}
\begin{equation}
    \pi \left( a | s \right) := \text{Pr} \lbrace a_t=a | s_t=s \rbrace,
\end{equation}
where  $\sum_{a \in \mathcal{A}} \pi \left( a | s \right) = 1$ holds for all $s \in \mathcal{A}$. The overall task of \gls{rl} is to derive an optimal policy $\pi^{*}$ w.r.t.\ some metric.

A suitable tool to define optimality and also to simplify updates is the notion of \emph{value functions}. The \emph{state value} function of state $s$ under the current policy $\pi$ is defined as
\begin{equation}
    V_{\pi}(s) := \mathbb{E}_{\pi} \left[ G_t | s_t=s \right].
\end{equation}
It describes the expected returns when starting in state $s$ and following policy $\pi$ from there on, with the value for a terminal state always zero by definition. It can be interpreted as a measure of \textit{how good} it is to be in a certain state, where quality is measured w.r.t.\ expected return. Explicitly separating the first step in the definition above gives rise to the \emph{Bellman (expectation) equation}
\begin{equation}
    V_{\pi}(s) = \sum\limits_{a \in \mathcal{A}} \pi \left( a | s \right) \sum\limits_{s' \in \mathcal{S}, r \in \mathcal{R}} p \left( s', r | s, a \right) \left[ r + \gamma \cdot V_{\pi}(s') \right],
\end{equation}
for all $s \in \mathcal{S}$. Consequently, the value function $V_{\pi}$ can be viewed as the unique solution to this Bellman equation. Alternatively, one can define the \emph{state-action value function} as the expected return when starting in state $s$, executing action $a$, and following policy $\pi$ from there on. It is defined as
\begin{equation}
    Q_{\pi} (s,a) := \mathbb{E}_{\pi} \left[ G_t | s_t=s, a_t=a \right],
\end{equation}
for all $s \in \mathcal{S}$ and $a \in \mathcal{A}$. It is straightforward to see that it holds $V_{\pi}(s) = \sum_{a \in \mathcal{A}} \pi \left( a | s \right) Q_{\pi} (s,a)$ for all $s \in \mathcal{S}$. This identity can be used to give the Bellman equation for the state-action value function as $Q_{\pi}(s,a) = \sum_{s' \in \mathcal{S}, r \in \mathcal{R}} p(s', r | s, a) \left[ r + \gamma \cdot \sum_{a' \in \mathcal{A}} \pi(a' | s') Q_{\pi}(s', a') \right]$.

The value function allows to explicitly define and evaluate the quality of policies, i.e., the policy $\pi$ is better or equal to another policy $\pi'$, iff $V_{\pi}(s) \geq V_{\pi'}(s)$ for all $s \in \mathcal{S}$. If a policy is better or equal to all others, it is considered an \emph{optimal policy} $\pi^{*}$. All optimal policies share the same \emph{optimal state-value} function
\begin{equation}
    V_{\pi^{*}}(s) := V^{*}(s) := \underset{\pi}{\max} ~V_{\pi}(s),
\end{equation}
for all $s \in \mathcal{S}$. A similar notion of optimality for the action-value function is given by
\begin{equation}
    Q^{*}(s,a) := \underset{\pi}{\max} ~Q_{\pi}(s,a),
\end{equation}
for all $s \in \mathcal{S}$ and $a \in \mathcal{A}$. It is straightforward to formulate the connection of both quantities as $V^{*}(s) = \underset{\pi}{\max} ~\left( \sum_{a \in \mathcal{A}} \pi \left( a | s \right) Q_{\pi}(s,a) \right) = \underset{a \in \mathcal{A}}{\max}~ Q^{*}(s,a)$. With this it is possible to derive the \emph{Bellman optimality equation} for the value function as
\begin{equation}
    V^{*}(s) = \underset{a \in \mathcal{A}}{\max} ~\sum\limits_{s' \in \mathcal{S}, r \in \mathcal{R}} p \left( s', r | s, a \right) \left[ r + \gamma \cdot V^{*}(s') \right],
\end{equation}
for all $s \in \mathcal{S}$. Using the stated connection this can be reformulated to extend to the state-action value function as $Q^{*}(s,a) = \sum_{s' \in \mathcal{S}, r \in \mathcal{R}} p(s', r | s, a) \left[ r + \gamma \cdot \underset{a' \in \mathcal{A}}{\max} ~Q^{*}(s',a') \right]$ for all $s \in \mathcal{S}$ and $a \in \mathcal{A}$.

\paragraph{Solving and Approximating the Bellman Equation}
One topic that has to be addressed is the actual representation of the policy and value functions. The most intuitive approach is to just store the values for all state-action pairs in a table, also referred to as the \emph{tabular} approach. While this formulation offers nice convergence and optimality guarantees for several scenarios, it has some serious drawbacks. Most prominently, it is intractable once the state-action space gets to large, which is the case for most real-world problems. A workaround is to use \emph{parametric function approximators}, which results in the parameterized functions $\pi_{\theta}$, $V_{\theta}$, or $Q_{\theta}$, respectively. The typical choice is a \gls{nn}~\cite{Hornik_1989}, in \cref{subsec:VQC_based} the usage of \glspl{vqc} for this task is considered from several angles. As there now is an approximation in the defining quantities, also convergence guarantees are much less straightforward than for the tabular case. The remaining parts of this section can be understood both for the tabular and parameterized case, although details might vary a bit.

The Bellman optimality equation offers a tool to derive an optimal policy. It has to be noted that the given formulation makes use of the environment dynamics $p$. Therefore, solution methods solving the equation with dynamic programming are referred to as \emph{model-based}. The two most prominent examples include \emph{value iteration}~\cite{Bellman_1957} and \emph{policy iteration}~\cite{Rummery_1994,Pashenkova_1996}.

There is also a whole range of \emph{model-free} approaches, where the agent does not make use of any model that represents the environment dynamics. Instead, all information is directly acquired by interaction with the environment. One prominent representative is the \emph{$Q$-learning} approach~\cite{Watkins_1992}, which basically is an approximation of $Q$-value iteration using samples. Starting with a random initialization, the update rule
\begin{equation}
    \label{eq:Qlearning_1}
    Q(s,a) \gets Q(s,a) + \alpha \left( r_t + \gamma \cdot \underset{a' \in \mathcal{A}}{\max} ~ Q(s',a') - Q(s,a) \right)
\end{equation}
directly derives from the Bellman equation, where $\alpha$ is a learning rate hyperparameter. The policy is usually defined to act \emph{epsilon-greedily} w.r.t.\ the current action-value function, i.e.\
\begin{equation}
    \label{eq:Qlearning_2}
    \pi(s) := \begin{cases}
     \underset{a \in \mathcal{A}}{\arg \max} ~ Q(s,a) & \text{with probability } 1-\varepsilon, \\
     \text{uniformly at random from } \mathcal{A} & \text{with probability } \varepsilon.
   \end{cases}
\end{equation}
An alternative approach is the \emph{policy gradient} idea~\cite{Sutton_1999}, which directly aims to learn the policy. Based on an parameterized policy $\pi_{\theta}$, it performs updates
\begin{equation}
    \label{eq:policygradients_1}
    \theta \gets \theta + \alpha \nabla_{\theta} J(\theta)
\end{equation}
via gradient ascent, where $J(\theta)$ is a performance measure, usually $J(\theta) = V_{\pi_{\theta}}(s_0)$. Unfortunately the desired gradient likely depends on some environment dynamics, which are not known. The \emph{policy gradient theorem}~\cite{Sutton_1999} describes a quantity proportional to $\nabla_{\theta} V_{\pi_{\theta}}$, which is easier to obtain. It is given by
\begin{equation}
    \label{eq:policygradients_2}
    \nabla_{\theta} V_{\pi_{\theta}}(s_0) \propto \sum\limits_{s \in \mathcal{S}} \mu(s) \sum\limits_{a \in \mathcal{A}} Q_{\pi_{\theta}}(s,a)\nabla_{\theta} \pi_{\theta} \left( a | s\right),
\end{equation}
where $\mu(s)$ is a function that expresses the fraction of time that is spend in state $s$. An concrete instance of this idea is the REINFORCE algorithm~\cite{Williams_1992}, where a Monte Carlo method is used to estimate the quantity described in the equation above. Furthermore, the training procedure can be stabilized by introducing a suitable baseline function that reduces the variance of the expected return~\cite{Zhao_2011}. 

Overall, there are several extensions and modifications of the described concepts. One method worth mentioning is the actor-critic approach~\cite{Konda_2003}, which combines ideas form policy gradient and value functions. As for smaller modifications, there is double $Q$-learning, which introduces an additional target action-value function to reduce some bias caused by the standard $Q$-learning procedure~\cite{Hasselt_2010}. Similarly, the introduction of an experience replay buffer~\cite{Lin_1992} should improve stability and sample efficiency. This finally leads to \emph{offline} or \emph{batch} \gls{rl}~\cite{Ernst_2005}, where the agent is not allowed to directly interact with the environment. Instead, it only has access to a set of previously collected experiences. This formulation is especially relevant in practice, as generating data is sometimes quite expensive. There is still a wide range of topics this summary did not touch. Where necessary, additional details will also be introduced in the upcoming chapters. For a more broad introduction to the topic one can refer to Ref.~\cite{Sutton_2019}, more recent developments are e.g.\ reviewed in Refs.~\cite{Arulkumaran_2017,Nian_2020}.

%------------------------------------------------------------
\section{The Quantum Computing Paradigm}
\label{sec:IntroQC}
%------------------------------------------------------------

The foundations of \gls{qc} were established at the beginning 20th century when the modern theory of quantum physics was developed. Benioff and Feynman proposed the idea of taking advantage of quantum mechanical systems for computing in the early 1980s \cite{Benioff_1980,Feynman_1982}. \Gls{qc} challenges the strong Church-Turing hypothesis, as it potentially provides efficient solutions to classically intractable problems \cite{Nielsen_2016}. This section gives a pragmatic introduction to the basics of \gls{qc}, and also provides an extension to \gls{qml} (here understood as \gls{ml} with \glspl{vqc} as a new class of models) with a focus on \gls{qrl}.

\paragraph{Single and Multi-Qubit Systems}
Similar to \gls{rl}, notation and conventions regarding quantum computing vary quite a bit throughout the literature. Regarding notation, we closely follow the textbook by Nielsen and Chuang \cite{Nielsen_2016}. 

For the moment, let us consider the basic unit of information for classical information processing. A single bit is either in state $0$ or state $1$, consequently, a sequence of $n$ bits can represent $2^n$ unique values. Obviously, the bit register can only be in one of these $2^n$ states at any point in time.

A qubit is the quantum version of a bit. We use the Dirac notation \cite{Nielsen_2016} to define $\ket{0}$ and $\ket{1}$ as two distinct, orthogonal states of the qubit system. These \emph{basis states} span a $2$-dimensional Hilbert space $\mathcal{H} \cong \mathbb{C}^2$, which contains all $1$-qubit (pure) quantum states. The qubits are subject to the laws of quantum mechanics and can be realized with, e.g., spin systems of subatomic particles \cite{Privman_2002}, ion traps \cite{Barry_2014}, neutral atoms \cite{Saffman_2010}, or superconducting circuits \cite{You_2006}. This gives rise to some interesting properties. In fact, a qubit can not only be in either state $\ket{0}$ or $\ket{1}$, but in a \emph{superposition} of both. An arbitrary $1$-qubit state is given as
\begin{align} \label{eq:arbitrary_1_qubit}
	\ket{\psi} = \alpha \ket{0} + \beta \ket{1}.
\end{align}

\begin{figure}[ht]
    \centering
    \includegraphics[width=0.4\textwidth]{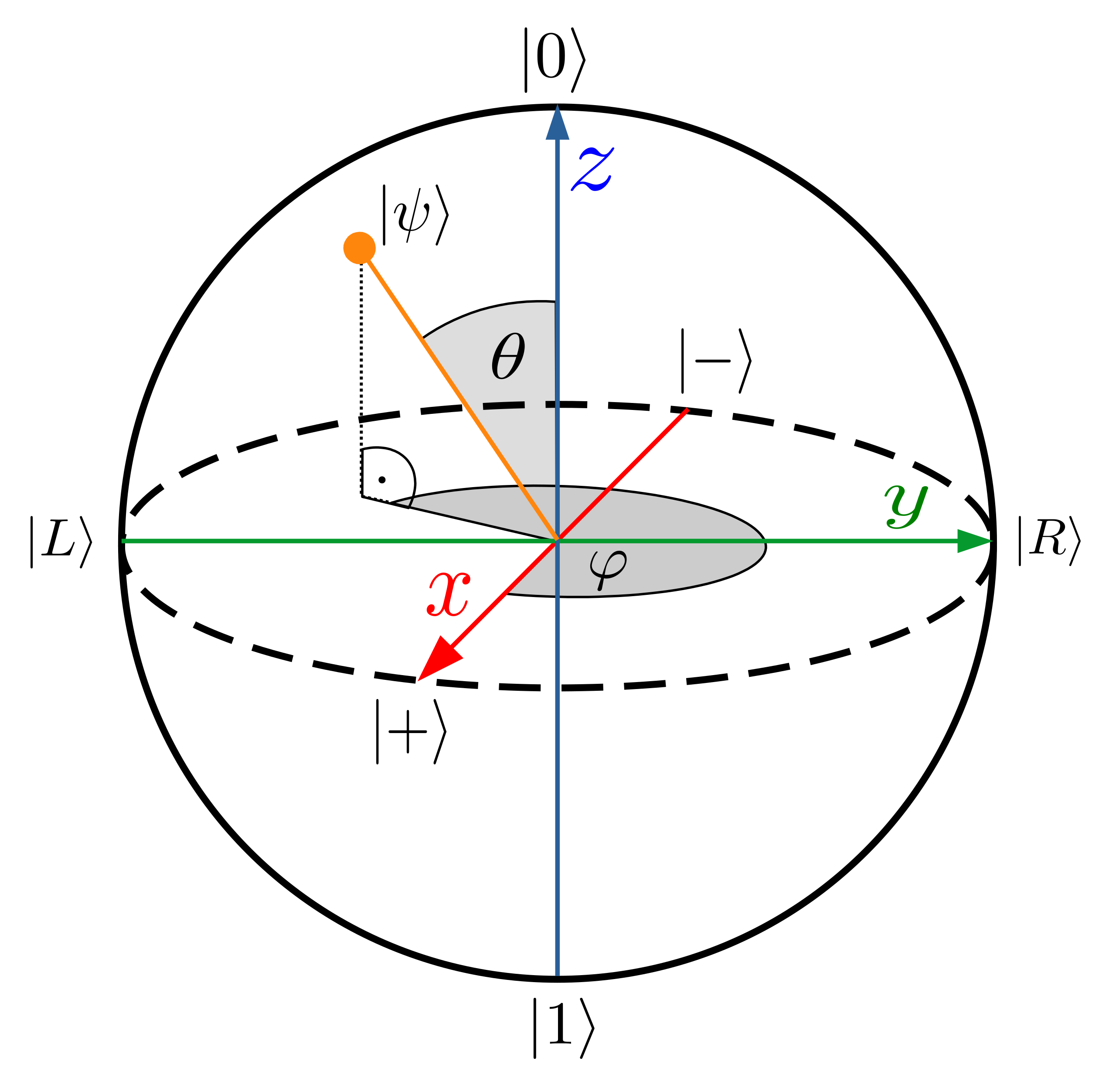}
    \caption{Bloch sphere representation of a $1$-qubit state.}
    \label{fig:bloch_1}
\end{figure}

The \emph{amplitudes} $\alpha$ and $\beta$ are complex numbers, which must satisfy $\abs{\alpha}^2 + \abs{\beta}^2 = 1$. To get a nice visual representation, \cref{eq:arbitrary_1_qubit} can be reformulated as
\begin{align} \label{eq:bloch_eq_1}
	\ket{\psi} = e^{i \gamma} \left( \cos \frac{\theta}{2} \ket{0} + e^{i \phi} \sin \frac{\theta}{2} \ket{1} \right),
\end{align}
with $\gamma,~\theta,~\phi \in \mathbb{R}$. As any global phase has no observable effect \cite{Nielsen_2016}, the prefactor $e^{i \gamma}$ in \cref{eq:bloch_eq_1} can be omitted. This representation makes it possible to visualize the state of a $1$-qubit system on the surface of the \emph{Bloch sphere}, see \cref{fig:bloch_1}. The north and south poles w.r.t.\ the $z$-axis correspond to the basis states $\ket{0}$ and $\ket{1}$, which are also referred to as \emph{computational basis states} of a single qubit. Another, less commonly used basis is given by the poles w.r.t.\ the $x$-axis, the elements are related by $\ket{+} = \frac{\ket{0} + \ket{1}}{\sqrt{2}}$ and $\ket{-} = \frac{\ket{0} - \ket{1}}{\sqrt{2}}$. Similarly, one could also use $\ket{R} = \frac{\ket{0} + i\ket{1}}{\sqrt{2}}$ and $\ket{L} = \frac{\ket{0} - i\ket{1}}{\sqrt{2}}$. An alternative representation associates quantum states with amplitude vectors:
\begin{align} \label{eq:vec_rep}
	\ket{0} \to \begin{bmatrix} 1 \\ 0 \end{bmatrix} \text{   and   } \ket{1} \to \begin{bmatrix} 0 \\ 1 \end{bmatrix}
\end{align}

Multiple-qubit systems are the point where things get interesting. An $n$-qubit system gives access to the $2^n$-dimensional \emph{Hilbert space}, in which an arbitrary \emph{pure} quantum state is defined as
\begin{equation} \label{eq:arbitrary_state}
	\ket{\psi} = c_0 \ket{00 \cdots 00} + c_1 \ket{00 \cdots 01} + \cdots + c_{2^n-1} \ket{11 \cdots 11},
\end{equation}
with $c_i \in \mathbb{C}$ and $\sum_{i=0}^{2^n-1} \abs{c_i}^2 = 1$. The basis states, e.g.\ $\ket{00 \cdots 01} = \ket{0} \otimes \ket{0} \otimes \cdots \otimes \ket{0} \otimes \ket{1}$, consist of tensor products of the individual qubits. The state $\ket{\psi} \to \left[ c_0, ~c_1, ~\cdots, ~c_{N-1} \right]^t$ possesses $N=2^n$ complex amplitudes, whose absolute squared values must sum up to one. Due to the principle of superposition, an $n$-qubit system is able to encode and process information scaling in $\mathcal{O}\left( 2^n \right)$, while for a classical setting, it is limited to $\mathcal{O}\left( n \right)$.

\paragraph{Evolution of Closed Quantum Systems}
In order for computation to be possible, there must be some method to manipulate quantum states. Exactly this is achieved by \emph{operators} acting on the Hilbert space $\mathcal{H}$. By definition, all operators, which describe the time evolution of a closed quantum system are reversible. Hence, they can be represented as unitary matrices, i.e.\@, for an operator $U$ it must hold that $U^{\dagger}U = I$. This constraint also conveys length preserving properties, i.e.\@, applying a unitary operator to a quantum state will again yield a valid quantum state satisfying \cref{eq:arbitrary_state}. 

In the following, explicit matrix representations of operators are specified in the computational basis. Starting simple, consider the bit-flip operator $\sigma_{x}$. This operator just flips the amplitudes of the $\ket{0}$ and $\ket{1}$ basis state, on the Bloch sphere this is equivalent to a rotation by $\pi$ about the $x$-axis. The corresponding operators also exist for $y$-axis and $z$-axis, in matrix notation those are given as

\begin{align} \label{eq:pauli_matrices}
	X := \sigma_x = \begin{bmatrix} 0 & 1 \\ 1 & 0 \end{bmatrix}, ~~Y := \sigma_y = \begin{bmatrix} 0 & -i \\ i & 0 \end{bmatrix}, ~~Z := \sigma_z= \begin{bmatrix} 1 & 0 \\ 0 & -1 \end{bmatrix}.
\end{align}

\begin{figure}[ht]
    \centering
    \includegraphics[width=0.4\textwidth]{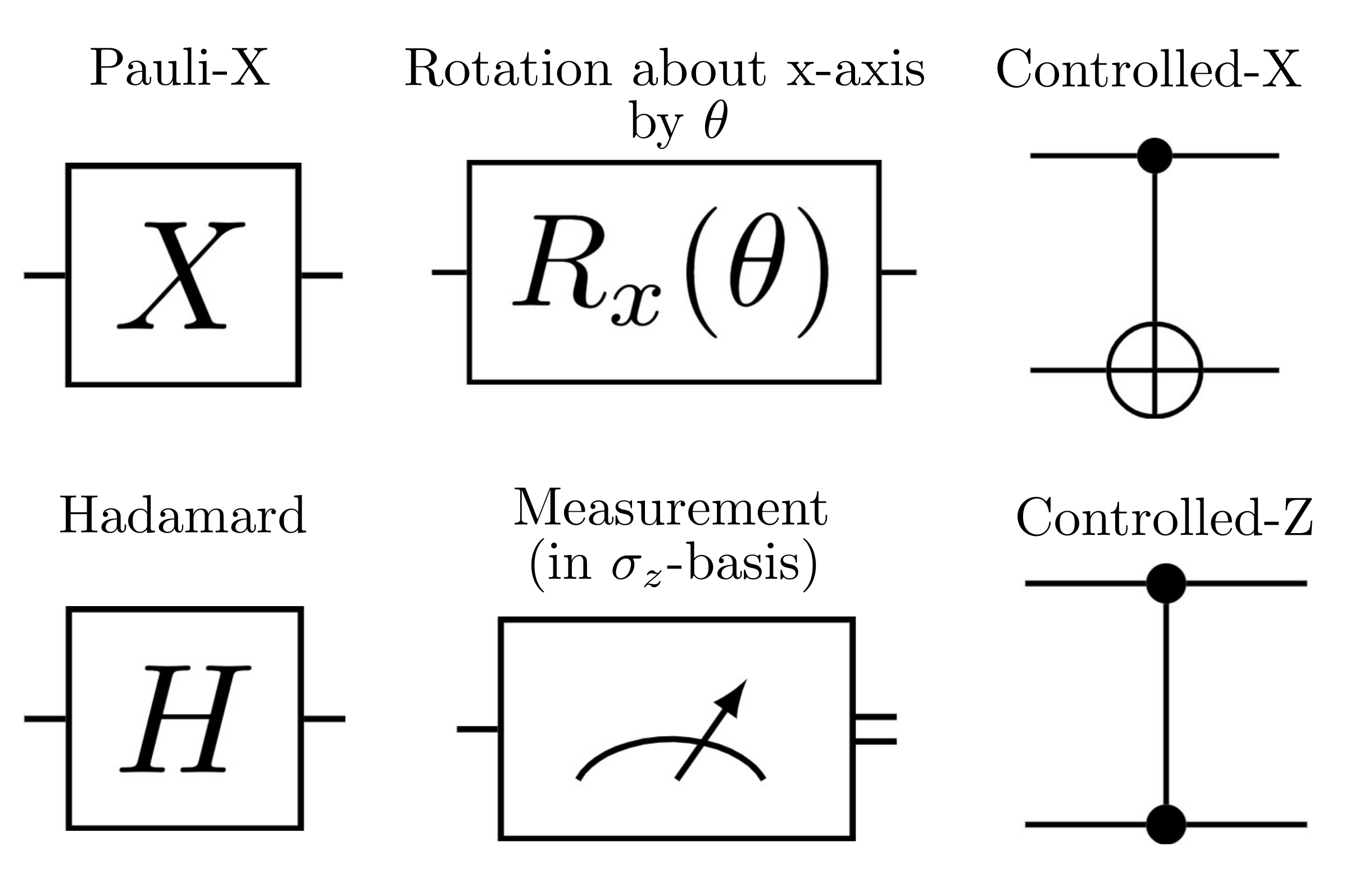}
    \caption{Circuit symbols of various quantum operators (gates).}
    \label{fig:qc_diagram}
\end{figure}

Allowing an additional degree of freedom, one can define an operator for arbitrary rotation with $\theta$ about axis $i$ as 
\begin{align} \label{eq:single_qubit_rotation}
	R_i(\theta) = e^{-i\frac{\theta}{2} \sigma_i}, ~~~\text{for } i \in \lbrace x, y, z \rbrace.
\end{align}
The last $1$-qubit operator we introduce is the \emph{Hadamard} matrix:
\begin{align}
	H = \frac{1}{\sqrt{2}} \begin{bmatrix} 1 & 1 \\ 1 & -1 \end{bmatrix},
\end{align}
which basically performs a change of basis with $H \ket{0} = \ket{+}$ and $H \ket{1} = \ket{-}$. By employing the tensor product for operators, we can extend $1$-qubit operators to act on single qubits comprising a multi-qubit system. We now move to genuine multi-qubit operators, acting non-trivially on two or more qubits. For our purposes, the most relevant $2$-qubit operators are the controlled $X$ ($CX$) and controlled $Z$ ($CZ$), where one qubit acts as the control and the other as the target. More concretely, the $CX$-gate flips the amplitudes of the target qubit, iff the control is in state $\ket{1}$. Similar to this, the $CZ$ operator performs a conditional phase flip. The matrix notations are given by

\begin{align}
	CX = \begin{bmatrix} 1 & 0 & 0 & 0 \\ 0 & 1 & 0 & 0 \\ 0 & 0 & 0 & 1 \\ 0 & 0 & 1 & 0 \end{bmatrix} \text{~~~ and ~~~} CZ = \begin{bmatrix} 1 & 0 & 0 & 0 \\ 0 & 1 & 0 & 0 \\ 0 & 0 & 1 & 0 \\ 0 & 0 & 0 & -1 \end{bmatrix}.
\end{align}

Quantum circuit diagrams are a nice way to visualize what is going on in a quantum algorithm. The individual qubits are represented as wires, where the order of operators, also called \emph{gates}, is defined by their relative position. To be more precise, the top wire gets associated with the leftmost qubit. A few common circuit symbols for the operators introduced so far are depicted in \cref{fig:qc_diagram}.

\paragraph{Extracting Classical Information via Measurements}
In classical computing, it is trivial to observe the exact states of all bits. For quantum systems, in order to extract information, an \emph{observable} quantity has to be measured. To build the bridge to quantum computing, for each physical observable there exists a Hermitian operator $O$ \cite{Nielsen_2016}, i.e., it holds $O^{\dagger} = O$. The eigenstates of $O$ define a basis of the quantum system's Hilbert space.

Once an observable $O$ is measured, the corresponding measurement device outputs an eigenvalue of $O$. The post-measurement state of the system is given by the eigenstate corresponding to the eigenvalue that is measured. The most commonly used observable might be Pauli-$Z$, which corresponds to a measurement in the computational basis for a single qubit, see also \cref{eq:pauli_matrices}. It has eigenvalues $\lambda_1 = +1$, $\lambda_2 = -1$ and corresponding eigenstates $v_1 = \left[ 1 ~~ 0 \right]^t$, $v_2 = \left[ 0 ~~ 1 \right]^t$.

The consequences for quantum computing are quite sobering, as observing superpositions w.r.t.\ the basis defined by the observable is impossible. Rather, one of the postulates of quantum mechanics states the \emph{Born rule}, which defines a probabilistic relationship between quantum state and measurement output. Let $\ket{0}, ~\ket{1}, ~..., ~\ket{N-1}$ be the basis defined by observable $O$ and $c_0, ~c_1, ~..., ~c_N$ the corresponding amplitudes of state $\ket{\psi}$ expressed in this basis. It holds, that measuring $O$ will result in the measurement outcome $\lambda_i$ with probability $\abs{c_i}^2$. Consequently, having obtained $\lambda_i$, the post-measurement state of the system is $\ket{i}$.

The first algorithm claiming provable \emph{quantum advantage}, i.e., an improvement w.r.t.\ some complexity metric compared to any classical approach, was published in 1992 by Deutsch and Josza \cite{Deutsch_1992} for a specially constructed problem. Most famous might be Shor's algorithm \cite{Shor_1997}, which provides an exponential speedup for tasks like prime factorization. Unfortunately, it requires large-scale, fault-tolerant and error-corrected quantum computers. All current hardware can be considered \gls{nisq} devices, which makes the execution of these algorithms infeasible. Despite this, the first claim of experimental quantum advantage was published just two years ago \cite{Arute_2019}. Yet, the considered problem was quite far from general practical applicability. A demonstration for achievable quantum supremacy on a practically relevant problem has still to be given. There are some promising candidates like quantum chemistry and material science. Recently, ideas have been put forward on combining quantum computing and machine learning \cite{Benedetti_2020,Schuld_2018}. These algorithms are expected to bypass at least some of the problems with execution on presently available \gls{nisq} hardware.

\paragraph{Quantum Machine Learning with Variational Quantum Circuits}

\begin{figure}[t]
    \centering
    \includegraphics[width=0.7\textwidth]{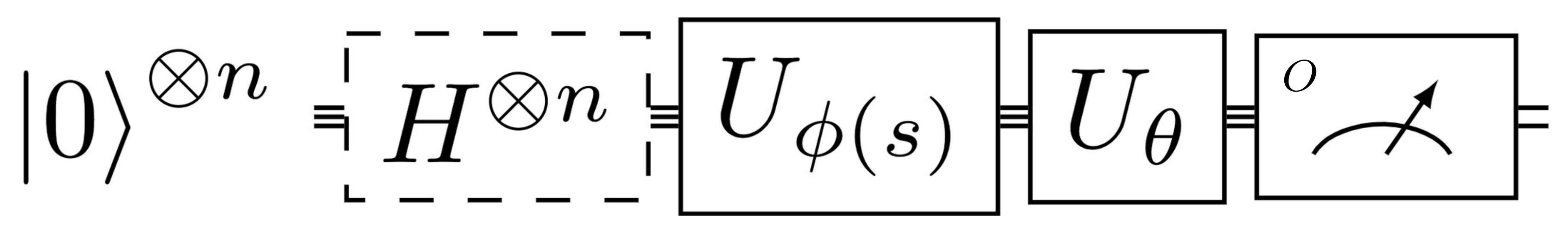}
    \caption{Variational quantum circuit consisting of feature map, variational layer, and measurement.}
    \label{fig:vqc_diagram}
\end{figure}

The research on \gls{qml} just really took off in the last two decades, yet there exists already a variety of approaches. As a rough clue, the hoped-for benefit of \gls{qml} relies, to a large extent, on the access to the high dimensional Hilbert space granted by quantum systems. Here, we want to briefly collect the background for the summaries of \gls{vqc}-based \gls{qrl} approaches in \cref{subsec:VQC_based}.

\Gls{qml} frequently deals with expectation values of quantum measurements. The expectation value of an observable $O$ w.r.t.\ the quantum state $\ket{\psi}$ is denoted as
\begin{align} \label{eq: expval_1}
	\expval{O}_{\psi} := \expval{\psi \left| O \right| \psi}.
\end{align}

While \glspl{vqc} define a new class of \gls{ml} models, one can make the case for the loose analogy to \glspl{nn}, where the relation of in- and output depends on a set of weights. An example for a parameterized quantum operator is given in \cref{eq:single_qubit_rotation}. The corresponding gate applies a rotation about a specific axis by some angle $\theta_0$. Multiple rotation gates form a quantum circuit, where $\boldsymbol{\theta}$ summarizes all free parameters. Varying these values gives the possibility to determine the evolution of the quantum system. Let  $U_{\theta}$ denote the corresponding unitary. An schematic example of a \gls{vqc} is displayed in \cref{fig:vqc_diagram}. Most \gls{rl} tasks use the concept of states, based on which an informed decision should be taken. This state information is encoded into the quantum system with an appropriate feature map. In general, the inputs $s$ are pre-processed with some mapping function $\Phi$. The results $\Phi(s)$ can be neatly integrated into the quantum circuit via the unitary $U_{\Phi(s)}$. To enhance the expressive power of the \gls{vqc}, one can use more sophisticated data encoding routines like \emph{data re-uploading}~\cite{Perez_2020} or \emph{incremental data-uploading}~\cite{Periyasamy_2022}. Eventually, some observable has to be measured. A common choice is the computational basis with $O = Z^{\otimes n}$. Overall, the output of the \gls{vqc}-model can be described as
\begin{equation}
	\begin{aligned}
		\expval{O}_{s,\theta} = & \expval{0 \left| \left( U_{\theta} U_{\Phi(s)} \right)^{\dagger} O U_{\theta} U_{\Phi(s)} \right| 0} \\
		:= & \expval{0 \left| U_{s,\theta}^{\dagger} O U_{s,\theta} \right| 0}.
	\end{aligned}
\end{equation}
For most tasks, this value is post-processed using some function $f$. Keeping things as general as possible, one can define a loss function $\mathcal{L}$ on $f \left( \expval{O}_{s,\theta} \right)$ (based on the concrete problem at hand). The update of the parameters can be performed using, e.g., gradient-based techniques:
\begin{align} \label{eq:grad_ascent_vqc}
	\theta \gets \theta + \alpha \cdot \nabla_{\theta} \mathcal{L}\left( f(\expval{O}_{s, \theta}) \right)
\end{align}
The required gradient can be obtained using the parameter-shift rule~\cite{Crooks_2019, Wierichs_2022}, or SPSA-based approximations~\cite{Wiedmann_2023}.

%------------------------------------------------------------
\section{Quantum Reinforcement Learning Algorithms}
\label{sec:QRLAlgs}
%------------------------------------------------------------

In \gls{qml}, there are approaches that either aim to stabilize the coherent function of the \gls{qpu} using \gls{ml} methods, or use the structure of a hybrid variational algorithm for \gls{ml} purposes. Very often, \gls{rl} is used to generate a solution for a quantum control problem, e.g., to learn quantum error correction strategies~\cite{Foesel_2018} or to generate control policies at a lower level~\cite{Zhang_2019,Dalgaard_2020}. Other work considers \gls{rl} as the optimizer of a \gls{vqa}~\cite{Khairy_2019, Khairy_2020}. While this represents a fascinating research topic in itself, here we will focus on the application of \gls{qrl} algorithms for solving specific tasks, be they classical or quantum. Research in the field of \gls{qml} has so far mostly focused on supervised and unsupervised learning. However, the literature already proposes quite a few theoretical concepts and even some small-scale experimental realizations for \gls{qrl}. Recent developments mostly focus on employing \glspl{vqc} as function approximators. When transferring from \gls{rl} to \gls{qrl}, i.e., the `quantization' of the \gls{rl} paradigm, there are various possibilities of how quantum computing enters the game. This has led to the development of different \gls{qrl} variants. A few works exist, that review current progress in \gls{qrl}~\cite{Kashyap_2021,MartinGuerrero_2022,Kunczik_2022,Lamata_2023,Neumann_2023} and the more general correspondence of \gls{rl} and \gls{qc}~\cite{Martin_2021}. There is also recent work towards a fair comparison of \gls{rl} and \gls{qrl} in restricted settings~\cite{Moll_2021,Franz_2022}.

\medbreak
\noindent
\textit{Quantum-Inspired Approaches.} The earliest idea for combining \gls{rl} with a quantum routine relies on the method of amplitude amplification, as it is used in Grover-type search algorithms~\cite{Chen_2006, Dong_2008,Dong_2006b,Chen_2006b,Dong_2006,Chen_2008,Dong_2008b,Chen_2010,Chunlin_2012,Fakhari_2013,Nuuman_2015,Li_2020a,Niraula_2021,Li_2020c,Yin_2021,Hu_2021,Ren_2022,Cho_2023}. Several qubit registers embed the states and actions relevant for the \gls{rl} system in a suitable Hilbert space. Starting from a uniform superposition, amplitudes favored by the reward or the value function are selectively amplified. The action selection is based on Born's rule, i.e., a measurement is carried out on the qubit register with regard to the `action-basis'. The algorithm was also investigated independently of \glspl{qpu}~\cite{Dong_2012} and recently further developed~\cite{Ganger_2019}. An introduction to this concept is also provided in Ref.~\cite{Rajagopal_2021}. As it turns out, these early variants should rather be considered a set of \gls{qirl} algorithms, that do not offer an intrinsic potential for quantum advantage. Recently, the technique was transferred to sampling from the experience replay buffer in $Q$-learning~\cite{Wei_2021}. A summary and review of this type of \gls{qirl} can be found in \cref{subsec:quantum_inspired}.

\medbreak
\noindent
\textit{\gls{vqc}-Based Function Approximation.} In \gls{drl}, \glspl{dnn} are employed as powerful function approximators. Typically, the approximation either happens in policy space (actor), in value space (critic), or both, resulting in so-called actor-critic approaches. Recently, \glspl{vqc} were proposed and analyzed in their role as function approximators in the \gls{rl} setting -- an extensive overview is provided in \cref{subsec:VQC_based}. On the one hand, this approach basically replaces a more or less well understood heuristic with a poorly understood heuristic. For the quantum heuristic many open questions regarding computational power, scalability and trainability remain. On the other hand, \glspl{vqc} nonetheless have spurred the hope for quantum advantage already with \gls{nisq} devices. 
The earliest work in this direction proposed \gls{vqc}-based approximation in value space, which is covered in \cref{subsec:VQC_based_ValueFunction}. This so-called \gls{vqc}-based $Q$-learning was introduced in Ref.~\cite{Chen_2020}, and extended in Refs.~\cite{Lockwood_2020,Lockwood_2021,Lokes_2022,Chen_2023f,Chen_2023e,FikaduTilaye_2023,Skolik_2022,Skolik_2023,LiuY_2023}. A method to efficiently evaluate the $Q$-function is discussed in Ref.~\cite{Sannia_2023}, which is however not entirely \gls{nisq}-feasible. The complimentary approach of approximation in policy space is discussed in \cref{subsec:VQC_based_Policy}. Originally proposed in Ref.~\cite{Jerbi_2021a}, several extensions have bee discussed in Refs.~\cite{Kunczik_2022,Quafu_2023,Sequeira_2023,Jerbi_2023,Meyer_2021,Meyer_2023a,Meyer_2023b}. Combinations of value and policy approximation are covered in \cref{subsec:VQC_based_CombinationApproximations}, with (soft) actor-critic approaches in Refs.~\cite{Wu_2023,Kwak_2021,Reers_2023,Chen_2023c,Lan_2021}, and multi-agent formulations in Refs.~\cite{Yun_2022,Yun_2023a}. The setting of offline quantum reinforcement learning is considered in \cref{subsec:VQC_based_Offline} by Refs.\cite{Periyasamy_2023,Cheng_2023a}. A collection of algorithmic and conceptual extensions that are relevant for a wide range of approaches is composed in \cref{subsec:VQC_based_Extensions}, based on Refs.~\cite{Chen_2023a,Chen_2023b,Kimura_2021,Hsiao_2022,Dragan_2022,Kruse_2023,Sun_2023,Andres_2023,Park_2020,Chen_2022a,Ding_2023,Koelle_2023}. A collection of application-focused work is summarized in~\cref{subsec:VQC_based_Application}, comprising Refs.~\cite{Acuto_2022,Heimann_2022,Cobussen_2023,Bar_2022,Sinha_2023,Hickmann_2023,Kim_2023,Correll_2023,Sanches_2022,Andres_2022,Liu_2023,Kumar_2023,Rainjonneau_2023,Shahid_2023,Rezazadeh_2022,Yan_2022,Park_2023b,Narottama_2023,Park_2023a,Park_2023c,Yun_2023b,Ansere_2023,Cherrat_2023b,Yang_2023,Chao_2023,Chen_2023d}.

\medbreak
\noindent
\textit{Projective Simulation.} Another \gls{qrl} method is based on \gls{ps}, which in the broadest sense is a particular learning paradigm and similar in spirit to \gls{rl}~\cite{Briegel_2012}. Based on experiences made through interaction with the environment, a memory network is created by the agent. The network has a directed structure with adaptive weights between the nodes of the network. The learning process and action selection are based on a random process (more precisely, a random walk) on the graph of the network, with the transition probabilities between nodes being given by the respective adaptive weights. \gls{ps} can be `quantized' by replacing the random walk with a so-called quantum random walk ~\cite{Paparo_2014,Teixeira_2021a,Teixeira_2021b,Melnikov_2017}. A formal analysis of convergence properties was given in Ref.~\cite{Boyajian_2020}. In fact, there is already work on a proof-of-principle implementation in the laboratory~\cite{Dunjko_2015a,Sriarunothai_2018} and proposals for quantum-optics implementations \cite{Flamini_2023}. Possible quantum advantages over classical \gls{ps} lie in the acceleration of the process of action selection, also referred to as deliberation in the literature. A more detailed summary is provided in \cref{subsec:projective_simulaton}.

\medbreak
\noindent
\textit{Quantum Boltzmann Machines.} Another line of research proposes to use Boltzmann machines as function approximators. These models are assumed to be advantageous compared to typical \glspl{nn} in environments with large action spaces. Ref.~\cite{Jerbi_2021b} demonstrates, that Boltzmann machines are closely related to energy-based models. For specific instances, those allow for a quantum representation, which enables potential quantum speed-up for post-\gls{nisq} devices. A similar concept is also proposed for the annealing-based \gls{qc} paradigm \cite{Crawford_2018,Schenk_2022, Levit_2017}. A summary of these ideas can be found in \cref{subsec:boltzman_machines}.

\medbreak
\noindent
\textit{Quantum Subroutines.} Another approach to go from \gls{rl} to \gls{qrl} replaces certain subroutines in existing \gls{rl} approaches. One idea is to replace policy or value iteration with some quantum-enhanced analogues. While this approach is limited to universal, fault-tolerant and error-corrected quantum hardware, several such algorithms have been proposed and analyzed~\cite{Wiedemann_2021,Wiedemann_2022, Wang_2021a,Cherrat_2023a,Ganguly_2023a,Zhong_2023,Ganguly_2023b}. Most importantly, these algorithms come with guarantees regarding speed-up, compared to their classical counterparts. \Gls{qrl} in these settings is often limited to the tabular case and assumes a quantum version of the \gls{rl} environment, i.e., oracle access. Our summaries and reviews can be found in \cref{subsec:QPI}.

\medbreak
\noindent
\textit{Full-Quantum Formulation.} An approach which not only `quantizes' certain subroutines, but all components of the pipeline, is considered in Refs.~\cite{Dunjko_2015b, Dunjko_2016, Dunjko_2017,Dunjko_2018}. Extensions~\cite{Hamann_2021,Hamann_2022}, applied to specific problems~\cite{Wang_2021b,Wan_2023}, and small-scale experimental realizations~\cite{Saggio_2021a} were presented. An alternative route to fully quantized \gls{qrl} was taken in \cite{Cornelissen_2018}. For our review of this line of research, see \cref{subsec:Oracles}.

\medbreak
\noindent
\textit{Various Concepts.} For the sake of completeness, we mention different approaches found in the literature. We note, however, that we did not pursue a detailed review for those works, typically because we focused on what we identified as the most considered lines of research. While some of the works listed in the following simply do not fit directly with the learning-based \gls{qrl} approach, for others it might not seem obvious how to generalize their particular setting to a broader class of problems. While quantum algorithms for dynamic programming have been discussed~\cite{Ronagh_2019, Ambainis_2019}, it currently remains unclear how to move from dynamic programming to a learning-based approach such as \gls{rl}. Similarly, quantum algorithms have been employed to solve planning tasks~\cite{Naguleswaran_2005}, but again the transfer to a learning-based approach is far from obvious. Closer related to the typical \gls{rl} setting is the task of imitation learning~\cite{Cheng_2023b}. A series of papers discussed \gls{qrl} in the setting of photonic circuits, see and Refs.~\cite{Hu_2019a,Hu_2019b,Hu_2019c,Hu_2019d,Sorensen_2020} and Refs.~\cite{Flamini_2020,Lamata_2021,Saggio_2021b,Nagy_2021,Shinkawa_2022}, with the connection to superconducting qubits established in Ref.~\cite{Lamata_2017,Cardenas_2018}. Another approach, which we did not review in detail, is given by combining \gls{rl} with the paradigm of quantum annealing~\cite{Neukart_2017,Ayanzadeh_2020,Neumann_2020,Muller_2021,FernandezVillaverde_2023,Nuzhin_2023}. Strategies have been developed to address the classical and quantum version of contextual bandits~\cite{Lumbreras_2022,Liu_2022,Brahmachari_2023,Buchholz_2023}. Furthermore, a quantum version of the classical \gls{rl} benchmark environment \textit{CartPole} has been formulated~\cite{Wang_2020,Meinerz_2023}. Similarly, various interpretations of \gls{qrl} for specialized tasks in the quantum domain exist~\cite{Alvarez_2016,Alvarez_2018,Bharti_2019,Albarran_2018,Albarran_2020,Shenoy_2020,Olivares_2020,LiuW_2022,Caglar_2023}. Different approaches have been proposed for combining \gls{rl} with quantum walks~\cite{Chen_2019,Dalla_2022,Mullor_2022}. Further work on optimization tasks rather than \gls{rl}, such as Ref.~\cite{Ramezanpour_2017,Javsek_2019,Beloborodov_2020}, have not been reviewed in detail. An interesting interpretation of self-learning physical machines is discussed in~\cite{Lopez_2023}, which potentially could be brought into line with \gls{qrl}.

%------------------------------------------------------------
\subsection{Quantum-Inspired Reinforcement Learning based on Amplitude Amplification}
\label{subsec:quantum_inspired}
%------------------------------------------------------------

\begin{table}[t!]
    \centering
    \begin{tabular}{p{\dimexpr 0.15\textwidth-2\tabcolsep-\arrayrulewidth}|p{\dimexpr 0.2\textwidth-2\tabcolsep-\arrayrulewidth}|p{\dimexpr 0.65\textwidth-2\tabcolsep}}
        \toprule
        \textbf{Citation} & \textbf{First Author} & \textbf{Title} \\
        \midrule
        \midrule
        \cite{Dong_2008} & D. Dong & \hyperref[subsubsec:Dong_2008]{Quantum reinforcement learning} \\
        \arrayrulecolor{black!30}\midrule
        \cite{Dong_2006} & D. Dong & \hyperref[subsubsec:Dong_2008]{Quantum mechanics helps in learning for more intelligent robots} \\
        \midrule
        \cite{Chen_2006} & C.-L. Chen & \hyperref[subsubsec:Dong_2008]{Quantum computation for action selection using reinforcement learning} \\
        \midrule
        \cite{Dong_2006b} & D. Dong & \hyperref[subsubsec:Dong_2008]{Quantum Robot: Structure, Algorithms and Applications } \\
        \midrule
        \cite{Chen_2006b} & C.-L. Chen & \hyperref[subsubsec:Dong_2008]{Superposition-Inspired Reinforcement Learning and Quantum Reinforcement Learning} \\
        \midrule
        \cite{Chen_2008} & C.-L. Chen & \hyperref[subsubsec:Dong_2008]{A Quantum Reinforcement Learning Method for Repeated Game Theory} \\
        \midrule
        \cite{Dong_2008b} & D. Dong & \hyperref[subsubsec:Dong_2008]{Incoherent Control of Quantum Systems With Wavefunction-Controllable Subspaces via Quantum Reinforcement Learning} \\
        \midrule
        \cite{Chen_2010} & C.-L. Chen & \hyperref[subsubsec:Dong_2008]{Complexity analysis of Quantum reinforcement learning} \\
        \midrule
        \cite{Dong_2012} & D. Dong & \hyperref[subsubsec:Dong_2008]{Robust Quantum-Inspired Reinforcement Learning for Robot Navigation} \\
        \midrule
        \cite{Chunlin_2012} & C. Chunlin & \hyperref[subsubsec:Dong_2008]{Hybrid control of uncertain quantum systems via fuzzy estimation and quantum reinforcement learning} \\
        \midrule
        \cite{Fakhari_2013} & P. Fakhari & \hyperref[subsubsec:Dong_2008]{Quantum inspired reinforcement learning in changing environment} \\
        \midrule
        \cite{Nuuman_2015} & S. Nuuman & \hyperref[subsubsec:Dong_2008]{A quantum inspired reinforcement learning technique for beyond next generation wireless networks} \\
        \arrayrulecolor{black}\bottomrule
    \end{tabular}
    \caption{[Part 1] Work considered for ``\gls{qirl} based on amplitude Amplification'' (\cref{subsec:quantum_inspired})}
\end{table}

\begin{table}[t!]
    \centering
    \begin{tabular}{p{\dimexpr 0.15\textwidth-2\tabcolsep-\arrayrulewidth}|p{\dimexpr 0.2\textwidth-2\tabcolsep-\arrayrulewidth}|p{\dimexpr 0.65\textwidth-2\tabcolsep}}
        \toprule
        \textbf{Citation} & \textbf{First Author} & \textbf{Title} \\
        \midrule
        \midrule
        \arrayrulecolor{black!30}
        \cite{Ganger_2019} & M. Ganger & \hyperref[subsubsec:Dong_2008]{Quantum Multiple Q-Learning} \\
        \midrule
        \cite{Li_2020a} & J.-A. Li & \hyperref[subsubsec:Dong_2008]{Quantum reinforcement learning during human decision-making} \\
        \midrule
        \cite{Li_2020c} & Y. Li & \hyperref[subsubsec:Dong_2008]{Intelligent Trajectory Planning in UAV-Mounted Wireless Networks: A Quantum-Inspired Reinforcement Learning Perspective} \\
        \midrule
        \cite{Rajagopal_2021} & K. Rajagopal & \hyperref[subsubsec:Dong_2008]{Quantum Amplitude Amplification for Reinforcement Learning} \\
        \midrule
        \cite{Niraula_2021} & D. Niraula & \hyperref[subsubsec:Dong_2008]{Quantum deep reinforcement learning for clinical decision support in oncology: application to adaptive radiotherapy} \\
        \midrule
        \cite{Wei_2021} & Q. Wei & \hyperref[subsubsec:Dong_2008]{Deep Reinforcement Learning With Quantum-Inspired Experience Replay} \\
        \midrule
        \cite{Yin_2021} & L. Yin & \hyperref[subsubsec:Dong_2008]{Quantum deep reinforcement learning for rotor side converter control of double-fed induction generator-based wind turbines} \\
        \midrule
        \cite{Hu_2021} & Y. Hu & \hyperref[subsubsec:Dong_2008]{Quantum‑enhanced reinforcement learning for control: a preliminary study} \\
        \midrule
        \cite{Ren_2022} & Y. Ren & \hyperref[subsubsec:Dong_2008]{NFT-Based Intelligence Networking for Connected and Autonomous Vehicles: A Quantum Reinforcement Learning Approach} \\
        \midrule
        \cite{Cho_2023} & B. Cho & \hyperref[subsubsec:Dong_2008]{Quantum bandit with amplitude amplification exploration in an adversarial environment} \\
        \arrayrulecolor{black}\bottomrule
    \end{tabular}
    \caption{[Part 2] Work considered for ``\gls{qirl} based on amplitude Amplification'' (\cref{subsec:quantum_inspired})}
\end{table}

%------------------------------------------------------------
\paragraph{\label{subsubsec:Dong_2008}Quantum reinforcement learning, Dong et al.~(2008) and related work}\mbox{}\\

%------------------------------------------------------------

\vspace{-1em}
\noindent\textit{Summary.} Ref.~\cite{Dong_2008} discusses a new \gls{rl} algorithm that is inspired by the superposition principle of quantum mechanics. The authors propose an algorithm that modifies the action-selection procedure and balances exploration and exploitation in a novel way. The authors present their ideas in modified form in a sequence of papers, see Refs.~\cite{Dong_2008,Dong_2006,Dong_2006b,Chen_2006b,Chen_2008,Chen_2006,Dong_2008b,Chen_2010,Dong_2012,Chunlin_2012,Fakhari_2013,Nuuman_2015,Ganger_2019,Li_2020a,Li_2020c,Hu_2021}, for an overview see also~\cite{Rajagopal_2021}. The original work~\cite{Dong_2008} discusses how to execute the proposed algorithm on actual quantum devices -- which, however, did not exist at this time. As discussed also below, it is not clear how to \emph{run the algorithm in quantum superposition}, and if this is possible in practice without taking away potential quantum advantage. Despite these doubts the proposed concepts enhance classical \gls{rl} with ideas from \gls{qc}, which leads us to view this approach as \gls{qirl}.
\medbreak
\noindent
\textit{Algorithmic Concepts and Extensions.} Initially, the algorithm is formulated as merely quantum inspired in Ref.~\cite{Chen_2008} (i.e., it is developed for a classical computer that simulates a quantum superposition). The motivation is to design an algorithm with better exploration-exploitation trade-off compared to e.g.\ $\epsilon$-greedy action selection. The underlying routine is a modification of \gls{td}, more concretely \gls{td}(0) in the following way: For each state the set of possible actions is in a ‘superposition’ and the agent (in state $s$) now selects an action with a given probability. The action is taken and the new state $s^\prime$ and reward $r$ is observed. Afterwards, the probability of the taken action is increased by $k(r+V(s^\prime))$, where $V(s^\prime)$ is the value function of state $s^\prime$, and $k$ is a hyperparameter. The term $r+V(s^\prime)$ samples a quantity similar to $Q(s,a)$. Consequently, the update creates a probability distribution, where for a given state the probability to select an action increases as the value of $Q(s,a)$ increases. Therefore, this action selection process corresponds to sampling from a stochastic policy dependent on the value of the state-action pairs.

Now the algorithm is translated to be run on a quantum computer. The stochastic policy is replaced by a quantum superposition. That is, for each state $s$ the possible actions are represented by the eigenstates of some observable and a superposition of these states is created. If the observable is measured, the state will collapse to an eigenstate associated with an action which will be taken by the agent and therefore constitutes the selection process. After receiving the reward and the new state, the Grover operator is applied $L=\mathrm{min}\{k(r+V(s^\prime)), L_\mathrm{max}\}$ times to a copy of the superposition state to enhance the amplitude corresponding to the previous selected action. The variable $L_\mathrm{max}$ guarantees that the Grover operator is not applied too many times. Note that repeated application of the procedure requires a new copy of the state after each measurement. Due to the no cloning theorem, this could be realized by many different independent copies of the initial memory, or by a purely classical representation of the states. The latter realization reduces the algorithm to the initial proposal of a quantum-inspired action selection process.

In Ref.~\cite{Dong_2012} the \gls{qirl} algorithm is applied to robot navigation. It is stated explicitly that \gls{qirl} is a classical action-selection method that differs from the ideas of \gls{qrl}, which in principle could benefit from a quantum computer. In Ref.~\cite{Ganger_2019} the algorithms are generalized to $Q$-learning and double- and multiple $Q$-learning. Also these approaches should be understood in the context of \gls{qirl}. Finally, Refs.~\cite{Li_2020a, Niraula_2021} apply \gls{qirl} to human decision making behavior, Ref.~\cite{Yin_2021} to a complex control task, and Ref.~\cite{Ren_2022} to autonomous vehicles. Recently, the quantum-inspired approach to action selection in \gls{rl} was transferred to experience replay buffer sampling in $Q$-learning \cite{Wei_2021}.
\medbreak
\noindent
\textit{Remarks.} Although it is mentioned in Ref.~\cite{Dong_2008,Dong_2006,Chen_2008,Chen_2006} that the whole algorithm could be run in quantum superposition on a quantum device, no details of such kind of genuine \gls{qrl} algorithm are given. Overall, it is unclear if such an algorithm might exist. Indeed, subsequent work focuses on the \gls{qirl} paradigm.

The claims made in Refs.~\cite{Dong_2008,Dong_2006,Dong_2006b,Chen_2008,Chen_2006,Dong_2008b,Dong_2012,Ganger_2019,Li_2020a,Li_2020c,Cho_2023} can be summarized as follows: speed-up in learning by better balancing exploration-exploitation; less GPU power needed on classical computer compared to algorithms like classical $Q$-learning; more robust against changes of learning rate. More experiments on larger environments for deeper insights into the scaling of the algorithm and a rigorous complexity analysis would be an interesting topic for future work.

%------------------------------------------------------------
\subsection{Quantum Reinforcement Learning with Variational Quantum Circuits}
\label{subsec:VQC_based}
%------------------------------------------------------------

This section summarizes the state-of-the-art on \gls{vqc}-based \gls{rl}. Several ideas have been proposed in this field, with extensions in different directions. Their common ground is the usage of a \gls{vqc} as parameterized function approximator.

\begin{figure}[ht]
    \centering
    \includegraphics[width=0.85\textwidth]{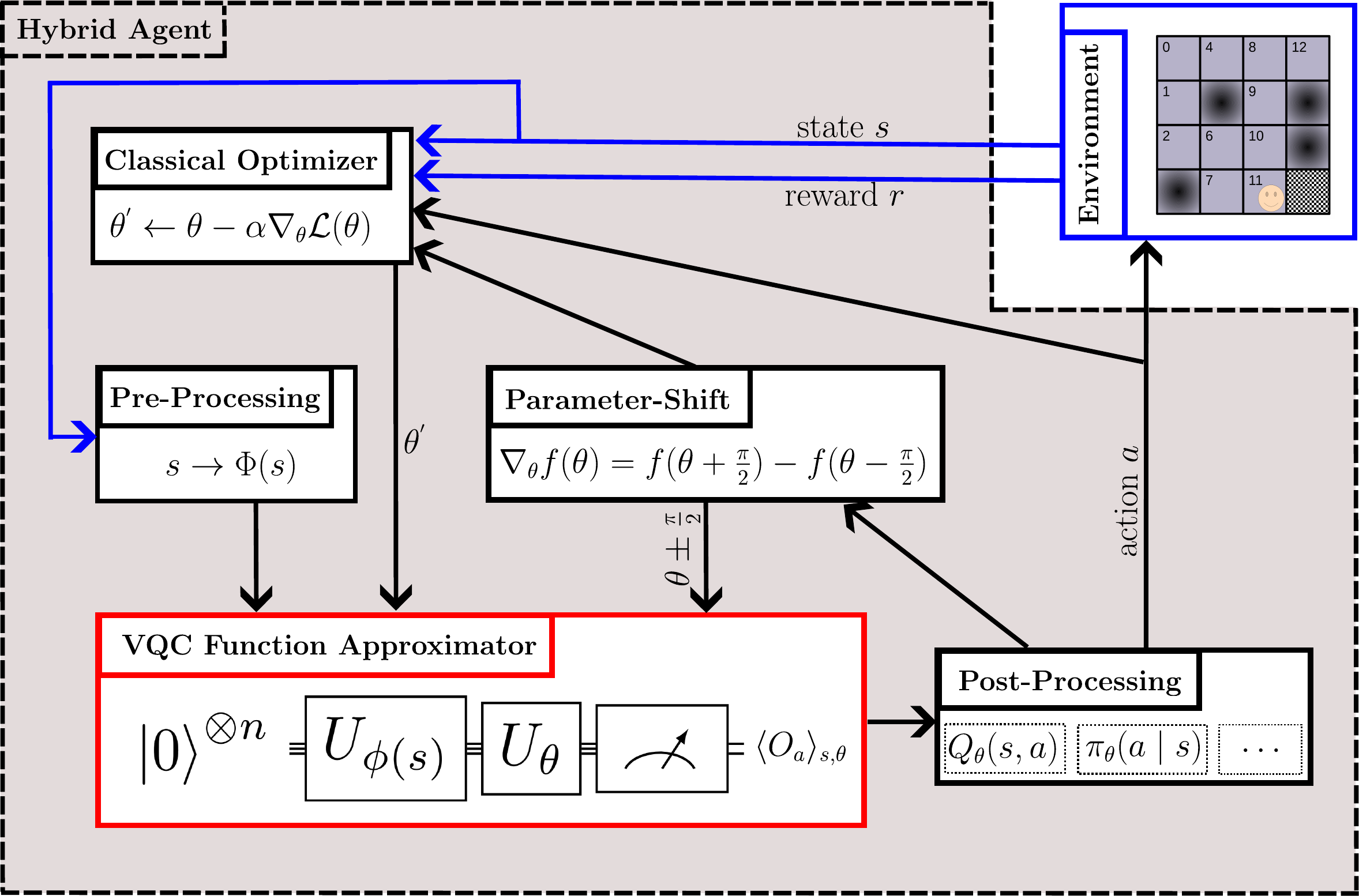}
    \caption{Hybrid quantum-classical agent in a typical \gls{vqc}-based \gls{rl} pipeline. This idea was first proposed by Chen et al.~\cite{Chen_2020} for $Q$-function approximation and extended by Jerbi et al.~\cite{Jerbi_2021a} to policy approximation. The \gls{qpu} is used to approximate the respective function, while pre- and post-processing and optimization happens on classical hardware. The interaction with the environment depends on the concrete problem instance (e.g.\ classical or quantum environment).}
    \label{fig:VQC_based_pipeline}
\end{figure}

The typical hybrid pipeline is summarized in \cref{fig:VQC_based_pipeline}. It was originally proposed for $Q$-function approximation by Chen et al.~\cite{Chen_2020} and extended to policy approximation by Jerbi et al.~\cite{Jerbi_2021a}. Other work proposes several modifications to this pipeline, which we will describe in the respective summaries. The algorithm must be understood as hybrid, as a lot of the work, especially the optimization, is executed on classical hardware. The agent observes the current state of the environment $s_t$, and applies some pre-processing $\phi$. The result is encoded using the feature map $U_{\phi(s)}$. With the current variational parameters $\theta_t$, a quantum state is prepared and a (potentially action-dependent) observable $O_a$ is measured. The expectation value $\expval{O_a}_{s,\theta}$ can be post-processed to represent, e.g., a state-action value function $Q_{\theta}(s,a)$, or the policy $\pi_{\theta}(a | s)$. Depending on the instance, the agent employs this function to sample an action $a_t$ and executes it in the environment. The reward $r_t$ (and potentially also the consecutive state $s_{t+1}$) is observed by the classical optimizer. To enable gradient-based parameter updates, an additional hybrid module uses the parameter-shift rule~\cite{Crooks_2019, Wierichs_2022} to compute the gradients of the \gls{vqc} outputs w.r.t.\ the variational parameters $\theta_t$. The classical optimizer determines the new parameter set $\theta_{t+1}$ and instantiates the \gls{vqc} with these updated parameters. This overall iterative procedure of environment interaction, function approximation, and parameter update is repeated for several episodes, in the same way as for, e.g., \gls{drl}.

Unfortunately, thus far there is no guaranteed quantum advantage for this approach, apart from some cryptography inspired artificial datasets~\cite{Jerbi_2021a,Skolik_2022}. However, several of the papers and preprints summarized in this section demonstrate promising experimental results.

\subsubsection{Value-Function Approximation}
\label{subsec:VQC_based_ValueFunction}
This section covers \gls{vqc}-based approximations in value space, as described for the instance of classical $Q$-learning in \cref{eq:Qlearning_1,eq:Qlearning_2}. The work by Chen et al.~\cite{Chen_2020} was indeed the first proposal of this type of approximation-based techniques, which was reproduced and extended in Refs.~\cite{Lokes_2022,Chen_2023e,Chen_2023f,FikaduTilaye_2023}. A modification of the state encoding procedure has been discussed in Lockwood and Si~\cite{Lockwood_2020}, and was up-scaled in another work by the same authors~\cite{Lockwood_2021}. A slight reformulation of the technique -- which comes with a provable advantage for very specific scenarios -- can be found in Skolik et al.~\cite{Skolik_2022}. An analysis of noise influence for this framework is discussed in Ref.~\cite{Skolik_2023}. An extension to environments with continuous action spaces is proposed in Ref.~\cite{LiuY_2023}. Ideas based on amplitude amplification to efficiently evaluate the approximated $Q$-function have been introduced in Ref.~\cite{Sannia_2023}, which however can not be realized given the current hardware restrictions.

\begin{table}[t!]
    \centering
    \begin{tabular}{p{\dimexpr 0.15\textwidth-2\tabcolsep-\arrayrulewidth}|p{\dimexpr 0.2\textwidth-2\tabcolsep-\arrayrulewidth}|p{\dimexpr 0.65\textwidth-2\tabcolsep}}
        \toprule
        \textbf{Citation} & \textbf{First Author} & \textbf{Title} \\
        \midrule
        \midrule
        \cite{Chen_2020} & S. Y.-C. Chen & \hyperref[subsubsec:Chen_2020]{Variational Quantum Circuits for Deep Reinforcement Learning} \\
        \arrayrulecolor{black!30}\midrule
        \cite{Lokes_2022} & S. Lokes & \hyperref[subsubsec:Chen_2020]{Implementation of Quantum Deep Reinforcement Learning Using Variational Quantum Circuits} \\
        \midrule
        \cite{Chen_2023f} & S. Y.-C. Chen & \hyperref[subsubsec:Chen_2020]{Quantum deep Q learning with distributed prioritized experience replay} \\
        \midrule
        \cite{Chen_2023e} & H.-Y. Chen & \hyperref[subsubsec:Chen_2020]{Deep-Q Learning with Hybrid Quantum Neural Network on Solving Maze Problems} \\
        \arrayrulecolor{black!30}\midrule
        \cite{FikaduTilaye_2023} & G. Fikadu Tilaye & \hyperref[subsubsec:Chen_2020]{Investigating the Effects of Hyperparameters in Quantum-Enhanced Deep Reinforcement Learning} \\
        \arrayrulecolor{black}\midrule
        \cite{Lockwood_2020} & O. Lockwood & \hyperref[subsubsec:Lockwood_2020]{Reinforcement Learning with Quantum Variational Circuits} \\
        \midrule
        \cite{Lockwood_2021} & O. Lockwood & \hyperref[subsubsec:Lockwood_2021]{Playing Atari with Hybrid Quantum-Classical Reinforcement Learning} \\
        \midrule
        \cite{Skolik_2022} & A. Skolik & \hyperref[subsubsec:Skolik_2022]{Quantum agents in the Gym: a variational quantum algorithm for deep $Q$-learning} \\
        \arrayrulecolor{black!30}\midrule
        \cite{Skolik_2023} & A. Skolik & \hyperref[subsubsec:Skolik_2022]{Robustness of quantum reinforcement learning under hardware errors} \\
        \arrayrulecolor{black!30}\midrule
        \cite{LiuY_2023} & Y. Liu & \hyperref[subsubsec:Skolik_2022]{Reinforcement Learning for Continuous Control: A Quantum Normalized Advantage Function Approach} \\
        \arrayrulecolor{black}\bottomrule
    \end{tabular}
    \caption{Work considered for ``\gls{qrl} with \glspl{vqc} -- Value-Function Approximation'' (\cref{subsec:VQC_based_ValueFunction})}
\end{table}

%------------------------------------------------------------
\paragraph{\label{subsubsec:Chen_2020}Variational Quantum Circuits for Deep Reinforcement Learning, Chen et al.~(2020) and related work}\mbox{}\\
% \cite{Chen_2020}
%------------------------------------------------------------
    
    \vspace{-1em}
    \noindent\textit{Summary.} This paper by Chen et al.~\cite{Chen_2020} represents the first attempt to utilize \glspl{vqc} for \gls{rl}. This is done in the context of using \glspl{vqc} as function approximators for the state-action value function. The authors perform simulations on simple benchmark environments and report.
    
    \medbreak
    
\begin{figure}[ht]
        \centering
        \includegraphics[width=0.95\textwidth]{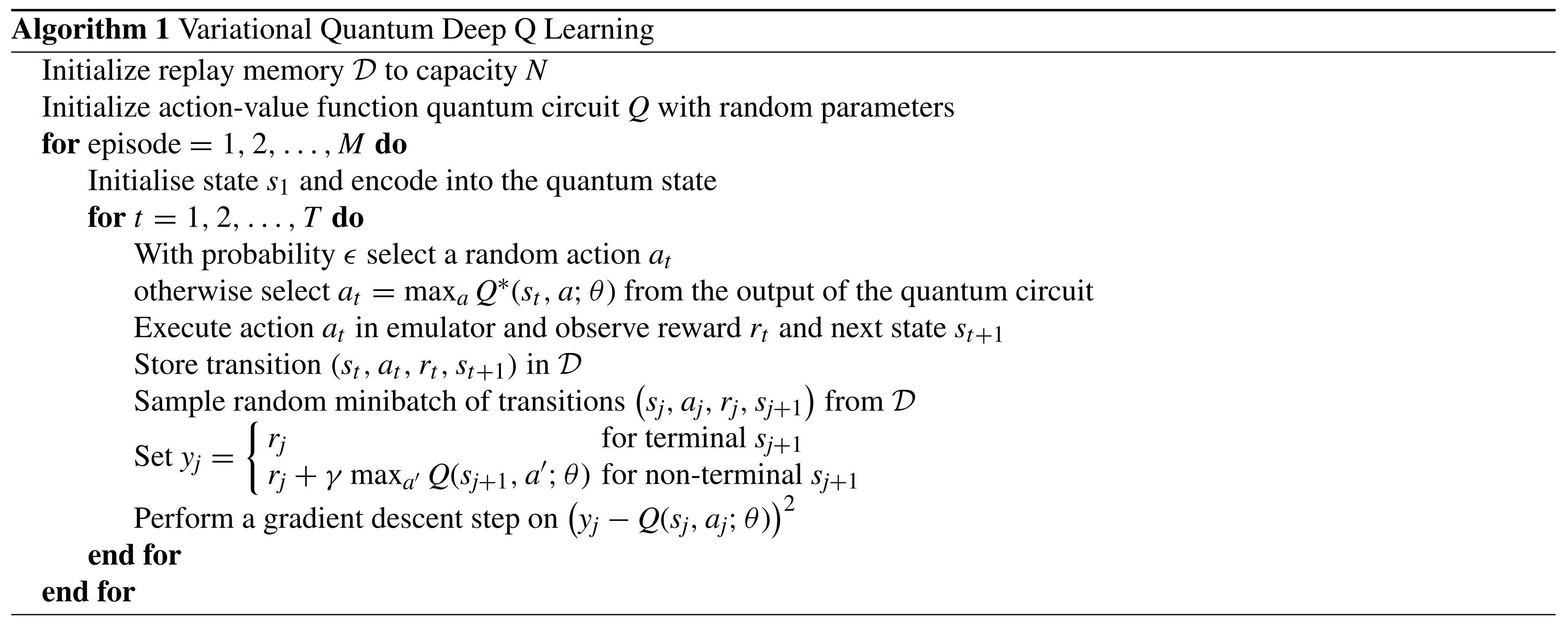}
        \caption{Hybrid algorithm proposed by and taken from Chen et al.~\cite{Chen_2020}; This algorithm uses a \gls{vqc} to approximate the state-action value function and follows the typical steps of \gls{dql}. Note, that the authors notation for the $Q$-function slightly deviates from our conventions.}
        \label{fig:algorithm_Chen_2020}
    \end{figure}

    \noindent\textit{Hybrid Algorithm.} The algorithm is inspired by \gls{dql}~\cite{Mnih_2015}, where a \gls{dnn} represents the $Q$-function. The authors replace the \gls{dnn} by a \gls{vqc}. The update is performed w.r.t.\ the \gls{mse} loss function $\mathcal{L}(\theta) = \mathbb{E} [ \left( r_t + \gamma \cdot \mathrm{max}_{a'} ~ Q_{\theta^{'}}(s_{t+1}, a') - Q_{\theta}(s_t, a_t) \right)^2 ]$ using, e.g., gradient descent. Additionally, experience replay and target networks (second set of parameters $\theta^{'}$) are employed to address the instabilities stemming from bootstrapping the value function, forming a \gls{ddql} algorithm. \cref{fig:algorithm_Chen_2020} gives the complete algorithm.

    \medbreak
    
\begin{figure}[ht]
        \centering
        \includegraphics[width=0.7\textwidth]{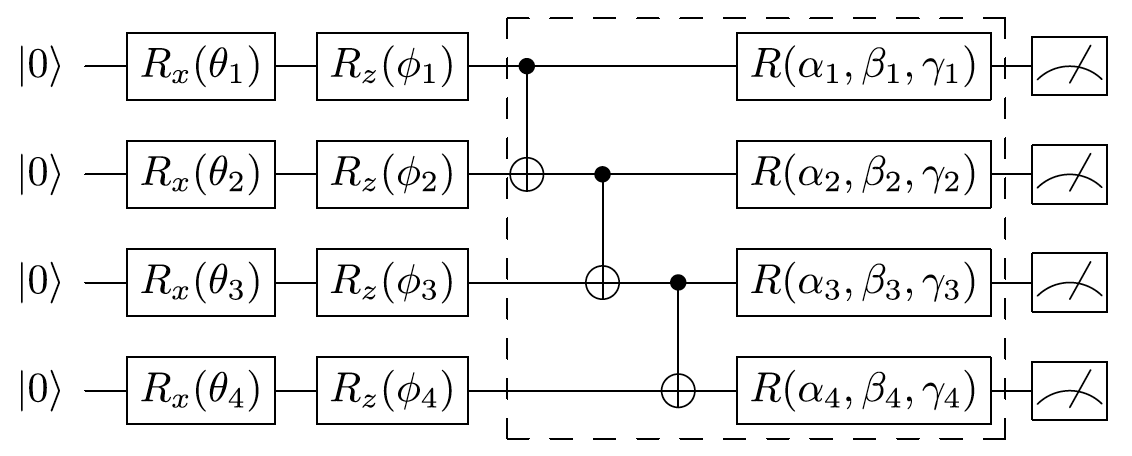}
        \caption{\gls{vqc} proposed by and taken from Chen et al.~\cite{Chen_2020}; The $R_x$ and $R_z$ gates are used for state encoding. Several parameterized layers (dashed box) are repeated to form the $Q$-function approximator. The values of the function are decoded using $1$-qubit Pauli-$Z$ observables.}
        \label{fig:vqc_Chen_2020}
    \end{figure}

    \noindent\textit{\gls{vqc} Architecture.} The feature map uses simple computational basis encoding on individual qubits. More concretely, the \gls{rl} state is interpreted as bitstring, which can be encoded using the identity $R_z(\pi)R_x(\pi) \ket{0} = \ket{1}$. The entanglement structure connects nearest neighbors with $CZ$ gates. The variational parameters are incorporated in single qubit rotations about the $x$, $y$, and $z$ axis. The state-action value is decoded by measuring Pauli-$Z$ observables on a number of qubits, that corresponds to the number of actions in the environment. The full \gls{vqc} is visualized in \cref{fig:vqc_Chen_2020}.
    
    \medbreak
    
    \noindent\textit{Experimental Results and Discussion.}
    The proposed \gls{vqc}-\gls{dql} algorithm is simulated for two environments. The first one is \texttt{FrozenLake}, with $16$ states and an $4$ actions. The second one is \texttt{CognitiveRadio}, which is adapted to \glspl{vqc} sizes of $2$ to $5$ qubits. The authors report that their \gls{vqc}-based agent performs at least equally well as a \gls{nn}. Moreover, they claim that this requires fewer parameters (about one order of magnitude compared to \glspl{dnn}), which points towards potential quantum advantage. The model is tested on actual quantum hardware with competitive results.
    
    \medbreak
    
    \noindent\textit{Remarks.} The employed encoding scheme (computational basis encoding) could be simplified by omitting the $R_{Z}$ rotations, as these only introduce a global phase. The \texttt{CognitiveRadio} environment might be oversimplified. We also note that the claim on reduced parameter count should be substantiated by experiments with environments of different scale.

    \medbreak

    \noindent\textit{Reproduction.} A reproduction study by Lokes et al.~\cite{Lokes_2022} conducts an extended hyperparameter search for the described setup. The results and claims are overall consistent with~\cite{Chen_2020}, but no novel findings could were reported.

    \medbreak

    \noindent\textit{Extension.} In the work by S. Y.-C. Chen~\cite{Chen_2023f} the quantum $Q$-learning framework introduced in~\cite{Chen_2020} is extended by incorporating prioritized experience replay. Additionally, an asynchronous training routine is employed, similar to the one discussed in~\cite{Chen_2023c}. Both techniques reduce the overall sampling complexity and therefore allow for solving more complex tasks with the same underlying quantum model. This is validated with numerical simulations on several versions of the \texttt{CartPole} environment.

    \medbreak

    \noindent\textit{Hybrid Model.} The work by Chen et al.~\cite{Chen_2023e} extends the quantum models used in~\cite{Chen_2020} with classical neural networks, to produce more expressive function approximators. With that extension, the quantum agent is able to solve a $20 \times 20$ gridworld maze, which should clearly be more complex than the originally considered $\texttt{FrozenLake}$ environment. However, with the provided analysis it in unclear to which extend the performance can be contributed to the quantum part of the model.

    \medbreak

    \noindent\textit{Hyperparameter Analysis.} A hyperparamter analysis is conducted by Fikadu Tilaye and Pandey~\cite{FikaduTilaye_2023}, with a focus on the $Q$-learning framework introduced in~\cite{Chen_2020}. The authors conclude, that deeper quantum circuits lead to a better overall performance, while a larger learning rate speeds up the overall process. However, the analysis is superficial and quite small-scale, so further investigations are necessary to allow for more general statements.

    \begin{table}[h!]
    \centering
    \begin{minipage}{\textwidth}
        \renewcommand\footnoterule{}
        \renewcommand{\thefootnote}{\alph{footnote}}
    \caption*{\textbf{Algorithmic Characteristics - Chen et al.}~\cite{Chen_2020}}
    \small
    \begin{tabular*}{\textwidth}{p{\dimexpr 0.19\textwidth-2\tabcolsep-\arrayrulewidth}|p{\dimexpr 0.14\textwidth-2\tabcolsep-\arrayrulewidth}|p{\dimexpr 0.15\textwidth-2\tabcolsep-\arrayrulewidth}|p{\dimexpr 0.1\textwidth-2\tabcolsep-\arrayrulewidth}|p{\dimexpr 0.1\textwidth-2\tabcolsep-\arrayrulewidth}|p{\dimexpr 0.1\textwidth-2\tabcolsep-\arrayrulewidth}|p{\dimexpr 0.22\textwidth-2\tabcolsep}}
        \toprule
        \centering\multirow{2}{*}{\textbf{Environment}} & \centering\textbf{Algorithm} & \centering\textbf{Quantum} & \centering\textbf{State} & \centering\textbf{Action} & \centering\multirow{2}{*}{\textbf{Qubits}} & \centering\textbf{Parameterized} \tabularnewline
        & \centering\textbf{Type} & \centering\textbf{Component} & \centering\textbf{Space} & \centering\textbf{Space} & & \centering\textbf{~Gates}\footnotemark[1] \tabularnewline
        \midrule
        \midrule
        %%%% ROW 1
        \centering\texttt{FrozenLake} &
        \centering\multirow{2}{*}{\gls{ddql}} &
        \centering\multirow{2}{*}{$Q$-function} &
        \centering{\tiny{discrete}} &
        \centering{\tiny{discrete}} &
        \centering\multirow{2}{*}{$4$} &
        \centering $4 \times 2$ (encoding) \tabularnewline
        \centering (OpenAI Gym) &
        &
        &
        \centering $16$ &
        \centering $4$ &
        & \centering $4 \times 4 \times 3$ (weights) \tabularnewline
        \midrule
        %%%% ROW 2
        \centering\texttt{CognitiveRadio} &
        \centering\multirow{2}{*}{\gls{ddql}} &
        \centering\multirow{2}{*}{$Q$-function} &
        \centering{\tiny{discrete}} &
        \centering{\tiny{discrete}} &
        \centering\multirow{2}{*}{$n$} &
        \centering $n \times 2$ (encoding) \tabularnewline
        \centering (see \cite{Chen_2020}) &
        &
        &
        \centering $n^2$ &
        \centering $n$ &
        & \centering $n \times 4 \times 3$ (weights) \tabularnewline
        \bottomrule
    \end{tabular*}
    \footnotetext[1]{~encoding gates: $qubits \times per\_qubit$; variational gates: $qubits \times layers \times per\_qubit\_per\_layer$;}
    \end{minipage}
\end{table}
    
%------------------------------------------------------------------------------------------------------%
%------------------------------------------------------------------------------------------------------%
%------------------------------------------------------------------------------------------------------%

%------------------------------------------------------------
\paragraph{\label{subsubsec:Lockwood_2020}Reinforcement Learning with Quantum Variational Circuits, Lockwood and Si (2020)}\mbox{}\\
% \cite{Lockwood_2020}
%------------------------------------------------------------

    \vspace{-1em}
    \noindent\textit{Summary.} The work by Lockwood and Si~\cite{Lockwood_2020} modifies several aspects of the routine proposed by Chen et al.~\cite{Chen_2020}. Most importantly, they introduce two new encoding schemes to deal with a continuous state space.
    
    \medbreak
    
    \noindent\textit{Modification of Architecture.} The first proposed encoding is denoted as \emph{scaled encoding}. It scales the \gls{rl} state values to the range $[0, 2 \pi)$, which are then encoded using some $1$-qubit parameterized rotations. The second on (so-called \emph{directional encoding}) only encodes the sign of the value. More concretely, if a state variable is positive, $R_x$ and $R_z$ rotations by $\pi$ are applied to the encoding qubit (following a similar idea as the computational state encoding \cite{Chen_2020}).

    The architecture for the variational layer consists of an entangling block (nearest-neighbor $CX$ gates) and parameterized $1$-qubit rotations about $x$, $y$, and $z$ axis. This block is repeated three times. For decoding the state-action value, the authors employ two different strategies. The first one feeds the measurement result into a classical fully-connected layer where the number of outputs corresponds to the number of possible actions. In the other case, a so-called quantum pooling operation, condenses the information of the quantum state into a subset of the qubits \cite{Cong_2019}. This allows for a more flexible architecture, independent of the number of actions in the environment.
    
    \medbreak
    
    \noindent\textit{Experimental Results.} The proposed algorithm and the encoding schemes are benchmarked on the \texttt{CartPole} and \texttt{Blackjack} environment. While the former one uses a combination of scaled and directional encoding, the second one only employs scaled encoding. Their findings agree with those reported previously in the literature, namely that \gls{vqc}-based models achieve similar performance to \gls{nn}-based function approximators. As also stated by Chen et al.~\cite{Chen_2020}, the usage of \glspl{vqc} reduces the required parameter complexity.
    
    \medbreak
    
    \noindent\textit{Remarks.} While the scaled encoding should be a sound choice, the directional encoding could be inappropriate for most environments. Usually, not only the sign of a specific state is relevant, but the concrete state contains relevant information. With this encoding, this information is lost, which should lead to a drop in performance for more complex environments. As stated previously, the reduced parameter complexity should be investigated for larger problem instances.
    
    \begin{table}[h!]
    \centering
    \begin{minipage}{\textwidth}
        \renewcommand\footnoterule{}
        \renewcommand{\thefootnote}{\alph{footnote}}
    \caption*{\textbf{Algorithmic Characteristics - Lockwood and Si}~\cite{Lockwood_2020}}
    \small
    \begin{tabular*}{\textwidth}{p{\dimexpr 0.19\textwidth-2\tabcolsep-\arrayrulewidth}|p{\dimexpr 0.14\textwidth-2\tabcolsep-\arrayrulewidth}|p{\dimexpr 0.15\textwidth-2\tabcolsep-\arrayrulewidth}|p{\dimexpr 0.1\textwidth-2\tabcolsep-\arrayrulewidth}|p{\dimexpr 0.1\textwidth-2\tabcolsep-\arrayrulewidth}|p{\dimexpr 0.1\textwidth-2\tabcolsep-\arrayrulewidth}|p{\dimexpr 0.22\textwidth-2\tabcolsep}}
        \toprule
        \centering\multirow{2}{*}{\textbf{Environment}} & \centering\textbf{Algorithm} & \centering\textbf{Quantum} & \centering\textbf{State} & \centering\textbf{Action} & \centering\multirow{2}{*}{\textbf{Qubits}} & \centering\textbf{Parameterized} \tabularnewline
        & \centering\textbf{Type} & \centering\textbf{Component} & \centering\textbf{Space} & \centering\textbf{Space} & & \centering\textbf{~Gates}\footnotemark[1] \tabularnewline
        \midrule
        \midrule
        %%%% ROW 1
        \centering\texttt{CartPole} &
        \centering\multirow{2}{*}{\gls{ddql}} &
        \centering\multirow{2}{*}{$Q$-function} &
        \centering{\tiny{continuous}} &
        \centering {\tiny{discrete}} &
        \centering\multirow{2}{*}{$4$} &
        \centering $4 \times 2$ (encoding) \tabularnewline
        \centering (OpenAI Gym) &
        &
        &
        \centering $4$-dim &
        \centering $2$ &
        & 
        \centering $4 \times 3 \times 3$ (weights) \tabularnewline
        \midrule
        %%%% ROW 2
        \centering\texttt{Blackjack} &
        \centering\multirow{2}{*}{\gls{ddql}} &
        \centering\multirow{2}{*}{$Q$-function} &
        \centering{\tiny{discrete}} &
        \centering{\tiny{discrete}} &
        \centering\multirow{2}{*}{$3$} &
        \centering $3 \times 2$ (encoding) \tabularnewline
        \centering (OpenAI Gym) &
        &
        &
        {\scriptsize{\centering $31 \times 11 \times 2$}} &
        \centering $2$ &
        &
        \centering $3 \times 3 \times 3$ (weights) \tabularnewline
        \bottomrule
    \end{tabular*}
    \footnotetext[1]{~encoding gates: $qubits \times per\_qubit$; variational gates: $qubits \times layers \times per\_qubit\_per\_layer$;}
    \end{minipage}
\end{table}
    
%------------------------------------------------------------------------------------------------------%
%------------------------------------------------------------------------------------------------------%
%------------------------------------------------------------------------------------------------------%

%------------------------------------------------------------
\paragraph{\label{subsubsec:Lockwood_2021}Playing Atari with Hybrid Quantum-Classical Reinforcement Learning, Lockwood and Si (2021)}\mbox{}\\
% \cite{Lockwood_2021}
%------------------------------------------------------------

    \vspace{-1em}
    \noindent\textit{Summary.} This work by Lockwood and Si~\cite{Lockwood_2021} extends their previous paper~\cite{Lockwood_2020}, which, in turn, was based on Chen et al.~\cite{Chen_2020}, where $Q$-learning with \gls{vqc} function approximation has been introduced. The paper considers the Atari environments \texttt{Pong} and \texttt{Breakout}, with continuous state space of dimensionality $28.224$ (the observations are cropped and converted to images with $84 \times 84 \times 4$ pixels). This environment complexity is not tractable with previously introduced encoding schemes, which require one qubit for each dimension. The proposed workaround uses a classical \gls{nn} to reduce the state dimensionality before encoding it into the \gls{vqc}.
    
    \medbreak
    
    \noindent\textit{Underlying Algorithm and Simulation.} Similar to Refs.~\cite{Chen_2020,Lockwood_2020}, the concept of \gls{ddql} is used. The pipeline is modified by replacing the pure \gls{vqc} function approximator with a hybrid model. Several different choices are considered, the most important details are highlighted below. The training is performed in an end-to-end manner, i.e., the gradients w.r.t.\ the \gls{vqc} parameters are propagated back through the classical encoding network.
    
    \medbreak
    
    \noindent\textit{Model Architecture.} The \gls{vqc} architecture is, as usually, composed of three parts (i.e.~state encoding, variational layers, and action decoding). To encode the state, the raw data is first fed through a classical \gls{nn}. This outputs a number of values equal to the number of parameters in the feature map, which itself consists of $1$-qubit parameterized rotations. The authors compare the performance of a densely connected and a \gls{cnn} for this task (the concrete architecture of these networks are not specified). Apart from that, encoding layers of different sizes (and therefore different number of parameters) from $5$ to $15$ qubits are compared.
    
    The variational layers itself consists of two parts, where the first one is a \gls{qcnn}~\cite{Cong_2019}. The authors state two motivations for this choice: First, it should help capture the spatial structure of the input images (but it is unclear, whether the encoding part retains the spatial structure). Second, \glspl{qcnn} help to avoid barren plateaus~\cite{Pesah_2021} (while the experiments show no sign of barren plateaus, it is not clear if this is due to this choice, or the limited size of the employed circuits). After this \gls{qcnn} there are three repetitions of entanglement gates and parameterized rotations, similar to those also used for state encoding.
    
    The paper proposes two methods to deal with the problem of measurement for unequal number of qubits and actions. The first method performs Pauli-$Z$ measurements on all qubits and uses an appended dense \gls{nn}. Alternatively, quantum pooling operations~\cite{Cong_2019} are used, which subsequently compress the measurement of two qubits into one.
    
    \medbreak
    
    \noindent\textit{Experimental Results and Discussion.} To demonstrate the basic functionality of the model, initial experiments are conducted on the \texttt{CartPole} environment. The results demonstrate a similar performance to Lockwood and Si~\cite{Lockwood_2020}. On the two Atari environments, the paper considers $12$ different hybrid architectures (dense vs.\ convolutional encoding, $5$ vs.\ $10$ vs.\ $15$ qubits, dense vs.\ pooling decoding), which are compared to a well-established classical architecture.
    
    It turns out, that the hybrid models are not able to learn at all. The authors state, that this is down to the lack of expressibility of the hybrid models, which only make use of about $10^4$ parameters, while the classical model uses about $10^6$. It is expected, that for more expressive models the performance improves, as learning on the much simpler \texttt{CartPole} environment was successful.
    
    \medbreak
    
    \noindent\textit{Remarks.} The experiments are conducted with a restricted set of hybrid models. Consequently, the claim that these results do not demonstrate the inapplicability of \gls{qrl} to more complex environments like Atari is reasonable. The assumption that this approach could be made to work on complex environments, as it succeeds on e.g.\ \texttt{CartPole}, should be sustained with additional experiments. For a modified architecture succeeding on the Atari environments, it is not completely clear, which part of the work is done by the classical and quantum part of the model. This is a typical caveat, whenever quantum and classical architectures are combined.
    
    % \medbreak
    
    \begin{table}[ht!]
    \centering
    \begin{minipage}{\textwidth}
        \renewcommand\footnoterule{}
        \renewcommand{\thefootnote}{\alph{footnote}}
    \caption*{\textbf{Algorithmic Characteristics - Lockwood and Si}~\cite{Lockwood_2021}}
    \small
    \begin{tabular*}{\textwidth}{p{\dimexpr 0.19\textwidth-2\tabcolsep-\arrayrulewidth}|p{\dimexpr 0.14\textwidth-2\tabcolsep-\arrayrulewidth}|p{\dimexpr 0.15\textwidth-2\tabcolsep-\arrayrulewidth}|p{\dimexpr 0.1\textwidth-2\tabcolsep-\arrayrulewidth}|p{\dimexpr 0.1\textwidth-2\tabcolsep-\arrayrulewidth}|p{\dimexpr 0.1\textwidth-2\tabcolsep-\arrayrulewidth}|p{\dimexpr 0.22\textwidth-2\tabcolsep}}
        \toprule
        \centering\multirow{2}{*}{\textbf{Environment}} & \centering\textbf{Algorithm} & \centering\textbf{Quantum} & \centering\textbf{State} & \centering\textbf{Action} & \centering\multirow{2}{*}{\textbf{Qubits}} & \centering\textbf{Parameterized} \tabularnewline
        & \centering\textbf{Type} & \centering\textbf{Component} & \centering\textbf{Space} & \centering\textbf{Space} & & \centering\textbf{Gates} \tabularnewline
        \midrule
        \midrule
        %%%% ROW 1
        \centering\multirow{3}{*}{\begin{tabular}{@{}c@{}}\texttt{CartPole}\\(OpenAI Gym)\end{tabular}} &
        \centering\multirow{3}{*}{\gls{ddql}} &
        \centering\multirow{3}{*}{$Q$-function} &
        \centering\multirow{3}{*}{\begin{tabular}{@{}c@{}}\tiny{continuous}\\$4$-dim\end{tabular}} &
        \centering\multirow{3}{*}{\begin{tabular}{@{}c@{}}\tiny{discrete}\\$2$\end{tabular}} &
        \centering\multirow{3}{*}{$5$} &
        \centering $N/A$ (classical)\footnotemark[1] \tabularnewline
        &
        &
        &
        &
        &
        & 
        \centering $\mathcal{O}\left( 10^1 \right)$ (encoding) \tabularnewline
        &
        &
        &
        &
        &
        & 
        \centering $\mathcal{O}\left( 10^2 \right)$ (weights) \tabularnewline
        \midrule
        %%%% ROW 2
        \centering\multirow{3}{*}{\begin{tabular}{@{}c@{}}\texttt{Pong-v0}\\(OpenAI Gym)\end{tabular}} &
        \centering\multirow{3}{*}{\gls{ddql}} &
        \centering\multirow{3}{*}{$Q$-function} &
        \centering\tiny{continuous} &
        \centering\multirow{3}{*}{\begin{tabular}{@{}c@{}}\tiny{discrete}\\$6$\end{tabular}} &
        \centering\multirow{3}{*}{$5$ to $15$} &
        \centering $\mathcal{O}\left( 10^6 \right)$ (classical) \tabularnewline
        &
        &
        &
        \centering $28224$- &
        &
        & 
        \centering $\mathcal{O}\left( 10^2 \right)$ (encoding) \tabularnewline
        &
        &
        &
        \centering dim\footnotemark[2] &
        &
        & 
        \centering $\mathcal{O}\left( 10^4 \right)$ (weights) \tabularnewline
        \midrule
        %%%% ROW 3
        \centering\multirow{3}{*}{\begin{tabular}{@{}c@{}}\texttt{Breakout-v0}\\(OpenAI Gym)\end{tabular}} &
        \centering\multirow{3}{*}{\gls{ddql}} &
        \centering\multirow{3}{*}{$Q$-function} &
        \centering\tiny{continuous} &
        \centering\multirow{3}{*}{\begin{tabular}{@{}c@{}}\tiny{discrete}\\$4$\end{tabular}} &
        \centering\multirow{3}{*}{$5$ to $15$} &
        \centering $\mathcal{O}\left( 10^6 \right)$ (classical) \tabularnewline
        &
        &
        &
        \centering $28224$- &
        &
        & 
        \centering $\mathcal{O}\left( 10^2 \right)$ (encoding) \tabularnewline
        &
        &
        &
        \centering dim\footnotemark[2] &
        &
        & 
        \centering $\mathcal{O}\left( 10^4 \right)$ (weights) \tabularnewline
        \bottomrule
    \end{tabular*}
    \footnotetext[1]{~potentially also uses a classical \gls{nn} for pre-processing, details are not stated;}
    \footnotetext[2]{~dimensionality of feature space is reduced with a \gls{nn} to fit size of feature map;}
    \end{minipage}
\end{table}
    
%------------------------------------------------------------------------------------------------------%
%------------------------------------------------------------------------------------------------------%
%------------------------------------------------------------------------------------------------------%

%------------------------------------------------------------
\paragraph{\label{subsubsec:Skolik_2022}Quantum agents in the Gym: a variational quantum algorithm for deep $Q$-learning, Skolik et al.~(2022)}\mbox{}\\
% \cite{Skolik_2022}
%------------------------------------------------------------

    \vspace{-1em}
    \noindent\textit{Summary.} This work by Skolik et al.~\cite{Skolik_2022} proposes another instance of $Q$-learning with \glspl{vqc} as function approximators. Being aware of preceding literature, the authors set out to analyze the role of architecture design, \gls{rl} state encoding schemes, and observables for action decoding. With regard to the previous work, the authors remark that the \texttt{CartPole} environment cannot be considered solved.
    
    \medbreak
    
    \noindent\textit{Importance of Architecture Design.} In terms of architecture choices, the problem of barren plateaus is emphasized: Architectures with many qubits and layers (which naively is required for high expressivity) are hard to train. Contrarily, over-parameterized architectures are easier to train, but probably less expressive and therefore less effective on a given task.
    
    The authors chose a hardware-efficient ansatz, despite being known to run into the barren plateau problem for large circuits. For the small circuit sizes considered in the present work, the barren-plateau problem does not appear to be relevant.

    \medbreak

    \noindent\textit{Encoding Schemes.} As for encoding schemes, discrete \gls{rl} states are encoded in the computational basis. Continuous states are scaled to the finite interval $[- \pi/2, + \pi/2]$ by applying $\arctan$ to the raw observations. The result serves as the rotation angle for an $R_x$ rotation, which is very similar to the scaled encoding proposed by Lockwood and Si \cite{Lockwood_2020}. In order to increase expressivity w.r.t.\ to the input, the encoding layer can be repeated through the circuit, forming a data re-uploading structure \cite{Perez_2020}. Effectively, this allows to learn and approximate a Fourier sum of a certain order, where the order is tied to the number of repetitions of the encoding layer~\cite{Schuld_2021}. The encoding is further modified by introducing learnable re-scaling parameters, that are multiplied with the raw states before computing the $\arctan$.
    
    \medbreak
    
    \noindent\textit{Experimental Results and Discussion.} The authors benchmark their architecture choices on the \texttt{FrozenLake} and \texttt{CartPole} environment. The performance on \texttt{CartPole} is compared to a small \gls{nn} with the same number of parameters, which seems to be inferior. Further, the range of $Q$-values that can be encountered in the two benchmark environments is investigated. For \texttt{FrozenLake}, representing the $Q$-value with the expectation values of $1$-qubit $Z$-operators is sufficient. For the \texttt{CartPole} environment, this strategy is found not to be adaptable enough. Instead, they chose the expectation values of the parities (of 2 non-overlapping pairs of qubits) and allow for additional trainable classical weights that set the scale for the $Q$-value approximation.

    \medbreak

    \noindent\textit{Remarks.} The authors emphasize the critical role of architectural choices at the outset of their manuscript. While they offer valuable insights into this topic, also open questions remain for future work in this direction. For the \texttt{CartPole} environment, several trainable classical weights are incorporated in the algorithm. Therefore, it is not completely clear, what part of the training is achieved by which part of the hybrid model.

    \medbreak

    \noindent\textit{Error Analysis.} The work by Skolik et al.~\cite{Skolik_2023} analysis the influence of hardware noise on the quantum $Q$-learning framework introduced in~\cite{Skolik_2022}, but also \gls{qpg} approaches discussed in \cref{subsec:VQC_based_Policy}. The results are numerically validated on the \texttt{CartPole} environment and a version of the Travelling Salesperson Problem. The results indicate, that the performance is very much dependent on the inherent structure of the noise. For some instances, the robustness of the learned policy is actually increased if noise is encountered during training. However, e.g. for strong incoherent noise the performance decreases quite substantially. Interesting from a practical point of view is especially the analysis of shot noise, which indicates that a low number of repetitions is enough to get a reliable estimate of the $Q$-function -- an explicit algorithm to exploit this property is proposed in this work.

    \medbreak

    \noindent\textit{Continuous Action Spaces.} A Q-learning approach based on \cite{Skolik_2022} that incorporates continuous action spaces is discussed by Liu et al.~\cite{LiuY_2023}. They use normalized advantage functions which allows for continuous action selection. An alternative would be to additionally use a policy function approximator to form an actor-critic approach, as discussed in~\cref{subsec:VQC_based_CombinationApproximations}.
    
    \begin{table}[h]
    \centering
    \begin{minipage}{\textwidth}
        \renewcommand\footnoterule{}
        \renewcommand{\thefootnote}{\alph{footnote}}
    \caption*{\textbf{Algorithmic Characteristics - Skolik et al.}~\cite{Skolik_2022}}
    \small
    \begin{tabular*}{\textwidth}{p{\dimexpr 0.19\textwidth-2\tabcolsep-\arrayrulewidth}|p{\dimexpr 0.14\textwidth-2\tabcolsep-\arrayrulewidth}|p{\dimexpr 0.15\textwidth-2\tabcolsep-\arrayrulewidth}|p{\dimexpr 0.1\textwidth-2\tabcolsep-\arrayrulewidth}|p{\dimexpr 0.1\textwidth-2\tabcolsep-\arrayrulewidth}|p{\dimexpr 0.1\textwidth-2\tabcolsep-\arrayrulewidth}|p{\dimexpr 0.22\textwidth-2\tabcolsep}}
        \toprule
        \centering\multirow{2}{*}{\textbf{Environment}} & \centering\textbf{Algorithm} & \centering\textbf{Quantum} & \centering\textbf{State} & \centering\textbf{Action} & \centering\multirow{2}{*}{\textbf{Qubits}} & \centering\textbf{Parameterized} \tabularnewline
        & \centering\textbf{Type} & \centering\textbf{Component} & \centering\textbf{Space} & \centering\textbf{Space} & & \centering\textbf{~Gates}\footnotemark[1] \tabularnewline
        \midrule
        \midrule
        %%%% ROW 1
        \centering\multirow{3}{*}{\begin{tabular}{@{}c@{}}\texttt{CartPole}\\(OpenAI Gym)\end{tabular}} &
        \centering\multirow{3}{*}{\gls{ddql}} &
        \centering\multirow{3}{*}{$Q$-function} &
        \centering\multirow{3}{*}{\begin{tabular}{@{}c@{}}\tiny{continuous}\\$4$-dim\end{tabular}} &
        \centering\multirow{3}{*}{\begin{tabular}{@{}c@{}}\tiny{discrete}\\$2$\end{tabular}} &
        \centering\multirow{3}{*}{$4$} &
        \centering $4 \times 1$ (encoding) \tabularnewline
        &
        &
        &
        &
        &
        & 
        \centering $4 \times 15 \times 2$ (weights) \tabularnewline
        &
        &
        &
        &
        &
        & 
        \centering $N/A$ (classical)\footnotemark[2] \tabularnewline
        \midrule
        %%%% ROW 2
         \centering\texttt{FrozenLake} &
        \centering\multirow{2}{*}{\gls{ddql}} &
        \centering\multirow{2}{*}{$Q$-function} &
        \centering{\tiny{discrete}} &
        \centering {\tiny{discrete}} &
        \centering\multirow{2}{*}{$4$} &
        \centering $4 \times 1$ (encoding) \tabularnewline
        \centering (OpenAI Gym) &
        &
        &
        \centering $16$ &
        \centering $4$ &
        & 
        \centering $4 \times 15 \times 2$ (weights) \tabularnewline
        \bottomrule
    \end{tabular*}
    \footnotetext[1]{~encoding gates: $qubits \times per\_qubit$; variational gates: $qubits \times layers \times per\_qubit\_per\_layer$;}
    \footnotetext[2]{~model incorporates classical weights after measurement, details are not stated;}
    \end{minipage}
\end{table}
    
%------------------------------------------------------------------------------------------------------%
%------------------------------------------------------------------------------------------------------%
%------------------------------------------------------------------------------------------------------%

\subsubsection{Policy Approximation}
\label{subsec:VQC_based_Policy}
This section covers \gls{vqc}-based approximations in policy space, as described for the instance of classical policy gradients in \cref{eq:policygradients_1,eq:policygradients_2}. The concept was introduced by Jerbi et al.~\cite{Jerbi_2021a}, shortly followed by a slight reformulations in Ref.~\cite{Kunczik_2022}, and an extension to allow for faster computation in Ref.~\cite{Quafu_2023}. Several modifications, including formulating full-quantum interaction with a quantum control environment, have been introduced in Sequeira et al.~\cite{Sequeira_2023} -- with a closer analysis of quantum-accessible environments revealing potential advantage compared to certain classical routines in Ref.~\cite{Jerbi_2023}. Algorithmic extensions to the \gls{qpg} setup were proposed in Ref.~\cite{Meyer_2021}. Details on a therein introduced classical post-processing function to improve \gls{rl} performance are discussed in Meyer et al.~\cite{Meyer_2023a}, and quantum natural gradients to enhance trainability are covered by the same authors in~\cite{Meyer_2023b}.

\begin{table}[t!]
    \centering
    \begin{tabular}{p{\dimexpr 0.15\textwidth-2\tabcolsep-\arrayrulewidth}|p{\dimexpr 0.2\textwidth-2\tabcolsep-\arrayrulewidth}|p{\dimexpr 0.65\textwidth-2\tabcolsep}}
        \toprule
        \textbf{Citation} & \textbf{First Author} & \textbf{Title} \\
        \midrule
        \midrule
        \cite{Jerbi_2021a} & S. Jerbi & \hyperref[subsubsec:Jerbi_2021a]{Parameterized Quantum Policies for Reinforcement Learning} \\
        \arrayrulecolor{black!30}\midrule
        \cite{Kunczik_2022} & L. Kunczik & \hyperref[subsubsec:Jerbi_2021a]{Reinforcement Learning with Hybrid Quantum Approximation in the NISQ Context} \\
        \midrule
        \cite{Quafu_2023} & Quafu Group & \hyperref[subsubsec:Jerbi_2021a]{Quafu-RL: The Cloud Quantum Computers based Quantum Reinforcement Learning} \\
        \arrayrulecolor{black}\midrule
        \cite{Sequeira_2023} & A. Sequeira & \hyperref[subsubsec:Sequeira_2023]{Policy gradients using variational quantum circuits} \\
        \arrayrulecolor{black!30}\midrule
        \cite{Jerbi_2023} & S. Jerbi & \hyperref[subsubsec:Sequeira_2023]{Quantum Policy Gradient Algorithms} \\
        \arrayrulecolor{black}\midrule
        \cite{Meyer_2023a} & N. Meyer & \hyperref[subsubsec:Meyer_2023a]{Quantum Policy Gradient Algorithm with Optimized Action Decoding} \\
        \midrule
        \cite{Meyer_2023b} & N. Meyer & \hyperref[subsubsec:Meyer_2023b]{Quantum Natural Policy Gradients: Towards Sample-Efficient Reinforcement Learning} \\
        \midrule
        \bottomrule
    \end{tabular}
    \caption{Work considered for ``\gls{qrl} with \glspl{vqc} -- Policy Approximation'' (\cref{subsec:VQC_based_Policy})}
\end{table}

%------------------------------------------------------------
\paragraph{\label{subsubsec:Jerbi_2021a}Parameterized Quantum Policies for Reinforcement Learning, Jerbi et al.~(2021) and related work}\mbox{}\\
% \cite{Jerbi_2021a}
%------------------------------------------------------------

    \vspace{-1em}
    \noindent\textit{Summary.} The paper by Jerbi et al.~\cite{Jerbi_2021a} starts out with a small summary of \gls{vqc}-based \gls{ml} models. They cite several reports of quantum advantage in the supervised and unsupervised \gls{qml}. This motivates their approach to go beyond the scope of $Q$-function approximation~\cite{Chen_2020,Lockwood_2020,Lockwood_2021,Skolik_2022}, and use the \gls{vqc} to directly approximated the policy.
    
    \medbreak
    
    \noindent\textit{Quantum Policy Gradient.} After a brief recap of policy gradient methods for solving \gls{rl} problems, the authors extend those ideas to a \gls{qpg} approach. More concretely, they quantize the REINFORCE algorithm~\cite{Williams_1992,} with value-function baselines by using \glspl{vqc} as function approximators for the (stochastic) policy. The define two families of \gls{vqc}-based policies: (1) A \texttt{RAW}-\gls{vqc} policy, where the action selection follows Born's rule. It is defined as $\pi_{\theta}(a | s) = \expval{P_a}_{s,\theta}$, where $P_a$ are the projectors on the elements of the computational basis. This allows action selection with only one evaluation of the quantum circuit; (2) A \texttt{SOFTMAX}-VQC policy, defined as $\pi_{\theta}(a | s)=e^{\beta \expval{O_a}_{s,\theta}}/\sum_{a'} e^{\beta \expval{O_{a'}}_{s,\theta}}$. The measurement result of an action-dependent observable $O_a$ is fed into a single-parameter softmax-function, to form a \gls{pdf}. The inverse-temperature parameter $\beta$ allows to adjust the peak-width of the distribution, i.e., the greediness of the policy.
    
    \medbreak
    
    \noindent\textit{Circuit Architecture.} The ansatz for the \gls{vqc} is chosen to be hardware-efficient, i.e., only single and two-qubit gates. The \gls{rl} state is encoded with $1$-qubit rotations. To increase the expressivity of the model, the authors introduce additional learnable state-scaling parameters $\lambda$. Those are multiplied to the rotational parameter denoting the state value, i.e., $\lambda_i \cdot s_i$ is the value of a $1$-qubit rotation. This also helps circumvent the problem of being restricted to a finite set of frequencies in such an encoding scheme~\cite{Schuld_2021}. The feature map is repeated several times, alternating with the variational layer, which forms a data re-uploading structure~\cite{Perez_2020}. A variational layer consists of $CZ$-gates for creating entanglement in an circular structure. The learnable parameters are used in $1$-qubit parameterized rotation gates. Depending on the policy type, measurements are either conducted in the computational basis, or more complex observables are measured.
    
    \medbreak
    
    \noindent\textit{Experimental Results.} Overall, all agents are able to learn meaningful behavior in the OpenAI Gym environments \texttt{CartPole}, \texttt{MountainCar}, and \texttt{Acrobot}. Further experiments are reported, which serve the purpose of assessing the importance of the various design choices: (1) Circuit depth increases performance and learning speed, where \texttt{SOFTMAX}-\gls{vqc} policies outperform \texttt{RAW}-\gls{vqc} policies in all instances; (2) Incorporating learnable state scaling parameters increases learning performance, trainable classical weights (in case of \texttt{SOFTMAX}-\gls{vqc}) multiplied to expectation values leads to increase in performance; (3) The performance gap between \texttt{RAW}-\gls{vqc} and \texttt{softmax}-\gls{vqc} policies seems to stem from the ability to adjust greediness.
    
    \medbreak
    
    \noindent\textit{Provable and Empirical Quantum Advantage.} To the best of our knowledge, this work is the first to corroborate the idea quantum advantage with \glspl{vqc} in the \gls{rl} setting. Therefore, the authors devise \gls{rl} environments (based on the \gls{dlp}), which are supposed to be classically intractable. Any classical algorithm would need a number of samples that scales exponential in the problem size to achieve a low generalization error. A \gls{vqc}-based algorithm with a very specific architecture only requires a polynomial amount of data. This implies an exponential advantage w.r.t.\ sample complexity, assuming it is infeasible to efficiently simulate the \gls{vqc} on classical hardware for large problem instances. The construction of the environment is inspired by previous results from \gls{qml}, where similar learning separations between classical and quantum models have been demonstrated~\cite{Liu_2021}.
    
    Further, the authors report numerical evidence of potential quantum advantage for environments based on expectation values sampled from \glspl{vqc}. The motivation lies in the (potential) intractability of simulating the given \gls{vqc} classically for large systems. More concretely, one uses a \gls{vqc} to define a labeling function (in the sense of a classification task) over the domain $[0,2\pi]^2$ (so-called \texttt{SL}-\gls{vqc}). This synthetic classification dataset is then rephrased as a \gls{rl} environment by incorporating some temporal structure (denoted as \texttt{Cliffwalk}-\gls{vqc}). Numerically, the authors observe a performance separation of models with classical \glspl{dnn} and \gls{vqc}-based policies. They claim, that this is likely due to the oscillatory structure in the labeling function.
    
    \medbreak
    
    \noindent\textit{Remarks.} While the proposal of provable quantum advantage is obviously quite encouraging, the practical realization is probably out of reach for the \gls{nisq}-era. The idea of solving the task efficiently on quantum hardware is based on Shor's algorithm. Formulated as a \gls{vqc}-based \gls{rl} problem, this would require circuits of complexity far beyond current scope. We think it requires also some more large-scale experiments, to support the empirical learning separation on the \texttt{SL}-\gls{vqc} and \texttt{Cliffwalk}-\gls{vqc} environments. A comparison to other hybrid models~\cite{Chen_2020,Lockwood_2020} shows, that the proposed \gls{qpg} approach is superior in terms of \gls{rl} performance on various environments.
    
    \medbreak

    \noindent\textit{Alternative Formulation.} In the PhD thesis by L. Kunczik~\cite{Kunczik_2022} a slightly different formulation of the \gls{qpg} framework is introduced, where the output of the quantum circuit is compounded with a classical weight vector. However, the underlying routine is very similar to~\cite{Jerbi_2021a}. Empirical results are reported to verify an desirable scaling of \gls{vqc}-based (as opposed to \gls{nn}-based) approaches. However, experiments are to small-scale for reliable statements regarding this correlation.

    \medbreak

    \noindent\textit{Cloud Computing.} The work by the BAQIS Quafu Group~\cite{Quafu_2023} realizes the framework introduced in~\cref{subsubsec:Jerbi_2021a} and executed it on the quantum devices provided via the Quafu cloud services. The results are ambiguous, as the agents trained on hardware are not really able to learn meaningful behaviour -- but are also only trained for a very limited number of timesteps, as also acknowledged by the authors.
    
    \begin{table}[h!]
    \centering
    \begin{minipage}{\textwidth}
        \renewcommand\footnoterule{}
        \renewcommand{\thefootnote}{\alph{footnote}}
    \caption*{\textbf{Algorithmic Characteristics - Jerbi et al.}~\cite{Jerbi_2021a}}
    \small
    \begin{tabular*}{\textwidth}{p{\dimexpr 0.19\textwidth-2\tabcolsep-\arrayrulewidth}|p{\dimexpr 0.14\textwidth-2\tabcolsep-\arrayrulewidth}|p{\dimexpr 0.15\textwidth-2\tabcolsep-\arrayrulewidth}|p{\dimexpr 0.1\textwidth-2\tabcolsep-\arrayrulewidth}|p{\dimexpr 0.1\textwidth-2\tabcolsep-\arrayrulewidth}|p{\dimexpr 0.1\textwidth-2\tabcolsep-\arrayrulewidth}|p{\dimexpr 0.22\textwidth-2\tabcolsep}}
        \toprule
        \centering\multirow{2}{*}{\textbf{Environment}} & \centering\textbf{Algorithm} & \centering\textbf{Quantum} & \centering\textbf{State} & \centering\textbf{Action} & \centering\multirow{2}{*}{\textbf{Qubits}} & \centering\textbf{Parameterized} \tabularnewline
        & \centering\textbf{Type} & \centering\textbf{Component} & \centering\textbf{Space} & \centering\textbf{Space} & & \centering\textbf{~Gates}\footnotemark[1] \tabularnewline
        \midrule
        \midrule
        %%%% ROW 1
        \centering\texttt{CartPole} &
        \centering\multirow{2}{*}{\scriptsize{REINFORCE}} &
        \centering\multirow{2}{*}{Policy} &
        \centering{\tiny{continuous}} &
        \centering {\tiny{discrete}} &
        \centering\multirow{2}{*}{$4$} &
        \centering\multirow{2}{*}{$30$} \tabularnewline
        \centering (OpenAI Gym) &
        &
        &
        \centering $4$-dim &
        \centering $2$ &
        & \tabularnewline
        \midrule
        %%%% ROW 2
        \centering\texttt{MountainCar} &
        \centering\multirow{2}{*}{\scriptsize{REINFORCE}} &
        \centering\multirow{2}{*}{Policy} &
        \centering{\tiny{continuous}} &
        \centering {\tiny{discrete}} &
        \centering\multirow{2}{*}{$2$} &
        \centering\multirow{2}{*}{$36$} \tabularnewline
        \centering (OpenAI Gym) &
        &
        &
        \centering $2$-dim &
        \centering $3$ &
        & 
        \tabularnewline
        \midrule
        %%%% ROW 3
        \centering\texttt{Acrobot} &
        \centering\multirow{2}{*}{\scriptsize{REINFORCE}} &
        \centering\multirow{2}{*}{Policy} &
        \centering{\tiny{continuous}} &
        \centering {\tiny{discrete}} &
        \centering\multirow{2}{*}{$6$} &
        \centering\multirow{2}{*}{$72$} \tabularnewline
        \centering (OpenAI Gym) &
        &
        &
        \centering $6$-dim &
        \centering $3$ &
        & \tabularnewline
        \midrule
        %%%% ROW 4
        \centering\texttt{SL}-\gls{vqc} &
        \centering\multirow{3}{*}{\scriptsize{REINFORCE}} &
        \centering\multirow{3}{*}{Policy} &
        \centering\multirow{3}{*}{\begin{tabular}{@{}c@{}}\tiny{continuous}\\$2$-dim\end{tabular}} &
        \centering\multirow{3}{*}{\begin{tabular}{@{}c@{}}\tiny{discrete}\\$2$\end{tabular}} &
        \centering\multirow{3}{*}{$2$} &
        \centering\multirow{3}{*}{37} \tabularnewline
        \centering\texttt{Cliffwalk}-\gls{vqc} &
        &
        &
        &
        &
        & \tabularnewline
        \centering (see \cite{Jerbi_2021a}) &
        &
        &
        &
        &
        & \tabularnewline
        \midrule
        %%%% ROW 5
        \centering\texttt{CognitiveRadio} &
        \centering\multirow{2}{*}{\scriptsize{REINFORCE}} &
        \centering\multirow{2}{*}{Policy} &
        \centering{\tiny{discrete}} &
        \centering {\tiny{discrete}} &
        \centering\multirow{2}{*}{$n$} &
        \centering $30$ to $75$ \tabularnewline
        \centering (see \cite{Chen_2020}) &
        &
        &
        \centering $n^2$ &
        \centering $n$ &
        &
        \centering for $n=2$ to $5$ \tabularnewline
        \bottomrule
    \end{tabular*}
    \footnotetext[1]{~this entails encoding, scaling, and variational parameters; the \texttt{SOFTMAX}-\gls{vqc} also uses classical parameters;}
    \end{minipage}
\end{table}

%------------------------------------------------------------------------------------------------------%
%------------------------------------------------------------------------------------------------------%
%------------------------------------------------------------------------------------------------------%

%------------------------------------------------------------
\paragraph{\label{subsubsec:Sequeira_2023}Policy gradients using variational quantum circuits, Sequeira et al.~(2023) and related work}\mbox{}\\
%\cite{Sequeira_2023}
%------------------------------------------------------------

    \vspace{-1em}
    \noindent\textit{Summary.} The article by Sequeira et al.~\cite{Sequeira_2023} proposes a quantum version of the REINFORCE algorithm with a \gls{vqc}-based function approximator, very similar to Jerbi et al.~\cite{Jerbi_2021a}. The methods are applied to the classical environments \texttt{CartPole} and \texttt{Acrobot} but also to a simple quantum control problem. It proposes an initialization technique for the variational parameters of a \gls{vqc}. Following the experimental results, a quantum advantage w.r.t.\ the number of required parameters and trainability of the models is claimed.
    
    \medbreak
    
    \noindent\textit{Underlying Reinforcement Learning Algorithm.} As in Jerbi et al.~\cite{Jerbi_2021a}, the policy is defined as $\pi_{\theta}(a | s) = e^{\beta \cdot \expval{O_a}_{\theta}}/\sum_{a'} e^{\beta \cdot \expval{O_{a'}}_{\theta}}$, and REINFORCE updates are performed. Hereby, the expectation values $\expval{O_a}_{\theta}$ for action $a$ is defined as the expectation  $\expval{\sigma_z^a}$, i.e., the expectation value of $1$-qubit Pauli-$Z$ observable measured on the $a$-th qubit.
    
    \medbreak
    
\begin{figure}[ht]
        \centering
        \includegraphics[width=0.7\textwidth]{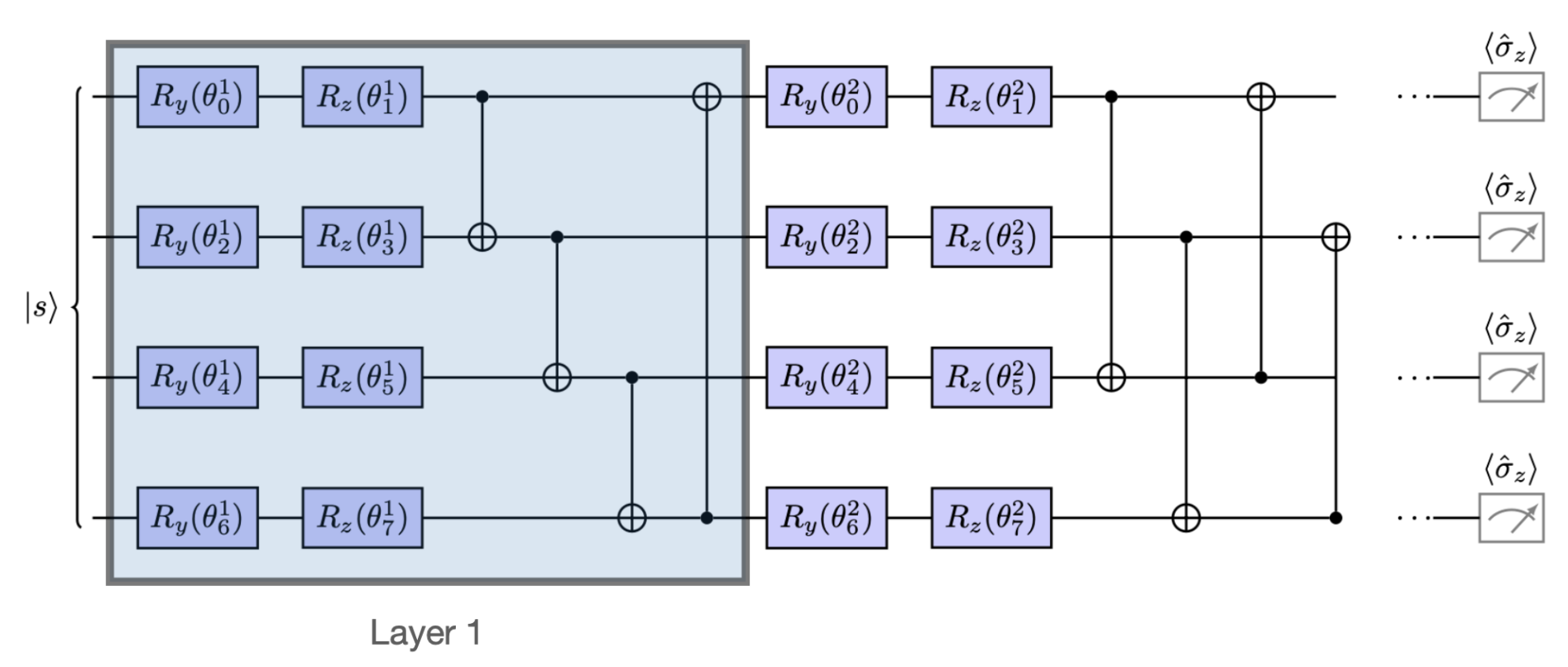}
        \caption{\Gls{vqc} architecture proposed by and taken from Sequeira et al.~\cite{Sequeira_2023}; It deviates from the typical circular entanglement structure.}
        \label{fig:ExtenstionsQPG_vqc}
    \end{figure}

    \noindent\textit{\gls{vqc} Architecture.} The architecture follows the typical three-part structure. In the beginning, the states are encoded with $R_x$ rotations, with the state values normalized to the range $[-\pi, \pi)$. Consequently, the number of qubits has to correspond to $\max \lbrace \abs{\mathcal{A}}, \abs{\mathcal{S}} \rbrace$. There are several parameterized layers (see \cref{fig:ExtenstionsQPG_vqc}) which incorporate variational parameters in $1$-qubit $R_y$ and $R_z$ rotations. The entanglement structure can be described as $CX[i, (i+l) \mod n]$,  where $n$ is the number of qubits, and $l$ the index of the layer. The measurement of $1$-qubit Pauli-$Z$ observables is a deviation to the procedure proposed by Jerbi et al.~\cite{Jerbi_2021a}, where multi-qubit observables were used.
    
    \medbreak
    
    \noindent\textit{Complexity of Gradient Estimation.} The paper gives an estimation of the required number of samples to get an $\epsilon$-approximation of the log-policy gradient. According to this consideration, for a success probability of $1-\delta$, the number of required measurements is bounded by $c \cdot \frac{(1-\epsilon)^2}{\epsilon^2} \cdot \log(\frac{k}{\delta})$. Hereby, $c$ is a constant depending on algorithmic hyperparameters and $k$ is the number of variational parameters. It is important to state, that this refers to the number of samples / data points required to get a good approximation of the true policy gradient, but not the explicit estimation of the gradients themself via e.g. the parameter-shift rule.
    
    \medbreak
    
    \noindent\textit{Initialization Technique.} There is some work proposing a technique for parameter initialization to avoid barren plateaus \cite{Grant_2019}. However, a technique to boost the overall performance has not yet been proposed. Inspired by classical \gls{ml}, the authors aim to break symmetries between different neurons (as usually initialization with constant values is a bad choice). A typical strategy is to select values uniformly at random from $[-\pi, \pi]$, or drawn them following a Gaussian distribution.
    
    Inspired by the classical \emph{Glorot} initialization scheme~\cite{Glorot_2010}, the paper proposed to use a normal distribution $\mathcal{N}(0, \mathrm{std}^2)$ with $\mathrm{std}=g \cdot \sqrt{2/\left( \mathrm{fan}_{in}+\mathrm{fan}_{out} \right)}$. Here, $g$ is a constant multiplicative factor, $\mathrm{fan}_{in}$ is the number of embedded features, and $\mathrm{fan}_{out}$ is the number of computational basis measurements. This technique demonstrates some promising experimental results, but no theoretical justification is given.
    
    \medbreak
    
    \noindent\textit{Analysis of Fisher Information Spectrum.} The paper analyzes the spectrum of the \gls{fim}, which serves as a tool to quantify the trainability of a model. The empirical \gls{fim} is computed as $F(\theta) = \frac{1}{T} \sum_{t=1}^{T} \nabla_{\theta} \log \pi(a_t | s_t,\theta) \nabla_{\theta} \log \pi(a_t | s_t,\theta)^t$. A similar analysis has also been proposed for \gls{qml}~\cite{Abbas_2021}.
    
    The results show, that the spectrum of the \gls{fim} associated with the quantum model exhibits significantly larger averaged eigenvalues. The compared \gls{nn} was optimized over several architectures, but not many details are provided in the paper. The authors conclude, that the quantum models are beneficial in terms of trainability, and might be resilient to barren plateaus.
    
    \medbreak
    
    \noindent\textit{Experimental Results and Discussion of Potential Quantum Advantage.} The proposed algorithm is tested on the classical benchmark environments \texttt{CartPole} and \texttt{Acrobot}. The performance is compared to the best classical \gls{nn} (it is not clear, what best means in this case, and to what extend this holds). The authors claim a significant advantage in terms of convergence speed.
    
    Additional experiments are conducted with the proposed Quantum-Glorot initialization technique. In the two environments \texttt{CartPole} and \texttt{Acrobot}, this technique demonstrates to be beneficial in terms of convergence speed and training stability.
    
    Finally, the experiments are extended to a \texttt{QuantumControl} environment. It requires to learn the mapping $\ket{0} \to \ket{1}$ via the time dependent Hamiltonian $H(t) = 4 J(t) \sigma_z + h \sigma_x$. This is converted to a set of unitary gates $U(t)$, such that $\ket{\psi_{t+1}} = U(t) \ket{\psi}$. The reward is defined as the overlap between the prepared state and $\ket{1}$, i.e.\ $r_t = \abs{\expval{\psi_t | 1}}^2$. The agent has to decide between the two actions $0 ~\hat{=}~ \text{no pulse}$ and $1 ~\hat{=}~ \text{apply pulse}$. The usage of a quantum environment removes the necessity of encoding classical states. Unfortunately, it is not described, how $\ket{\psi_t}$ is incorporated in the \gls{vqc} (a $1$-qubit parameterized circuit is apparently used to solve the task). The results on this environment suggest, that the agent is able to learn the optimal pulses in a low number of epochs.
    
    Summarizing the experiments, the authors claim an advantage in convergence speed compared to classical approaches (questionable, as there should be \glspl{nn} which perform much better). Additionally, there seems to be a clear advantage in terms of parameter complexity.
    
    \medbreak
    
    \noindent\textit{Remarks.} The authors claim, that it is possible to estimate the log-policy gradient with only an logarithmic amount of samples (in the number of variational parameters). While this certainly holds for simulation, it is not clear, if such a technique can be applied on quantum hardware (e.g., some kind of sparse or perturbed gradients). The introduced initialization strategy gives some good experimental results, although some additional experiments and theoretical justifications would be desirable. The formulation of the empirical \gls{fim} drops the dependency on the prior state distribution, which potentially renders the considered spectrum less representative of the model than for a generic supervised learning problem. The claim of quantum advantage w.r.t.\ parameter complexity and absence of barren plateaus should be supported with experiments on larger-scale environments.

    \medbreak

    \noindent\textit{Quantum-Accessible Environments.} An explicit analysis of quantum-accessible environments is conducted in Jerbi et al.~\cite{Jerbi_2023}. One instance of such an environment is considered in~\cite{Sequeira_2023}, but also~\cite{Wu_2023} uses a related formulation. The paper derives explicit quadratic advantages in sampling complexity, if the learned policy satisfies certain regularity conditions. We consider this to be a very important step toward identifying the actual potential of \gls{qrl}. Interestingly, the stated results suggest that most of the scenarios studied in literature actually satisfy the smoothness conditions. An open problem is the identification of practically relevant problems that can be formulated in the described quantum-accessible setting.

    %---------------------------------------------
    % Algorithmic characteristics
    % ------------------------------------------------------------
    
     \begin{table}[h!]
    \centering
    \begin{minipage}{\textwidth}
        \renewcommand\footnoterule{}
        \renewcommand{\thefootnote}{\alph{footnote}}
    \caption*{\textbf{Algorithmic Characteristics - Sequeira et al.}~\cite{Sequeira_2023}}
    \small
    \begin{tabular*}{\textwidth}{p{\dimexpr 0.19\textwidth-2\tabcolsep-\arrayrulewidth}|p{\dimexpr 0.14\textwidth-2\tabcolsep-\arrayrulewidth}|p{\dimexpr 0.15\textwidth-2\tabcolsep-\arrayrulewidth}|p{\dimexpr 0.1\textwidth-2\tabcolsep-\arrayrulewidth}|p{\dimexpr 0.1\textwidth-2\tabcolsep-\arrayrulewidth}|p{\dimexpr 0.1\textwidth-2\tabcolsep-\arrayrulewidth}|p{\dimexpr 0.22\textwidth-2\tabcolsep}}
        \toprule
        \centering\multirow{2}{*}{\textbf{Environment}} & \centering\textbf{Algorithm} & \centering\textbf{Quantum} & \centering\textbf{State} & \centering\textbf{Action} & \centering\multirow{2}{*}{\textbf{Qubits}} & \centering\textbf{Parameterized} \tabularnewline
        & \centering\textbf{Type} & \centering\textbf{Component} & \centering\textbf{Space} & \centering\textbf{Space} & & \centering\textbf{Gates} \tabularnewline
        \midrule
        \midrule
        %%%% ROW 1
        \centering\texttt{CartPole} &
        \centering\multirow{2}{*}{\scriptsize{REINFORCE}} &
        \centering\multirow{2}{*}{Policy} &
        \centering{\tiny{continuous}} &
        \centering {\tiny{discrete}} &
        \centering\multirow{2}{*}{$4$} &
        \centering $4$ (encoding) \tabularnewline
        \centering (OpenAI Gym) &
        &
        &
        \centering $4$-dim &
        \centering $2$ &
        &
        \centering $24$ (weights) \tabularnewline
        \midrule
        %%%% ROW 2
        \centering\texttt{Acrobot} &
        \centering\multirow{2}{*}{\scriptsize{REINFORCE}} &
        \centering\multirow{2}{*}{Policy} &
        \centering{\tiny{continuous}} &
        \centering {\tiny{discrete}} &
        \centering\multirow{2}{*}{$6$} &
        \centering $6$ (encoding) \tabularnewline
        \centering (OpenAI Gym) &
        &
        &
        \centering $6$-dim &
        \centering $3$ &
        &
        \centering $36$ (weights) \tabularnewline
        \midrule
        %%%% ROW 3
        \centering\texttt{QuantumControl} &
        \centering\multirow{2}{*}{\scriptsize{REINFORCE}} &
        \centering Policy, &
        \centering\multirow{2}{*}{\scriptsize{quantum}} &
        \centering {\tiny{discrete}} &
        \centering\multirow{2}{*}{$N/A$} &
        \centering $0$ (encoding)\footnotemark[1] \tabularnewline
        \centering (see \cite{Sequeira_2023}) &
        &
        Environment&
        &
        \centering $2$ &
        &
        \centering $N/A$ (weights) \tabularnewline
        \bottomrule
    \end{tabular*}
    \footnotetext[1]{~the \gls{rl} state is a quantum state, i.\@e.\ no classical information has to be encoded;}
    \end{minipage}
\end{table}

%------------------------------------------------------------------------------------------------------%
%------------------------------------------------------------------------------------------------------%
%------------------------------------------------------------------------------------------------------%

%------------------------------------------------------------
\paragraph{\label{subsubsec:Meyer_2023a}Quantum Policy Gradient Algorithm with Optimized Action Decoding, Meyer et al.\\(2023)}\mbox{}\\
%\cite{Meyer_2023b}
%------------------------------------------------------------

    \vspace{-1em}
    \noindent\textit{Summary.} The work by Meyer et al.~\cite{Meyer_2023a} builds upon the \gls{qpg} framework introduced in~\cite{Jerbi_2021a}. It takes a closer look at the introduced \texttt{RAW}-\gls{vqc} policy and -- based on measurements in the computational basis -- introduces a classical post-processing function for action selection. By optimizing this function w.r.t.\ a novel quality measure, significant performance improvements can be made. The introduced procedure is also suited for problems with large action spaces. Experiments on a $5$-qubit quantum device represent the first successful training of a \gls{vqc}-based \gls{rl} routine on actual quantum hardware.

    \medbreak

    \noindent\textit{Classical Post-Processing.} The work focuses on the \texttt{RAW}-\gls{vqc} policy, i.e.\ $\pi_{\boldsymbol{\theta}}(a | \boldsymbol{s}) = \expval{P_a}_{\boldsymbol{s},\boldsymbol{\theta}}$. For measurements in the computational basis, this can be viewed as a partitioning of all possible bitstrings $\mathcal{C}$. This allows the definition of a classical post-processing function $f_{\mathcal{C}}: \lbrace 0, 1 \rbrace^n \to \lbrace 0, 1, \cdots, \abs{\mathcal{A}}-1 \rbrace$, such that $f_{\mathcal{C}}(\boldsymbol{b}) = a$, iff $\boldsymbol{b} \in \mathcal{C}_a$. The policy can therefore be expressed as $\pi_{\boldsymbol{\theta}}(a | \boldsymbol{s}) \approx \frac{1}{K} \cdot \sum_{k=0}^{K-1} \delta_{f_{\mathcal{C}}(\boldsymbol{b}^{(k)})=a}$ where $\boldsymbol{b}^{(k)}$ is the bitstring observed in the $k$-th shot.

    \medbreak

    \noindent\textit{Globality Measure.} The formulation in terms of a classical post-processing function allows for the definition of a quality measure on the explicitly used partitioning of $\mathcal{C}$. The authors start out with the \emph{extracted information} $\text{EI}_{f_{\mathcal{C}}}(\boldsymbol{b})$, which denotes the number of bits necessary to get an unambiguous assignment of the bitstring $\boldsymbol{b}$ to the set $\mathcal{C}_a$ it is contained in. This is extended to a \emph{globality measure} by averaging over all possible bitstrings, i.e. $G_{f_{\mathcal{C}}} := \frac{1}{2^n} \sum_{\boldsymbol{b} \in \lbrace 0, 1 \rbrace^n} EI_{f_{\mathcal{C}}}(\boldsymbol{b})$. This measure quantifies, how much information is used on average to make an decision for an action. While this measure is hard to compute in general, the authors discuss an explicit construction of a post-processing function, that guarantees saturating the globality measure (which is trivially upper-bounded by the number of involved qubits). Based on that construction, an optimal post-processing function is given by $f_{\mathcal{C}}(\boldsymbol{b}) = \left[ b_0 \cdots b_{m-1} \left( \bigoplus_{i=m}^{n-1} b_i \right) \right]_{10}$, where $\left[ \cdot \right]_{10}$ refers to the decimal representation and $m = \log_{2}(\abs{\mathcal{A}})-1$.

    \medbreak

    \noindent\textit{Experimental Results.} The claim that a high value of the globality measure correlates with a good \gls{rl} performance is experimentally demonstrated on several environments. Experiments on the \texttt{CartPole} benchmark with globality values ranging from $G_{f_{\mathcal{C}}} = 1.0$ to the maximum possible $G_{f_{\mathcal{C}}} = 4.0$ show a clear correlation between the measure and the actual performance of the resulting algorithm. It is noted, that the construction of the post-processing function explicitly is detached from the complexity of the actual quantum model, and therefore is a very efficient way to improve the performance. The \gls{qrl} agents with $G_{f_{\mathcal{C}}} > 3.0$ also outperform the $\texttt{SOFTMAX}$-\gls{vqc} policy, which was originally conjectured to be superior in~\cite{Jerbi_2021a}. These results are strengthened by experiments on \texttt{FrozenLake} and \texttt{ContextualBandits} environments. Empirical results regarding effective dimension and the Fisher information spectrum~\cite{Abbas_2021} also demonstrate an improved expressivity and trainability of models with high globality measure.

    \medbreak

    \noindent\textit{Training on Quantum Hardware.} Using this enhanced \gls{qpg} algorithm, the authors execute a full training routine on an $8$-state and $2$-action \texttt{ContextualBandits} environment on quantum hardware. They employ a $3$-qubit sub-topology of the $5$-qubit IBM quantum device \texttt{ibmq\_manila}~\cite{IBMquantum_2023}. The results confirm, that training \gls{vqc}-based \gls{qrl} algorithms on actual hardware is indeed possible. However, there is still a deterioration of performance compared to the noise-free simulation, which is explained by the currently inevitable hardware noise. Verification of the learned parameters demonstrates, that the agent actually identifies the optimal action in all cases, only the certainty of that decision is less pronounced compared to simulation.

    \medbreak

    \noindent\textit{Remarks.} The described action decoding procedure is easy to extend to problems with large action spaces. However, some additional engineering is necessary to account for action spaces of size that cannot be expressed as a power of two. It is left open, at which point the benefit of using a post-processing function with high globality is out-weighted by the likely occurrence of barren plateaus~\cite{Cerezo_2021}. Potentially the flexible definition of the post-processing function can be used to balance those two objectives. While the demonstration of trainability on quantum hardware is certainly pretty small-scale, it can be considered an important step towards the practical usability of these type of algorithms.

    \medbreak

    \begin{table}[h!]
    \centering
    \begin{minipage}{\textwidth}
        \renewcommand\footnoterule{}
        \renewcommand{\thefootnote}{\alph{footnote}}
    \caption*{\textbf{Algorithmic Characteristics - Meyer et al.}~\cite{Meyer_2023a}}
    \small
    \begin{tabular*}{\textwidth}{p{\dimexpr 0.19\textwidth-2\tabcolsep-\arrayrulewidth}|p{\dimexpr 0.14\textwidth-2\tabcolsep-\arrayrulewidth}|p{\dimexpr 0.15\textwidth-2\tabcolsep-\arrayrulewidth}|p{\dimexpr 0.1\textwidth-2\tabcolsep-\arrayrulewidth}|p{\dimexpr 0.1\textwidth-2\tabcolsep-\arrayrulewidth}|p{\dimexpr 0.1\textwidth-2\tabcolsep-\arrayrulewidth}|p{\dimexpr 0.22\textwidth-2\tabcolsep}}
        \toprule
        \centering\multirow{2}{*}{\textbf{Environment}} & \centering\textbf{Algorithm} & \centering\textbf{Quantum} & \centering\textbf{State} & \centering\textbf{Action} & \centering\multirow{2}{*}{\textbf{Qubits}} & \centering\textbf{Parameterized} \tabularnewline
        & \centering\textbf{Type} & \centering\textbf{Component} & \centering\textbf{Space} & \centering\textbf{Space} & & \centering\textbf{~Gates}\footnotemark[1] \tabularnewline
        \midrule
        \midrule
        %%%% ROW 1
        \centering\texttt{CartPole} &
        \centering\multirow{2}{*}{\scriptsize{REINFORCE}} &
        \centering\multirow{2}{*}{Policy} &
        \centering{\tiny{continuous}} &
        \centering {\tiny{discrete}} &
        \centering\multirow{2}{*}{$4$} &
        \centering\multirow{2}{*}{$24$ to $40$\footnotemark[2]} \tabularnewline
        \centering (OpenAI Gym) &
        &
        &
        \centering $4$-dim &
        \centering $2$ &
        & \tabularnewline
        \midrule
        %%%% ROW 2
        \centering\texttt{FrozenLake} &
        \centering\multirow{2}{*}{\scriptsize{REINFORCE}} &
        \centering\multirow{2}{*}{Policy} &
        \centering{\tiny{discrete}} &
        \centering {\tiny{discrete}} &
        \centering\multirow{2}{*}{$4$} &
        \centering\multirow{2}{*}{$24$ to $40$} \tabularnewline
        \centering (OpenAI Gym) &
        &
        &
        \centering $16$ &
        \centering $4$ &
        & 
        \tabularnewline
        \midrule
        %%%% ROW 3
        \centering\texttt{\scriptsize{ContextualBandits}} &
        \centering\multirow{2}{*}{\scriptsize{REINFORCE}} &
        \centering\multirow{2}{*}{Policy} &
        \centering{\tiny{discrete}} &
        \centering {\tiny{discrete}} &
        \centering\multirow{2}{*}{$5$} &
        \centering\multirow{2}{*}{$70$} \tabularnewline
        \centering (see \cite{Sutton_2019}) &
        &
        &
        \centering $32$ &
        \centering $8$ &
        & \tabularnewline
        \midrule
       %%%% ROW 4
        \centering\texttt{\scriptsize{ContextualBandits}} &
        \centering\multirow{2}{*}{\scriptsize{REINFORCE}} &
        \centering\multirow{2}{*}{Policy} &
        \centering{\tiny{discrete}} &
        \centering {\tiny{discrete}} &
        \centering\multirow{2}{*}{$3$} &
        \centering\multirow{2}{*}{$30$} \tabularnewline
        \centering (see \cite{Sutton_2019})\footnotemark[3] &
        &
        &
        \centering $8$ &
        \centering $2$ &
        & \tabularnewline
        \bottomrule
    \end{tabular*}
    \footnotetext[1]{~this entails encoding, scaling, and variational parameters;}
    \footnotetext[2]{~the \texttt{SOFTMAX}-\gls{vqc} also uses additional classical parameters;}
    \footnotetext[3]{~hardware experiment: modified circuit structure to reduce transpilation overhead, details in~\cite{Meyer_2023a};}
    \end{minipage}
\end{table}

%------------------------------------------------------------------------------------------------------%
%------------------------------------------------------------------------------------------------------%
%------------------------------------------------------------------------------------------------------%

%------------------------------------------------------------
\paragraph{\label{subsubsec:Meyer_2023b}Quantum Natural Policy Gradients: Towards Sample-Efficient Reinforcement Learning, Meyer et al.~(2023)}\mbox{}\\
%\cite{Meyer_2023b}
%------------------------------------------------------------

    \vspace{-1em}
    \noindent\textit{Summary.} The paper by Meyer et al.~\cite{Meyer_2023b} proposes an enhanced training routine for the framework proposed in~\cite{Jerbi_2021a} and extended in~\cite{Meyer_2023a}. A second-order extension -- based on so-called quantum natural gradients -- is employed to define the \gls{qnpg} algorithm. The modified technique is experimentally demonstrated to have preferable properties regarding trainability, and is also verified on actual quantum hardware.

    \medbreak

    \noindent\textit{Natural Gradients.} The original \gls{qpg} algorithm is trained based on first-order updates, i.e.\ $\Delta \boldsymbol{\theta} = \alpha \nabla_{\boldsymbol{\theta}}\mathcal{L}(\boldsymbol{\theta})$. This update structure has the shortcoming, that it is closely tied to the Euclidean geometry and does not take into account the actual curvature of the loss landscape. This can be mitigated by using the \gls{fim} $F(\boldsymbol{\theta})$, which describes the local curvature of the parameter space around a given point. This can be used to define a natural gradient update as $\Delta \boldsymbol{\theta} = \alpha F^{-1}(\boldsymbol{\theta}) \nabla_{\boldsymbol{\theta}}\mathcal{L}(\boldsymbol{\theta})$~\cite{Amari_1998}.

    \begin{figure}[ht]
        \centering
        \includegraphics[width=0.7\textwidth]{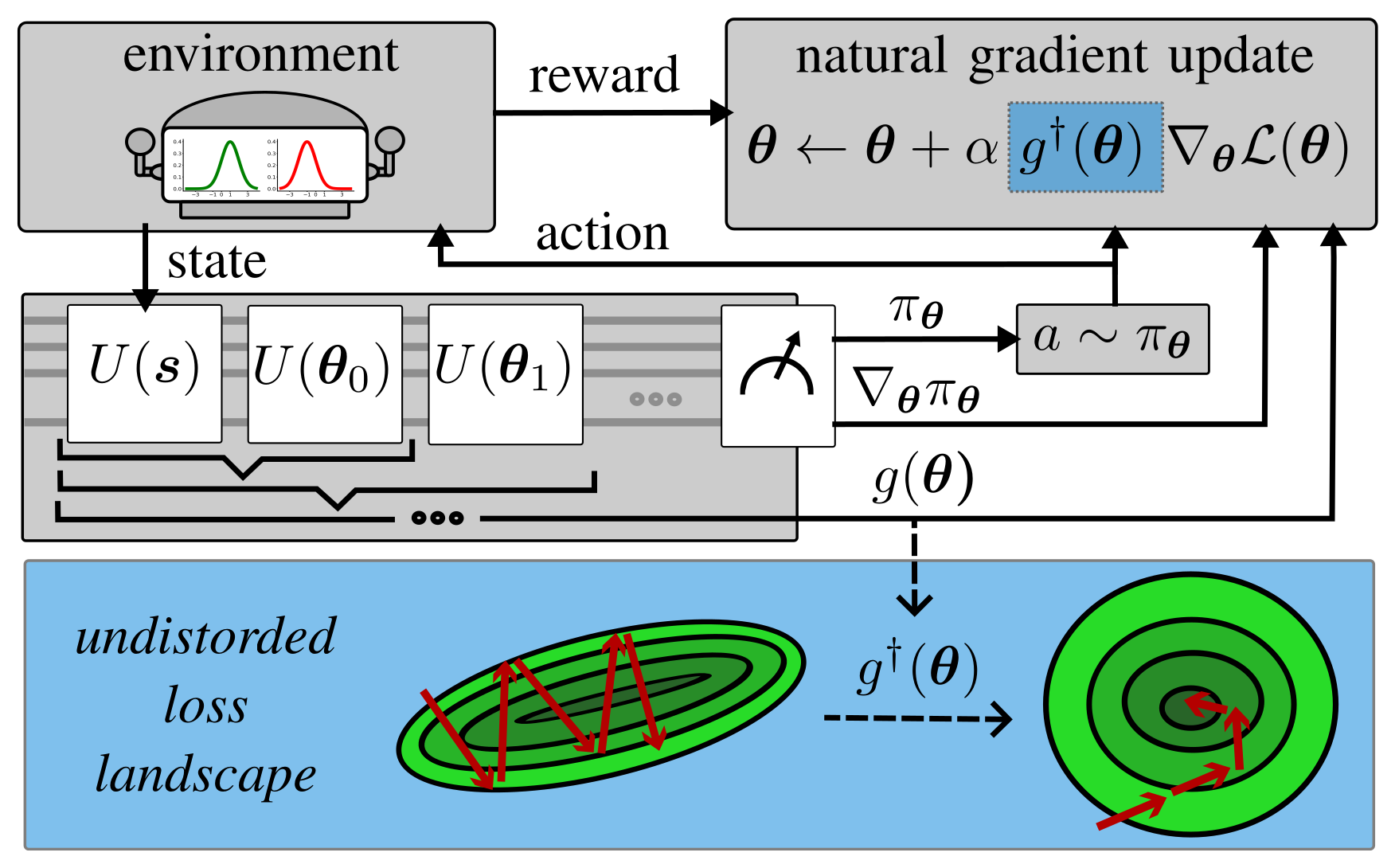}
        \caption{\gls{qnpg} pipeline proposed by Meyer et al.~\cite{Meyer_2023b}; The pseudoinverse of the quantum \gls{fim} is used to perform training in an undistorted neighborhood of the loss landscape.}
        \label{fig:VQC_Meyer_2023b}
    \end{figure}

    \noindent\textit{Quantum Natural Policy Gradients.} In order to employ this concept for training in the quantum realm, the paper employs a generalization of the classical \gls{fim}. This quantum \gls{fim} (derived from the Fubini-Study metric tensor~\cite{Cheng_2010}) $g(\boldsymbol{\theta})$ is hard to compute in general -- however, a block-diagonal approximation can be estimated efficiently in hardware~\cite{Stokes_2020}. Based on that the paper defines the \gls{qnpg} update rule as $\Delta \boldsymbol{\theta} = \alpha g^{\dagger}(\boldsymbol{\theta}) \nabla_{\boldsymbol{\theta}}\mathcal{L}(\boldsymbol{\theta})$. Additionally, a regularized version of the \gls{qnpg} algorithm is introduced, to counter instabilities encountered during inverting the quantum \gls{fim}. It has to be highlighted, that the overhead of incorporating these second-order update rule is almost negligible compared to the anyways necessary computation of first-order gradients. The pipeline of the overall algorithm is visualized in~\cref{fig:VQC_Meyer_2023b}.
    
    \medbreak

    \noindent\textit{Experimental Results.} The effectiveness of the training routine is demonstrated on different instances of \texttt{ContextualBandits}. On a small-scale setting with only a single qubit and two trainable parameters, it is shown that the (regularized) \gls{qnpg} algorithm converges significantly faster for random initializations compared to the original \gls{qpg} formulation. For specific initializations it is moreover validated, that the second-order extension does what it was designed for and helps to traverse distorted regions of the loss landscape. An up-scaled experiment with a $12$-qubit \gls{vqc} underlines the efficiency of the introduced routine.

    \medbreak

    \noindent\textit{Training on Quantum Hardware.} To demonstrate the practical feasibility of the \gls{qpg} approach the authors train an medium-scale instance on actual quantum hardware. The experiment employs a $12$-qubit sub-topology of the $27$-qubit system \texttt{ibmq\_ehningen}~\cite{IBMquantum_2023}. The results demonstrate, that the quantum agent is actually able to learn meaningful behavior in the $4096$-state \texttt{ContextualBandits} environment. There is some deterioration of the performance compared to noise-free simulation, which is not caused by the training routine itself, as demonstrated by experiments with analytically optimal parameters. However, the learned policy identifies the correct action in a majority of the cases, similar to the hardware results in~\cite{Meyer_2023a}.

    \medbreak

    \noindent\textit{Remarks.} The paper demonstrates the effectiveness of the \gls{qnpg} routine for \texttt{ContextualBandits} environment, the extension to more generic problems is however left for future work. A very interesting consideration is the influence of quantum natural gradients on the barren plateau problem, which is discussed with different results in Refs.~\cite{Haug_2021,Thanasilp_2023}. The hardware experiment using $12$ qubits is a big improvement upon the results in~\cite{Meyer_2023a} and can be considered as the currently largest-scale practical demonstration of \gls{vqc}-based \gls{qrl}.

    \medbreak

    \begin{table}[h!]
    \centering
    \begin{minipage}{\textwidth}
        \renewcommand\footnoterule{}
        \renewcommand{\thefootnote}{\alph{footnote}}
    \caption*{\textbf{Algorithmic Characteristics - Meyer et al.}~\cite{Meyer_2023b}}
    \small
    \begin{tabular*}{\textwidth}{p{\dimexpr 0.19\textwidth-2\tabcolsep-\arrayrulewidth}|p{\dimexpr 0.14\textwidth-2\tabcolsep-\arrayrulewidth}|p{\dimexpr 0.15\textwidth-2\tabcolsep-\arrayrulewidth}|p{\dimexpr 0.1\textwidth-2\tabcolsep-\arrayrulewidth}|p{\dimexpr 0.1\textwidth-2\tabcolsep-\arrayrulewidth}|p{\dimexpr 0.1\textwidth-2\tabcolsep-\arrayrulewidth}|p{\dimexpr 0.22\textwidth-2\tabcolsep}}
        \toprule
        \centering\multirow{2}{*}{\textbf{Environment}} & \centering\textbf{Algorithm} & \centering\textbf{Quantum} & \centering\textbf{State} & \centering\textbf{Action} & \centering\multirow{2}{*}{\textbf{Qubits}} & \centering\textbf{Parameterized} \tabularnewline
        & \centering\textbf{Type} & \centering\textbf{Component} & \centering\textbf{Space} & \centering\textbf{Space} & & \centering\textbf{~Gates} \tabularnewline
        \midrule
        \midrule
        %%%% ROW 1
        \centering\texttt{\scriptsize{ContextualBandits}} &
        \centering\multirow{2}{*}{\gls{qnpg}} &
        \centering\multirow{2}{*}{Policy} &
        \centering{\tiny{discrete}} &
        \centering {\tiny{discrete}} &
        \centering\multirow{2}{*}{$1$} &
        \centering $1$ (encoding) \tabularnewline
        \centering (see \cite{Sutton_2019}) &
        &
        &
        \centering $2$ &
        \centering $2$ &
        & \centering $2$ (weights) \tabularnewline
        \midrule
       %%%% ROW 2
        \centering\texttt{\scriptsize{ContextualBandits}} &
        \centering\multirow{2}{*}{\gls{qnpg}} &
        \centering\multirow{2}{*}{Policy} &
        \centering{\tiny{discrete}} &
        \centering {\tiny{discrete}} &
        \centering\multirow{2}{*}{$12$} &
        \centering $12$ (encoding) \tabularnewline
        \centering (see \cite{Sutton_2019})\footnotemark[1] &
        &
        &
        \centering $4096$ &
        \centering $2$ &
        & \centering $36$ (weights) \tabularnewline
        \bottomrule
    \end{tabular*}
    \footnotetext[1]{~hardware experiment: hardware-native circuit structure, details in~\cite{Meyer_2023b};}
    \end{minipage}
\end{table}

\subsubsection{Combined Approximations}
\label{subsec:VQC_based_CombinationApproximations}
It is possible to combine the approach of approximation in value space from \cref{subsec:VQC_based_ValueFunction} and in policy space from \cref{subsec:VQC_based_Policy}. This is formulated in an actor-critic approach in Wu et al.~\cite{Wu_2023}, which is re-implemented and extended in Refs.~\cite{Kwak_2021,Reers_2023}. An asynchronous training routine is proposed by S. Y.-C. Chen~\cite{Chen_2023c}. A soft actor-critic formulation is described by Q. Lan~\cite{Lan_2021}. An extension to multiple agents is proposed in Yun et al.~\cite{Yun_2022} and extended in Ref.~\cite{Yun_2023a}.

An overview of progress in the field of quantum multi-agent \gls{rl} can be found in Ref.~\cite{Zhao_2023}.

\begin{table}[ht!]
    \centering
    \begin{tabular}{p{\dimexpr 0.15\textwidth-2\tabcolsep-\arrayrulewidth}|p{\dimexpr 0.2\textwidth-2\tabcolsep-\arrayrulewidth}|p{\dimexpr 0.65\textwidth-2\tabcolsep}}
        \toprule
        \textbf{Citation} & \textbf{First Author} & \textbf{Title} \\
        \midrule
        \midrule
        \cite{Wu_2023} & S. Wu & \hyperref[subsubsec:Wu_2023]{Quantum reinforcement learning in continuous action space} \\
        \midrule
        \cite{Kwak_2021} & Y. Kwak & \hyperref[subsubsec:Kwak_2021]{Introduction to Quantum Reinforcement Learning: Theory and PennyLane-based Implementation} \\
        \arrayrulecolor{black!30}\midrule
        \cite{Reers_2023} & V. Reers & \hyperref[subsubsec:Kwak_2021]{Towards Performance Benchmarking for Quantum Reinforcement Learning} \\
        \arrayrulecolor{black}\midrule
        \cite{Chen_2023c} & S. Y.-C. Chen & \hyperref[subsubsec:Chen_2023c]{Asynchronous training of quantum reinforcement learning} \\
        \midrule
        \cite{Lan_2021} & Q. Lan & \hyperref[subsubsec:Lan_2021]{Variational Quantum Soft Actor-Critic} \\
        \midrule
        \cite{Yun_2022} & W. J. Yun & \hyperref[subsubsec:Yun_2022]{Quantum Multi-Agent Reinforcement Learning via Variational Quantum Circuit Design} \\
        \midrule
        \cite{Yun_2023a} & W. J. Yun & \hyperref[subsubsec:Yun_2023a]{Quantum Multi-Agent Meta Reinforcement Learning} \\
        \arrayrulecolor{black}\bottomrule
    \end{tabular}
    \caption{Work considered for ``\gls{qrl} with \glspl{vqc} -- Combined Approximations'' (\cref{subsec:VQC_based_CombinationApproximations})}
\end{table}

%------------------------------------------------------------
\paragraph{\label{subsubsec:Wu_2023}Quantum reinforcement learning in continuous action space, Wu et al.~(2023)}\mbox{}\\
%\cite{Wu_2023}
%------------------------------------------------------------

    \vspace{-1em}
    \noindent\textit{Summary.} This paper by Wu et al.~\cite{Wu_2023} extends the concept of \gls{vqc}-based \gls{rl} to continuous action spaces. The authors choose a quantum control environment, more concretely one that encodes an eigenvalue problem. This allows to interpret the action as a (parameterized) unitary. The experimental results suggest an exponential reduction in model complexity compared to classical approaches.
    
    \medbreak
    
    \noindent\textit{Eigenvalue Problem as \gls{rl} Environment.} The \gls{rl} agent has to solve an eigenvalue problem, i.e., find the eigenvalue of a given Hamiltonian. This should be done in the following iterative procedure: Let $H$ be the Hamiltonian of an $n$-qubit quantum system $E$ and $\ket{s_0}$ an initial state from $E$. The system should be driven towards the eigenstate of $H$, denoted as $\ket{u_0}$. Also the corresponding eigenvalue $\lambda_0$ should be returned. Although not explicitly stated in the paper, we assume the agent should search for the eigenstate with the associated smallest eigenvalue, as this corresponds to the ground state.
    
    The observation for this environment is the current quantum state $\ket{s_t}$, which is provided to the agent via some quantum channel. The actions the agent can execute correspond to parameterized unitaries $U(\theta_t)$, where $\theta_t$ are classical parameters sampled from the VQC via measurements. Once instantiated, this unitary is applied to $\ket{s_t}$ to evolve the state $U(\theta_t) \ket{s_t} = \ket{s_{t+1}}$. The agent receives a classical reward, which describes the closeness of the current state to the searched eigenstate of the Hamiltonian.
    
    The authors state, that their proposed technique has some parallels to Grover's search. More concretely, the trained agent provides an alternative to the amplitude amplification procedure, which could alternatively be used to solve the task at hand.
    
    \medbreak
    
\begin{figure}[ht]
        \centering
            \subfloat[\centering The QRL model. Each iterative step can be described by the following loop: (1) at step $t$, the agent receives $\ket{s_t}$ and generates the action parameter $\boldsymbol{\theta}_t$ according to the current policy; (2) the agent generates $\ket{s_{t+1}} = U(\boldsymbol{\theta}_t)\ket{s_t}$ (3) based on $\ket{s_t}$ and $\ket{s_{t+1}}$, a reward $r_{t+1}$ is calculated and fed back to the agent, together with $\ket{s_{t+1}}$; (4) based on $s_{t+1}$ and $r_{t+1}$, the policy is updated and then used to generate $\boldsymbol{\theta}_{t+1}$.]{\includegraphics[width=0.45\textwidth]{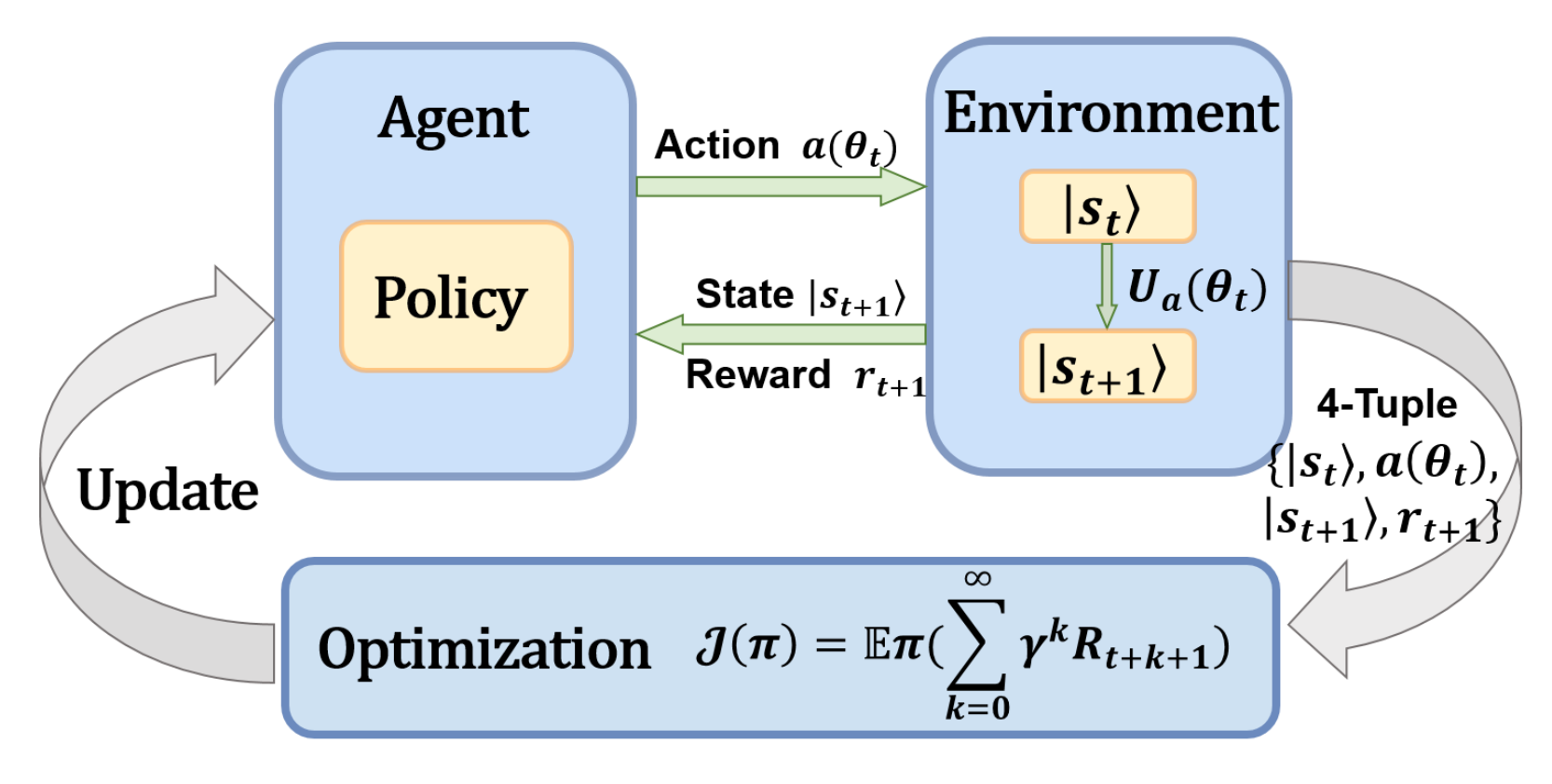}}
        \qquad
            \subfloat[\centering The quantum circuit for our QRL framework at each iteration. The entire \gls{qrl} process includes two stages, so we give the circuit separately. In stage 1, the circuit includes two registers: the reward register, initialized $\ket{0}$, and the environment register $\ket{s_t}$. $U_{policy}$ is generated by the quantum neural network, and determines the action unitary $U(\boldsymbol{\theta}_t)$. $U_r$ and $M$ are designed to generate the reward $r_{t+1}$. In stage 2, the circuit has only environment register and does not need to feedback the reward value and update the policy.]{\includegraphics[width=0.45\textwidth]{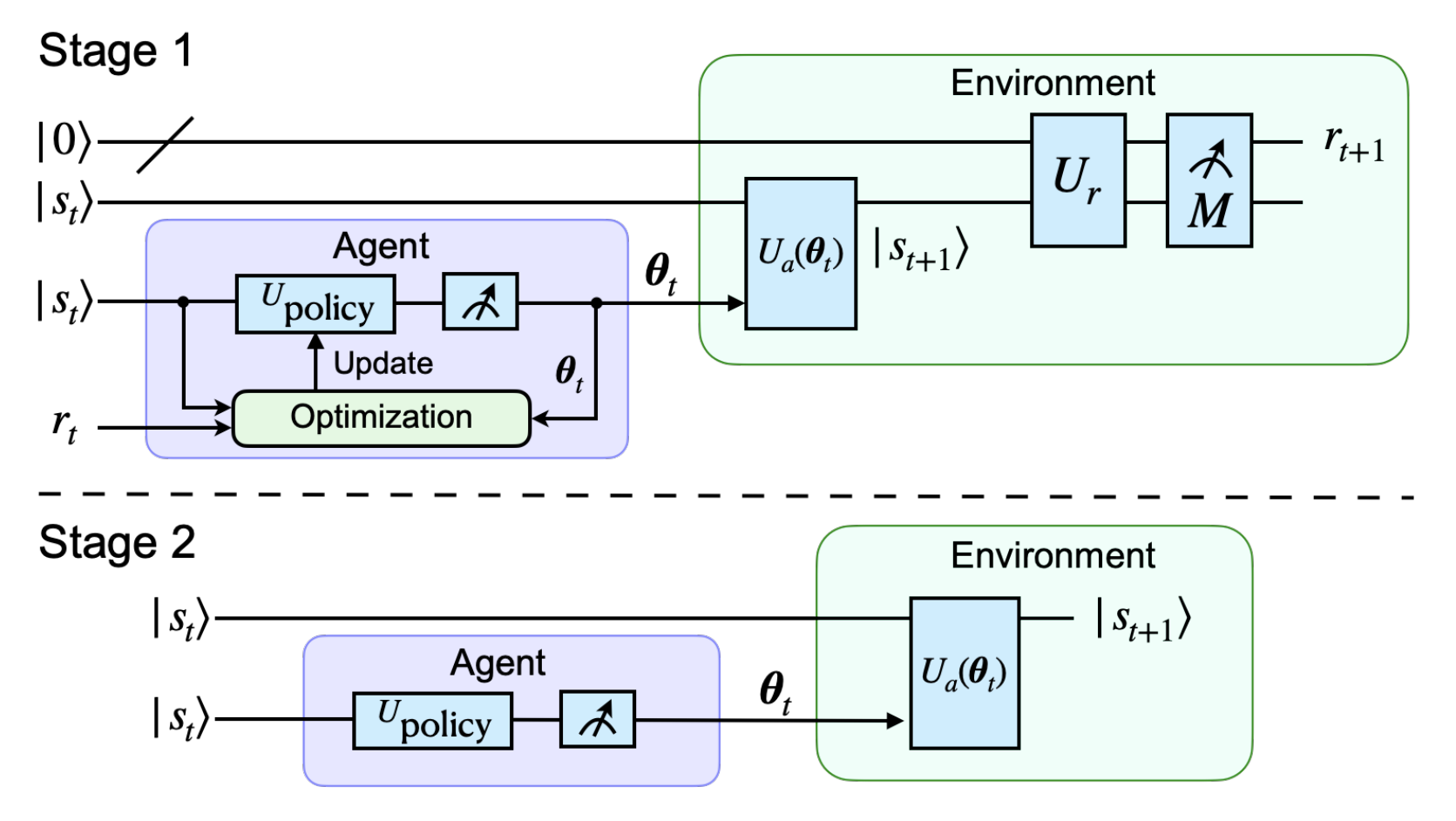}}
        \caption{Hybrid model for quantum environment proposed by and taken from Wu et al.~\cite{Wu_2023} (including subcaptions); We note an ambiguity in notation, as the parameters $\theta_{t,i}$ must not be confused with the parameters of the action unitary $U(\theta)$. The first set are the ones updated by the \gls{rl} algorithm, the other ones are extracted via measurements from the quantum state prepared by the \gls{vqc}.}
        \label{fig:ContinuousAction_model}
    \end{figure}

    \noindent\textit{Model Architecture and Underlying \gls{rl} Algorithm.} The overall approach can be considered hybrid, as the optimization of the \gls{vqc} parameters is still conducted on classical hardware. A schematic description of the approach is given in \cref{fig:ContinuousAction_model}. The agent observes a quantum state from the environment, which is used as the initial state $\ket{s_t}$ of the \gls{vqc} function approximator. Measurements on the prepared quantum state determine the parameters $\ket{\theta_t}$. Those are then fed into the unitary operator $\ket{U(\theta_t)}$ and applied to the environment state. The new state $\ket{s_{t+1}}$, combined with an ancilla reward qubit initialized to $\ket{0}$, is then evolved using some user-defined reward unitary $U_r$. Measurements are performed on this state to determine the reward produced by the executed action. This procedure repeats several timesteps, with the objective to approximate the eigenstate $\ket{u_0}$.
    
    The \gls{vqc} architecture does not incorporate a feature map, as the observation $\ket{s_t}$ is used as the initial state $\ket{\Phi}$. Each parameterized layer consists of $1$-qubit rotations and a circular entanglement structure. For every element of the action parameters $\theta_j$, there is an associated observable $B_j$, which is measured on the prepared quantum state. (The paper does not mention, how the action unitary $U(\theta)$ is explicitly constructed.) Following this step, a phase estimation circuit implements the reward unitary $U_r=U_{PE}$. This transforms the state to the basis of eigenstates, i.e., $U_{PE} \ket{0}\ket{s_{t+1}} = \sum_{k=1}^{n} \alpha_{t+1,k} \ket{\lambda_k}\ket{u_k}$. With a measurement of the eigenvalue phase register, the desired eigenvalue $\lambda_0$ is observed with a probability of $p_{t+1}=|\alpha_{t+1,0}|^2=|\expval{s_{t+1} | u_0}|^2$. The reward can then be defined as e.g.\ $r_{t+1} = p_{t+1}-p_t$. Obviously, for $p_{t+1} \to 1$, the state $\ket{s_{t+1}}$ converges to $\ket{u_0}$.
    
    The underlying \gls{rl} routine is an actor-critic method. Therefore, the paper combines a policy-\gls{vqc} as actor and a $Q$-function-\gls{vqc} as critic to a so-called \gls{qddpg} algorithm. The experience of the agent, i.e., tuples $(\ket{s_t}, \theta_t, r_t, \ket{s_{t+1}})$, are stored in a replay buffer to prevent overfitting. Additionally, target networks are employed for both, the actor and the critic.
    
    \medbreak
    
    \noindent\textit{Experimental Results and Model Complexity.} All experimental results in the paper are based on classical simulations. For training, the Hamiltonian $H=\frac{1}{4}(s_x \sigma_x + s_y \sigma_y + s_z \sigma_z + I)$ is instantiated with the coefficients $(s_x, s_y, s_z) = (0.13, 0.28, 0.95)$. Concrete details on the training procedure, e.g., the number of episodes, are not stated. The trained model is applied to $1000$ random initial states. The overlap with the respective $\ket{u_0}$ is approaching one, consequently the agent is able to get quite close to the desired eigenstate in all cases. The trained model shows good generalization capabilities, i.e., it can be applied to various initial states. This is in contrast to e.g.\ a \gls{vqe}, where the control pulse for one initial state is meaningless for other ones.
    
    The overall gate complexity for one \gls{rl} episode is stated as $\mathcal{O}(m\cdot \mathrm{polylog}(N))$. Here, $m$ is the number of shots for sampling expectation values and $N$ denotes the number of qubits. This statement assumes that $H$ can be efficiently simulated as otherwise the complexity of $U_{PE}$ would exceed $\mathcal{O}(\mathrm{polylog}(N))$. Additionally, all \glspl{vqc} in the method are also assumed to have a gate complexity of at most $\mathcal{O}(\mathrm{polylog}(N))$. With this perquisites, the authors claims an exponential advantage in model complexity compared to classical approaches.
    
    \medbreak
    
    \noindent\textit{Generalization to Discrete Action Spaces.} The paper also generalizes the presented concept to discrete action spaces, with the \texttt{FrozenLake} environment as an example. The observations are encoded as basis states into the \gls{vqc} via computational encoding, similar to Chen et al.~\cite{Chen_2020}. The movements applied by the actions are formulated as unitaries acting on the \gls{vqc} state. A slight generalization of Chen et al.~\cite{Chen_2020} is used for this, which allows to perform the transforms $\ket{0} \to \ket{1}$ and $\ket{1} \to \ket{0}$ in a parameterized manner. The reward unitary is formulated in a similar fashion. It is stated that experiments with this configuration were successful, but no concrete results are provided.
    
    \medbreak
    
    \noindent\textit{Remarks.} There are some caveats and ambiguities we identified regarding the proposed approach. First, the algorithm requires knowledge of and ability to prepare the desired eigenstate $\ket{u_0}$ for the training procedure. With this state already known, the whole procedure of reproducing it is a somewhat circular task. However, as the learned model seems to generalize to different input states, the technique offers clear advantage over approaches like quantum phase estimation. Second, the model requires repeated preparation of the environment state $\ket{s_t}$, as it is disturbed by measurements to extract the reward information. This should be doable, as one knows the state preparation routine $\ket{s_t} = U(\theta_{t-1})\cdots U(\theta_0)\ket{s_0}$. The influence of this additional overhead is unfortunately not considered in the complexity considerations discussed above. Third, the claim of exponential quantum advantage w.r.t.\ model complexity (i.e.\ $\mathcal{O}(\mathrm{polylog}(N))$ for all \glspl{vqc}) should be supported by larger-scale experiments.
    
    \medbreak
    
     \begin{table}[ht]
    \centering
    \begin{minipage}{\textwidth}
        \renewcommand\footnoterule{}
        \renewcommand{\thefootnote}{\alph{footnote}}
    \caption*{\textbf{Algorithmic Characteristics - Wu et al.}~\cite{Wu_2023}}
    \small
    \begin{tabular*}{\textwidth}{p{\dimexpr 0.19\textwidth-2\tabcolsep-\arrayrulewidth}|p{\dimexpr 0.14\textwidth-2\tabcolsep-\arrayrulewidth}|p{\dimexpr 0.15\textwidth-2\tabcolsep-\arrayrulewidth}|p{\dimexpr 0.1\textwidth-2\tabcolsep-\arrayrulewidth}|p{\dimexpr 0.1\textwidth-2\tabcolsep-\arrayrulewidth}|p{\dimexpr 0.1\textwidth-2\tabcolsep-\arrayrulewidth}|p{\dimexpr 0.22\textwidth-2\tabcolsep}}
        \toprule
        \centering\multirow{2}{*}{\textbf{Environment}} & \centering\textbf{Algorithm} & \centering\textbf{Quantum} & \centering\textbf{State} & \centering\textbf{Action} & \centering\multirow{2}{*}{\textbf{Qubits}} & \centering\textbf{Parameterized} \tabularnewline
        & \centering\textbf{Type} & \centering\textbf{Component} & \centering\textbf{Space} & \centering\textbf{Space} & & \centering\textbf{Gates} \tabularnewline
        \midrule
        \midrule
        %%%% ROW 1
        \centering\texttt{Quantum} &
        \centering\multirow{3}{*}{\begin{tabular}{@{}c@{}}Actor-Critic\\``\gls{qddpg}''\end{tabular}} &
        \centering $Q$-function, &
        \centering\multirow{3}{*}{\scriptsize{quantum}} &
        \centering\multirow{3}{*}{\begin{tabular}{@{}c@{}}{\scriptsize{conti-}}\\{\scriptsize{nuous}}\footnotemark[1]\end{tabular}} &
        \centering\multirow{3}{*}{$n$} &
        \centering\multirow{3}{*}{\begin{tabular}{@{}c@{}}$0$ (encoding)\footnotemark[2] \\$n \times d \times 3$ (weights)\footnotemark[3]\end{tabular}} \tabularnewline
        \centering\texttt{Eigenvalues} &
        &
        \centering Policy, &
        &
        &
        &
        \tabularnewline
        \centering (see \cite{Wu_2023}) &
        &
        \centering Environment &
        &
        &
        &
        \tabularnewline
        \midrule
        %%%% ROW 2
        \centering\texttt{FrozenLake} &
        \centering Actor-Critic &
        \centering $Q$-function, &
        \centering{\tiny{discrete}}\footnotemark[4] &
        \centering {\tiny{discrete}}\footnotemark[4] &
        \centering\multirow{2}{*}{$n$} &
        \centering $0$ (encoding)\footnotemark[2] \tabularnewline
        \centering (OpenAI Gym) &
        \centering ``\gls{qddpg}'' &
        \centering Policy &
        \centering $16$ &
        \centering $4$ &
        &
        \centering $N/A$ (weights) \tabularnewline
        \bottomrule
    \end{tabular*}
    \footnotetext[1]{~output is interpreted as parameters of a unitary, i.\@e.\ a quantum operation applied to the environment;}
    \footnotetext[2]{~the \gls{rl} state is a quantum state, i.\@e.\ no classical information has to be encoded;}
    \footnotetext[3]{~variational gates: $qubits \times layers \times per\_qubit\_per\_layer$; details are not specified;}
    \footnotetext[4]{~state and action space are encoded into the quantum realm for a neat integration into the pipeline;}
    \end{minipage}
\end{table}
    
%------------------------------------------------------------------------------------------------------%
%------------------------------------------------------------------------------------------------------%
%------------------------------------------------------------------------------------------------------%

%------------------------------------------------------------
\paragraph{\label{subsubsec:Kwak_2021}Introduction to Quantum Reinforcement Learning: Theory and PennyLane-based Implementation, Kwak et al.~(2021)}\mbox{}\\
% \cite{Kwak_2021}
%------------------------------------------------------------

    \vspace{-1em}
    \noindent\textit{Summary.} The paper by Kwak et al.~\cite{Kwak_2021} gives a short introduction to both \gls{rl} and (variational) \gls{qc}. This is followed up by a tutorial on how to implement a \gls{vqc}-enhanced \gls{rl} algorithm with \texttt{PennyLane} to solve the \texttt{CartPole} environment.
    
    \medbreak
    
    \noindent\textit{Hybrid \gls{rl} Agent.} The paper employs the typical hybrid structure, with the \gls{vqc} as a function approximator. The optimization of the parameters and the interaction with the \texttt{CartPole} environment is executed on classical hardware. The underlying algorithm uses an actor-critic approach, where the actor is quantum and the critic is classical. A set of $1$-qubit rotations is used to encode the state of the \texttt{CartPole} environment into the four-qubit system. This encoding layer is followed by $4$ layers with learnable $1$-qubit rotations and an unspecified entanglement structure. The result is extracted from the measurement of $2$ qubits in the computational basis and the respective expectation values are interpreted as the action-value function.
    
    \medbreak
    
    \noindent\textit{Remarks.} The agent is able to surpass random behavior, but lacks behind other hybrid approaches~\cite{Lockwood_2020,Jerbi_2021a}. To the best of our understanding, the implemented quantum actor-critic approach deviates in some details from previously considered approaches. Most importantly, a hybrid approach is used, where the actor is represented with a \gls{vqc} and the critic employs a classical \gls{dnn}. A benchmark analysis of the described setup is proposed and conducted by V. Reers~\cite{Reers_2023}.
    
    \medbreak
    
     \begin{table}[h]
    \centering
    \begin{minipage}{\textwidth}
        \renewcommand\footnoterule{}
        \renewcommand{\thefootnote}{\alph{footnote}}
    \caption*{\textbf{Algorithmic Characteristics - Kwak et al.}~\cite{Kwak_2021}}
    \small
    \begin{tabular*}{\textwidth}{p{\dimexpr 0.19\textwidth-2\tabcolsep-\arrayrulewidth}|p{\dimexpr 0.14\textwidth-2\tabcolsep-\arrayrulewidth}|p{\dimexpr 0.15\textwidth-2\tabcolsep-\arrayrulewidth}|p{\dimexpr 0.1\textwidth-2\tabcolsep-\arrayrulewidth}|p{\dimexpr 0.1\textwidth-2\tabcolsep-\arrayrulewidth}|p{\dimexpr 0.1\textwidth-2\tabcolsep-\arrayrulewidth}|p{\dimexpr 0.22\textwidth-2\tabcolsep}}
        \toprule
        \centering\multirow{2}{*}{\textbf{Environment}} & \centering\textbf{Algorithm} & \centering\textbf{Quantum} & \centering\textbf{State} & \centering\textbf{Action} & \centering\multirow{2}{*}{\textbf{Qubits}} & \centering\textbf{Parameterized} \tabularnewline
        & \centering\textbf{Type} & \centering\textbf{Component} & \centering\textbf{Space} & \centering\textbf{Space} & & \centering\textbf{~Gates}\footnotemark[1] \tabularnewline
        \midrule
        \midrule
        %%%% ROW 1
        \centering\texttt{CartPole} &
        \centering\multirow{2}{*}{{\footnotesize{Actor-Critic}}\footnotemark[2]} &
        \centering\multirow{2}{*}{$Q$-function} &
        \centering{\tiny{continuous}} &
        \centering {\tiny{discrete}} &
        \centering\multirow{2}{*}{$4$} &
        \centering $4 \times 1$ (encoding) \tabularnewline
        \centering (OpenAI Gym) &
        &
        &
        \centering $4$-dim &
        \centering $2$ &
        &
        \centering $4 \times 4 \times 3$ (weights) \tabularnewline
        \bottomrule
    \end{tabular*}
    \footnotetext[1]{~encoding gates: $qubits \times per\_qubit$; variational gates: $qubits \times layers \times per\_qubit\_per\_layer$;}
    \footnotetext[2]{~only the actor employs a \gls{vqc}, the critic uses a classical \gls{dnn};}
    \end{minipage}
\end{table}
    
%------------------------------------------------------------------------------------------------------%
%------------------------------------------------------------------------------------------------------%
%------------------------------------------------------------------------------------------------------%

%------------------------------------------------------------
\paragraph{\label{subsubsec:Chen_2023c}Asynchronous training of quantum reinforcement learning, S. Y.-C. Chen~(2023)}\mbox{}\\
%\cite{Chen_2023c}
%------------------------------------------------------------

    \vspace{-1em}
    \noindent\textit{Summary.} This work by S. Y.-C. Chen~\cite{Chen_2023c} introduces an actor-critic approach, that is trainable in an asynchronous fashion. This yields the big advantage, that training could be spread out over several classical simulators or quantum hardware devices. The efficiency of the introduced \gls{qa3c} algorithm compared to previous formulations is demonstrated on several benchmark environments. 

    \medbreak

    \noindent\textit{Quantum A3C.} The underlying concept is based on the classical A3C algorithm~\cite{Mnih_2016}. This framework makes use of a global shared memory and a process-specific memory for each individual agent. Each agent interacts with the environment independently, and only once certain criteria are met the global model is updated using the information provided by the local agents. This enables a distributed and therefore easy parallelizable training routine. The approximator for $Q$-function and policy both are realized using \glspl{vqc} with classical neural networks pre- and appended to form a hybrid model.

    \medbreak

    \noindent\textit{Experimental Results.} The proposed \gls{qa3c} algorithm is executed on the environments \texttt{Acrobot}, \texttt{CartPole}, and \texttt{MiniGrid-SimpleCrossing}. It is observed over all instances, that the hybrid quantum model is competitive with a much larger classical model. Moreover it is demonstrated, that \gls{qa3c} outperforms classical A3C employing classical models of comparable complexity.

    \medbreak

    \noindent\textit{Remarks.} The distribution of the training among several workers is certainly an important consideration taking the current access modalities of quantum hardware providers into account. However, it is not clear if training practically can be distributed considering the long queue waiting times. Moreover, it has to be taken into account, that it is not clear what actually is the role of the \gls{vqc}, due to the appended neural networks. However, the comparison to full-classical agents of similar size is an interesting consideration. As usually it has to be highlighted that the experiments were to small-scale to make meaningful statements on potential quantum advantage.

    \medbreak
    
    \begin{table}[ht]
    \centering
    \begin{minipage}{\textwidth}
        \renewcommand\footnoterule{}
        \renewcommand{\thefootnote}{\alph{footnote}}
    \caption*{\textbf{Algorithmic Characteristics - S. Y.-C. Chen et al.}~\cite{Chen_2023c}}
    \small
    \begin{tabular*}{\textwidth}{p{\dimexpr 0.19\textwidth-2\tabcolsep-\arrayrulewidth}|p{\dimexpr 0.14\textwidth-2\tabcolsep-\arrayrulewidth}|p{\dimexpr 0.15\textwidth-2\tabcolsep-\arrayrulewidth}|p{\dimexpr 0.1\textwidth-2\tabcolsep-\arrayrulewidth}|p{\dimexpr 0.1\textwidth-2\tabcolsep-\arrayrulewidth}|p{\dimexpr 0.1\textwidth-2\tabcolsep-\arrayrulewidth}|p{\dimexpr 0.22\textwidth-2\tabcolsep}}
        \toprule
        \centering\multirow{2}{*}{\textbf{Environment}} & \centering\textbf{Algorithm} & \centering\textbf{Quantum} & \centering\textbf{State} & \centering\textbf{Action} & \centering\multirow{2}{*}{\textbf{Qubits}} & \centering\textbf{Parameterized} \tabularnewline
        & \centering\textbf{Type} & \centering\textbf{Component} & \centering\textbf{Space} & \centering\textbf{Space} & & \centering\textbf{~Gates}\footnotemark[1] \tabularnewline
        \midrule
        \midrule
        %%%% ROW 1
        \centering\multirow{3}{*}{\begin{tabular}{@{}c@{}}\texttt{Acrobot}\\(OpenAI Gym)\end{tabular}} &
        \centering\multirow{3}{*}{\begin{tabular}
        {@{}c@{}}Actor-critic\\''\gls{qa3c}``\end{tabular}} &
        \centering\multirow{3}{*}{\begin{tabular}
        {@{}c@{}}$Q$-function,\\Policy\end{tabular}} &
        \centering\multirow{3}{*}{\begin{tabular}{@{}c@{}}\tiny{continuous}\\$6$-dim\footnotemark[3]\end{tabular}} &
        \centering\multirow{3}{*}{\begin{tabular}{@{}c@{}}\tiny{discrete}\\$3$\footnotemark[3]\end{tabular}} &
        \centering\multirow{3}{*}{$8$} &
        \centering $N/A$ (encoding) \tabularnewline
        &
        &
        &
        &
        &
        & 
        \centering $48$ (weights)\footnotemark[2] \tabularnewline
        &
        &
        &
        &
        &
        & 
        \centering $148$ (classical) \tabularnewline
        \midrule
        %%%% ROW 2
        \centering\multirow{3}{*}{\begin{tabular}{@{}c@{}}\texttt{CartPole}\\(OpenAI Gym)\end{tabular}} &
        \centering\multirow{3}{*}{\begin{tabular}
        {@{}c@{}}Actor-critic\\''\gls{qa3c}``\end{tabular}} &
        \centering\multirow{3}{*}{\begin{tabular}
        {@{}c@{}}$Q$-function,\\Policy\end{tabular}} &
        \centering\multirow{3}{*}{\begin{tabular}{@{}c@{}}\tiny{continuous}\\$4$-dim\footnotemark[3]\end{tabular}} &
        \centering\multirow{3}{*}{\begin{tabular}{@{}c@{}}\tiny{discrete}\\$2$\footnotemark[3]\end{tabular}} &
        \centering\multirow{3}{*}{$8$} &
        \centering $N/A$ (encoding) \tabularnewline
        &
        &
        &
        &
        &
        & 
        \centering $48$ (weights)\footnotemark[2] \tabularnewline
        &
        &
        &
        &
        &
        & 
        \centering $107$ (classical) \tabularnewline
        \midrule
         %%%% ROW 3
        \centering\multirow{3}{*}{\begin{tabular}{@{}c@{}}\texttt{SimpleCrossing}\\(OpenAI Gym)\end{tabular}} &
        \centering\multirow{3}{*}{\begin{tabular}
        {@{}c@{}}Actor-critic\\''\gls{qa3c}``\end{tabular}} &
        \centering\multirow{3}{*}{\begin{tabular}
        {@{}c@{}}$Q$-function,\\Policy\end{tabular}} &
        \centering\multirow{3}{*}{\begin{tabular}{@{}c@{}}\tiny{continuous}\\$127$-dim\footnotemark[3]\end{tabular}} &
        \centering\multirow{3}{*}{\begin{tabular}{@{}c@{}}\tiny{discrete}\\$6$\footnotemark[3]\end{tabular}} &
        \centering\multirow{3}{*}{$8$} &
        \centering $N/A$ (encoding) \tabularnewline
        &
        &
        &
        &
        &
        & 
        \centering $48$ (weights)\footnotemark[2] \tabularnewline
        &
        &
        &
        &
        &
        & 
        \centering $2431$ (classical) \tabularnewline
        \bottomrule
    \end{tabular*}
    \footnotetext[1]{~the training process is distributed over $80$ workers, which incorporate a local copy of the parameters;}
    \footnotetext[2]{~actor and the critic are composed of an individual hybrid model, i.e. the number of weights are doubled;}
    \footnotetext[3]{~action and state-spaces are mapped to the required dimensionality by using classical neural networks;}
    \end{minipage}
\end{table}

%------------------------------------------------------------------------------------------------------%
%------------------------------------------------------------------------------------------------------%
%------------------------------------------------------------------------------------------------------%

%------------------------------------------------------------
\paragraph{\label{subsubsec:Lan_2021}Variational Quantum Soft Actor-Critic, Q.\ Lan (2021)}\mbox{}\\
% \cite{Lan_2021}
%------------------------------------------------------------

    \vspace{-1em}
    \noindent\textit{Summary.} The paper by Q.\ Lan~\cite{Lan_2021} introduces a quantum version of a \gls{sac} approach. The advantage of this algorithm, compared to previous suggestions, is the possibility to work with a continuous action space. The algorithm is tested on the \texttt{Pendulum} environment.
    
    \medbreak
    
    \noindent\textit{Soft Actor-Critic for Continuous Control.} The term continuous control refers to a setup, in which the agent acts in a continuous action space. Most publications in the context of \gls{qrl} deal with discrete action spaces, while a few others discuss continuous control for quantum environments~\cite{Wu_2023,Sequeira_2023}. This work focuses on classical environments, which requires some kind of action decoding based on measurements of the quantum state. Instead of directly selecting the actions based on measurement results, the parameterized hybrid model learns the parameters of a distribution, from which the action is sampled. The \gls{vqc}, and a downstream \gls{nn}, are used to represent mean $\mu$ and variance $\sigma$ of a Gaussian distribution. This allows the agent to act in a continuous action space in a straightforward manner.
    
    In contrast to the standard \gls{rl} setup, \gls{sac}~\cite{Haarnoja_2018} not only aims to optimize the expected return, but also the policy entropy \cite{Ziebart_2008,Haarnoja_2017}. Therefore, the expected return is defined as $G_t = \sum_{i=t}^{\infty} \gamma^{i-t} (r(s_i, a_i) + \alpha \mathcal{H}[\pi_{\theta}(\cdot | s_i)])$, where $\mathcal{H}[p] = - \int_{\mathbb{R}} p(x) \log p(x) \mathrm{d}x$ is the differential entropy for the probability density function $p(x)$. Among other advantages, this entropy normalization potentially enhances exploration by encouraging more stochastic policies~\cite{Haarnoja_2017}.
    
    \medbreak
    
    \noindent\textit{\gls{vqc} Architecture.} The paper considers two different \gls{vqc} architectures. The first one uses the typical three-part structure of rotational encoding, variational layers, and measurements. The second architecture is more complex, as it uses data re-uploading~\cite{Perez_2020}, and a more complex encoding structure~\cite{Skolik_2022,Jerbi_2021a}. It can be expected, that the second choice gives rise to more expressive models, which usually correlates with \gls{rl} performance.
    
    \medbreak
    
    \noindent\textit{Experimental Results.} The experimental section of the paper compares the performance of the two resulting quantum \gls{sac} approaches to a classical \gls{nn} on the \texttt{Pendulum} environment. On the one hand, the quantum model with the simple \gls{vqc} architecture is inferior to the other two approaches. On the other hand, the quantum model with data re-uploading performs similar to the classical model, and both are able to learn near-optimal behavior. The quantum model incorporates only $41$ parameters, while the classical one uses $1250$. This is interpreted as an quantum advantage w.r.t.\ parameter complexity.
    
    Some additional architecture experiments are conducted, mainly focusing one the depth of the underlying \glspl{vqc}. It is observed, that a certain number of variational layers is required to enable training. Overall, the performance is strongly correlated with the concrete architecture choice, which is in line with the results known from literature~\cite{Franz_2022}.
    
    \medbreak
    
    \noindent\textit{Remarks.} To substantiate the claim of quantum advances w.r.t.\ parameter complexity, more experiments with increasing environment size should be performed. By using \glspl{nn} in combination with \glspl{vqc}, it is not completely clear, which part of the learning is actually conducted by the quantum part. The differing performance of the two architecture choices highlight the importance of designing a sophisticated data encoding scheme.
    
    \medbreak
    
     \begin{table}[ht]
    \centering
    \begin{minipage}{\textwidth}
        \renewcommand\footnoterule{}
        \renewcommand{\thefootnote}{\alph{footnote}}
    \caption*{\textbf{Algorithmic Characteristics - Q. Lan}~\cite{Lan_2021}}
    \small
    \begin{tabular*}{\textwidth}{p{\dimexpr 0.19\textwidth-2\tabcolsep-\arrayrulewidth}|p{\dimexpr 0.14\textwidth-2\tabcolsep-\arrayrulewidth}|p{\dimexpr 0.15\textwidth-2\tabcolsep-\arrayrulewidth}|p{\dimexpr 0.1\textwidth-2\tabcolsep-\arrayrulewidth}|p{\dimexpr 0.1\textwidth-2\tabcolsep-\arrayrulewidth}|p{\dimexpr 0.1\textwidth-2\tabcolsep-\arrayrulewidth}|p{\dimexpr 0.22\textwidth-2\tabcolsep}}
        \toprule
        \centering\multirow{2}{*}{\textbf{Environment}} & \centering\textbf{Algorithm} & \centering\textbf{Quantum} & \centering\textbf{State} & \centering\textbf{Action} & \centering\multirow{2}{*}{\textbf{Qubits}} & \centering\textbf{Parameterized} \tabularnewline
        & \centering\textbf{Type} & \centering\textbf{Component} & \centering\textbf{Space} & \centering\textbf{Space} & & \centering\textbf{~Gates}\footnotemark[1] \tabularnewline
        \midrule
        \midrule
        %%%% ROW 1
        \centering\texttt{Pendulum} &
        \centering Quantum- &
        \centering\multirow{2}{*}{$Q$-function} &
        \centering\tiny{continuous} &
        \centering\multirow{2}{*}{\begin{tabular}{@{}c@{}}{\scriptsize{conti-}}\\{\scriptsize{nuous}}\end{tabular}} &
        \centering\multirow{2}{*}{$3$} &
        \centering $3$ to $12$ (encoding) \tabularnewline
        \centering (OpenAI Gym) &
        \centering\gls{sac} &
        &
        \centering $3$-dim &
        &
        &
        \centering $36$ (weights) \tabularnewline
        \bottomrule
    \end{tabular*}
    \footnotetext[1]{~the hybrid model also incorporates additional classical parameters in an appended \gls{nn};}
    \end{minipage}
\end{table}

%------------------------------------------------------------------------------------------------------%
%------------------------------------------------------------------------------------------------------%
%------------------------------------------------------------------------------------------------------%

%------------------------------------------------------------
\paragraph{\label{subsubsec:Yun_2022}Quantum Multi-Agent Reinforcement Learning via Variational Quantum Circuit Design, Yun et al.~(2022)}\mbox{}\\
% \cite{Yun_2022}
%------------------------------------------------------------

    \vspace{-1em}
    \noindent\textit{Summary.} This paper by Yun et al.~\cite{Yun_2022} introduces a \gls{qmarl} approach. It is applied to an environment inspired by wireless communication. The authors achieve results that are competitive with classical \glspl{nn} with higher parameter complexity.
    
    \medbreak
    
    \noindent\textit{\gls{qmarl} Framework and \gls{vqc} Architecture.} The approach is inspired by the classical method of \gls{ctde}. This approach deals with the problems introduced by a non-stationary reward structure, caused by the interaction of multiple agents~\cite{Lowe_2017}.
    
    The actor-critic structure employs only a single critic (i.e.\ represented by a single \gls{vqc}), which receives the rewards. A naive implementation would would increase the qubit count with the number of agents. To resolve this problem, the state encoding routine is modified, such that only one qubit is required for each agent.
    
    The general \gls{vqc} architecture follows the typical three-part structure. The states are encoded using a feature map with $1$-qubit rotations.  The state space of the environment is four-dimensional. Consequently, four qubits are used to represent the actor associated to each of the four agents. For the critic, all rotations for the state of one agent are applied to a single qubit. This implies a qubit count equal to the number of agents (i.e.\ implemented for $4$ qubits in the article). The following learnable layer(s) consist of $1$-qubit rotations and some unspecified entanglement structure. The choice of the measured observables $M$ are not explicitly stated.
    
    \medbreak
    
    \noindent\textit{Experimental Results and Discussion.} The \gls{qmarl} algorithm is applied to a communication task referred to as \texttt{Single-Hop Offloading} environment. It simulates two clouds, between which packages have to be distributed along four edges. Each cloud and edge has a queue with a certain capacity. One agent is used to learn the actions of its associated edge. The objective is to minimize the overflow and underflow of queues.
    
    The paper compares four different \gls{marl} and \gls{qmarl} frameworks: (1) The described version, where actor and critic are represented with a \gls{vqc}; (2) A modified pipeline, where the critic is represented with a classical \gls{nn}; (3) A small-scale classical \gls{marl} approach; All three setups contain $50$ trainable parameters each. (4) A large-scale classical \gls{marl} algorithm with over $40000$ trainable parameters.
    
    The results demonstrate, that the \gls{qmarl} approach (1) is competitive with the large-scale \gls{marl} algorithm (4). In contrast, the hybrid \gls{qmarl} method (2) and also the small-scale classical \gls{marl} seem to lack expressivity to solve this task. The authors conclude, that \gls{qmarl} yields some quantum advantage, as the parameter complexity is drastically reduced.
    
    \medbreak
    
    \noindent\textit{Remarks.} Potentially, compressing all observations of one agent into one qubit is not sufficient to represent the information in a lossless manner. Therefore, larger-scale experiments should be conducted to get more insights into the proposed quantum multi-agent architecture. The same holds for the reduced parameter complexity compared to classical models.
    
    \medbreak
    
     \begin{table}[ht]
    \centering
    \begin{minipage}{\textwidth}
        \renewcommand\footnoterule{}
        \renewcommand{\thefootnote}{\alph{footnote}}
    \caption*{\textbf{Algorithmic Characteristics - Yun et al.}~\cite{Yun_2022}}
    \small
    \begin{tabular*}{\textwidth}{p{\dimexpr 0.19\textwidth-2\tabcolsep-\arrayrulewidth}|p{\dimexpr 0.14\textwidth-2\tabcolsep-\arrayrulewidth}|p{\dimexpr 0.15\textwidth-2\tabcolsep-\arrayrulewidth}|p{\dimexpr 0.1\textwidth-2\tabcolsep-\arrayrulewidth}|p{\dimexpr 0.1\textwidth-2\tabcolsep-\arrayrulewidth}|p{\dimexpr 0.1\textwidth-2\tabcolsep-\arrayrulewidth}|p{\dimexpr 0.22\textwidth-2\tabcolsep}}
        \toprule
        \centering\multirow{2}{*}{\textbf{Environment}} & \centering\textbf{Algorithm} & \centering\textbf{Quantum} & \centering\textbf{State} & \centering\textbf{Action} & \centering\multirow{2}{*}{\textbf{Qubits}} & \centering\textbf{Parameterized} \tabularnewline
        & \centering\textbf{Type} & \centering\textbf{Component} & \centering\textbf{Space} & \centering\textbf{Space} & & \centering\textbf{Gates} \tabularnewline
        \midrule
        \midrule
        %%%% ROW 1
        \centering\texttt{Single-Hop} &
        \centering \footnotesize{Multi-Agent} &
        \centering\multirow{3}{*}{\begin{tabular}{@{}c@{}}$Q$-function\\Policy\end{tabular}} &
        \centering\multirow{3}{*}{\begin{tabular}{@{}c@{}}{\tiny{continuous}}\\$4$-dim\end{tabular}} &
        \centering\multirow{3}{*}{\begin{tabular}{@{}c@{}}{\tiny{discrete}}\\{$4$}\end{tabular}} &
        \centering\multirow{3}{*}{$4$} &
        \centering\multirow{3}{*}{\begin{tabular}{@{}c@{}}$4$ or $16$ (encoding)\footnotemark[1]\\$N/A$ (weights)\end{tabular}} \tabularnewline
        \centering\texttt{Offloading} &
        \centering \footnotesize{Actor-Critic} &
        &
        &
        &
        &
        \tabularnewline
        \centering (see \cite{Yun_2022}) &
        \centering\footnotesize ``\gls{qmarl}'' &
        &
        &
        &
        &
        \tabularnewline
        \bottomrule
    \end{tabular*}
    \footnotetext[1]{~the $4$ quantum actors use $4$ encoding parameters each; the quantum centralized critic contains $16$;}
    \end{minipage}
\end{table}
    
%------------------------------------------------------------------------------------------------------%
%------------------------------------------------------------------------------------------------------%
%------------------------------------------------------------------------------------------------------%

%------------------------------------------------------------
\paragraph{\label{subsubsec:Yun_2023a}Quantum Multi-Agent Meta Reinforcement Learning, Yun et al.~(2023)}\mbox{}\\
% \cite{Yun_2023a}
%------------------------------------------------------------

    \vspace{-1em}
    \noindent\textit{Summary.} The second paper by Yun et al.~\cite{Yun_2023a} extends their previous approach~\cite{Yun_2022} with various new techniques for \gls{qmarl}. It proposes to use meta-learning by pre-training only one individual agent. This is followed by a fine-tuning the multi-agent scenario. Therefore, two different types of trainable parameters are used, i.e.\ trainable measurements are introduced to complement the typical variational parameters. The approach is also extended to continual learning, where meta-learning is performed on multiple environments at once.
    
    \medbreak
    
    \noindent\textit{\gls{vqc} Architecture and meta-\gls{qmarl}.} The underlying \gls{rl} algorithm employs an \gls{sac} approach with the \gls{vqc} as function approximator for the action-value function. The quantum circuit uses the three-layer structure of $1$-qubit rotation data encoding, variational layers with entanglement gates, and measurement. The paper applies \gls{qrl} to multi-agent problems and extends the original proposal on quantum \gls{ctde} by Yun et al.~\cite{Yun_2022}. An additional step is introduced for the training procedure, resulting in a meta-learning approach.
    
    In order to realize these concepts, the authors define two different sets of parameters. First, there are the typical variational parameters $\boldsymbol{\phi}$, usually parameterizing $1$-qubit rotations. Second, it is also possible to parameterize and train the measurement observables. The paper proposes to use $M_{\theta_{1,2}^{(m)}} = R_{x}^{\dagger}(\theta_1) \cdot R_{y}^{\dagger}(\theta_2) \cdot Z \cdot R_{y}(\theta_2) \cdot R_{x}(\theta_1)$ as observable on the $m$-th qubit, i.e.\ two trainable parameters for each $1$-qubit observable. Basically, this trainable observable introduces a change of basis, as final measurements are always performed in the computational basis. The instantiated observable can be visualized on the Bloch sphere as the angle w.r.t.\ which the measurement is performed.
    
    Both parameter sets are trained in alternating steps, where the first one is referred to as \emph{meta \gls{qnn} angle training}, and focuses exclusively on the variational parameters $\boldsymbol{\phi}$. This step trains only a single quantum agent, which interacts with several other agents in the multi-agent environment. Unfortunately, the authors do not state how this interaction is actually realized. We assume, that the quantum agent interacts with other classical agents in this initial training phase. During training, the pole parameters $\boldsymbol{\theta}$ are not updated, but they can be varied with some randomly selected value to form a kind of angle-to-pole regularization. The second phase, the \emph{local \gls{qnn} pole training}, focuses on the parameterized observables. Those are fine-tuned individually for each copy of the meta-trained \gls{qnn}, corresponding to the all-quantum agents interacting in the multi-agent environment. The authors propose, that by meta-training the network, it is more efficient to fine-tune the individual agents. This is justified with the lower parameter complexity, as the variational parameters remain constant in the second training phase. The loss function is the sum of all $Q$-learning losses of the individual agents.
    
    Additionally, the paper introduces the concept of \emph{pole memory}, which refers to storing the trained pole parameters for the individual agents. As these sets are much smaller than the set of variational parameters, it is more efficient to store the full configuration.
    
    \medbreak
    
    \noindent\textit{Experimental Results.} The introduced training routing is executed on a two-step two-agent environment. It is observed, that the meta-training convergence is slower than direct training of a \gls{qmarl} agent. However, once this training has converged, finetuning is much more efficient. Overall, the authors conclude, that the additional step of meta-training enhances convergence in a multi-agent environment.
    
    \medbreak
    
    \noindent\textit{Extension to Continual Learning.}  The above setting is also extended to continual learning, i.e.\ training in more than one environment (or typically the same environment with slightly altered dynamics). 
    
    The investigation focuses on the difference in performance with and without the use of pole memory. The results suggest that resetting the poles to the initial state (i.e.~the parameter setting with which meta training was conducted) benefits convergence speed and stability in an environment with alternating dynamics. Meta training with a higher degree of angle-to-pole regularization seems to enhance the generalization performance of the meta-\gls{qnn}.
    
    \medbreak
    
    \noindent\textit{Remarks.} The paper does not state explicitly how exactly the initial meta training is conducted. Considering the results, the \glspl{vqc} seem to have some capability w.r.t.\ transfer learning, as which the meta-learning and continual training can be interpreted. The idea of employing trainable observables has also potential for other approaches, as it partially avoids the necessity to explicitly pre-select an action decoding scheme. Practically, these trainable observables are introduced by adding an additional layer to the \gls{vqc} which learns a specific measurement. A significant difference to pre-existing procedures is that these parameters are not trained simultaneously with the typical variational parameters. It is not completely clear, whether this two-step training procedure is beneficial in a general setup.
    
    \medbreak
    
     \begin{table}[ht]
    \centering
    \begin{minipage}{\textwidth}
        \renewcommand\footnoterule{}
        \renewcommand{\thefootnote}{\alph{footnote}}
    \caption*{\textbf{Algorithmic Characteristics - Yun et al.}~\cite{Yun_2023a}}
    \small
    \begin{tabular*}{\textwidth}{p{\dimexpr 0.19\textwidth-2\tabcolsep-\arrayrulewidth}|p{\dimexpr 0.14\textwidth-2\tabcolsep-\arrayrulewidth}|p{\dimexpr 0.15\textwidth-2\tabcolsep-\arrayrulewidth}|p{\dimexpr 0.1\textwidth-2\tabcolsep-\arrayrulewidth}|p{\dimexpr 0.1\textwidth-2\tabcolsep-\arrayrulewidth}|p{\dimexpr 0.1\textwidth-2\tabcolsep-\arrayrulewidth}|p{\dimexpr 0.22\textwidth-2\tabcolsep}}
        \toprule
        \centering\multirow{2}{*}{\textbf{Environment}} & \centering\textbf{Algorithm} & \centering\textbf{Quantum} & \centering\textbf{State} & \centering\textbf{Action} & \centering\multirow{2}{*}{\textbf{Qubits}} & \centering\textbf{Parameterized} \tabularnewline
        & \centering\textbf{Type} & \centering\textbf{Component} & \centering\textbf{Space} & \centering\textbf{Space} & & \centering\textbf{~Gates}\footnotemark[1] \tabularnewline
        \midrule
        \midrule
        %%%% ROW 1
        \centering\texttt{Single-Hop} &
        \centering \footnotesize{Meta-Multi-} &
        \centering\multirow{3}{*}{$Q$-function} &
        \centering\multirow{3}{*}{\begin{tabular}{@{}c@{}}{\tiny{continuous}}\\$4$-dim\end{tabular}} &
        \centering\multirow{3}{*}{\begin{tabular}{@{}c@{}}{\tiny{discrete}}\\{$4$}\end{tabular}} &
        \centering\multirow{3}{*}{$4$} &
        \centering\multirow{3}{*}{\begin{tabular}{@{}c@{}}$4$ (encoding)\\$N/A$ (weights)\end{tabular}} \tabularnewline
        \centering\texttt{Offloading} &
        \centering \footnotesize{Agent \gls{sac}} &
        &
        &
        &
        &
        \tabularnewline
        \centering (see \cite{Yun_2022}) &
        \centering\tiny ``Meta-\gls{qmarl}'' &
        &
        &
        &
        &
        \tabularnewline
        \midrule
        %%%%% ROW 2
        \centering\texttt{Two-Step} &
        \centering \footnotesize{Meta-Multi-} &
        \centering\multirow{3}{*}{$Q$-function} &
        \centering\multirow{3}{*}{\begin{tabular}{@{}c@{}}{\tiny{continuous}}\\$4$-dim\end{tabular}} &
        \centering\multirow{3}{*}{\begin{tabular}{@{}c@{}}{\tiny{discrete}}\\{$2$}\end{tabular}} &
        \centering\multirow{3}{*}{$2$} &
        \centering\multirow{3}{*}{\begin{tabular}{@{}c@{}}$2$ (encoding)\\$N/A$ (weights)\end{tabular}} \tabularnewline
        \centering\texttt{Game} &
        \centering \footnotesize{Agent \gls{sac}} &
        &
        &
        &
        &
        \tabularnewline
        \centering (see \cite{Yun_2023a}) &
        \centering\tiny ``Meta-\gls{qmarl}'' &
        &
        &
        &
        &
        \tabularnewline
        \bottomrule
    \end{tabular*}
    \footnotetext[1]{~the parameter counts are denoted for a single agent;}
    \end{minipage}
\end{table}

\subsubsection{Offline Methods}
\label{subsec:VQC_based_Offline}
Offline reinforcement learning~\cite{Levine_2020} deals with the setting, when no direct interaction with the environment is possible. Instead, the agent is trained on a set of pre-acquired data. Two alternative formulations for the quantum realm have been proposed in Periyasamy et al.~\cite{Periyasamy_2023} and Cheng et al.~\cite{Cheng_2023a}.

\begin{table}[ht!]
    \centering
    \begin{tabular}{p{\dimexpr 0.15\textwidth-2\tabcolsep-\arrayrulewidth}|p{\dimexpr 0.2\textwidth-2\tabcolsep-\arrayrulewidth}|p{\dimexpr 0.65\textwidth-2\tabcolsep}}
        \toprule
        \textbf{Citation} & \textbf{First Author} & \textbf{Title} \\
        \midrule
        \midrule
        \cite{Periyasamy_2023} & M. Periyasamy & \hyperref[subsubsec:Periyasamy_2023]{Batch Quantum Reinforcement Learning} \\
        \arrayrulecolor{black!30}\midrule
        \cite{Cheng_2023a} & Z. Cheng & \hyperref[subsubsec:Cheng_2023a]{Offline Quantum Reinforcement Learning in a Conservative Manner} \\
        \arrayrulecolor{black}\bottomrule
    \end{tabular}
    \caption{Work considered for ``\gls{qrl} with \glspl{vqc} -- Offline Methods'' (\cref{subsec:VQC_based_Offline})}
\end{table}

%------------------------------------------------------------
\paragraph{\label{subsubsec:Periyasamy_2023}Batch Quantum Reinforcement Learning, Periyasamy et al.~(2023)}\mbox{}\\
%\cite{Periyasamy_2023}
%------------------------------------------------------------

    \vspace{-1em}
\noindent\textit{Summary.}
In this work, Periyasamy et al.~\cite{Periyasamy_2023} propose \gls{bcqq}, a offline \gls{qrl} algorithm based on the classical discrete \gls{bcq} algorithm by Fujimoto et al.~\cite{Fujimoto_2019}. Furthermore, the authors introduce a novel \gls{dru} scheme, which they call cyclic \gls{dru}. Experiments are executed in the OpenAI \texttt{CartPole} environment.
    
\medbreak
    
\noindent\textit{Algorithm.}
The key idea in \gls{bcq} is that in order to avoid a distributional shift from training to testing, a trained policy should induce at test time a similar state-action visitation to that observed in the the offline training data, the so-called batch. Hence, the name \textit{batch-constrained}.

To achieve this, \gls{bcq} trains a generative model \(G_{\omega}\) to pre-select likely actions based on the batch. Through this selection, the policy is constrained to only choose from a subset of actions. In the case of a discrete action space, the generative model can be understood as a map \(G_{\omega}: \mathcal{S} \rightarrow \Delta \left(\mathcal{A}\right)\) that takes the current environment state as input and outputs the probability with which each action would occur in the batch. In particular, if the batch is filled using transitions from a policy \(\pi_b\) then the generative model should imitated this policy, i.e. \(G_\omega(a|s) \approx \pi_b(a|s)\). Therefore, \(G_{\omega}\) is called imitator. 

Through using this imitator, actions can be pre-selected by discarding actions whose probability relative to the most likely one is below a threshold \(\tau\)
\begin{equation}
\label{eq:BatchConstraintActions}
    \tilde{\mathcal{A}}(s) = \left\{a \in \mathcal{A} \Bigg{|} \frac{G_\omega(a|s)}{\text{max}_{\hat{a} \in \mathcal{A}} G_\omega(\hat{a}|s)} > \tau\right\}.
\end{equation}
The actions selected by the imitator are then evaluated by a $Q$-network, which is trained by only considering the selected actions in the loss computation.
The imitator itself is trained with a standard cross-entropy loss
\begin{equation*}
    l(\omega) = - \sum_{(s, a) \in \mathcal{B}}\text{log}\left(G_\omega (a|s)\right).
\end{equation*}
Additionally, to address the overestimation bias of $Q$-learning towards state transitions that are underrepresented in the batch, double DQN \cite{Van_2016} is employed.

Finally, the \gls{bcqq} algorithm is obtained by applying the \gls{vqdqn} proposed by Franz et al.~\cite{Franz_2022} as function approximators for both the imitator and $Q$-network. Moreover, for the model training the authors use the AMSgrad optimizer in combination with gradients approximated via SPSA. Wiedmann et al.~\cite{Wiedmann_2023} demonstrated that SPSA can be used to efficiently train medium-sized \glspl{vqc} with a reduced number of circuit runs, compared to the commonly used parameter-shift rule. 
% The overall \gls{bcqq} training procedure is summarized in \cref{algo:DiscreteBCQQ}.

% \begin{algorithm*}[htbp!]
% \caption{Discrete Batch-Constraint Quantum $Q$-learning training algorithm}\label{algo:DiscreteBCQQ}
% \begin{algorithmic}
%     \STATE Normalize dataset \(\mathcal{B}\) between \([-\pi, \pi]\)
%     \STATE Initialize encoding unitary \(U(.)\) which encodes state \(s\) into \(|\psi(s)\rangle\)
%     \STATE Initialize Q value approximator using VQC with \(\theta\) and  \(\theta'\)
%     \STATE Initialize generative model \(G_\omega\) using VQC with \(\omega\)
%     \WHILE{training not converged}
%         \STATE Sample mini-batch \(M\) from \(\mathcal{B}\)
%         \FORALL{\((s, a, r, s') \in M\)}
%             \STATE Get batch-constraint actions \(\tilde{\mathcal{A}}(s')\) from \eqref{eq:BatchConstraintActions}
%             \STATE Choose \(a' \in \underset{\tilde{a} \in \tilde{\mathcal{A}}(s')}{\text{argmax}} Q_\theta(|\psi(s')\rangle, \tilde{a})\)
%         \ENDFOR
%         \STATE Optimize \(\theta\) w.r.t. \(l(\theta) = \underset{M}{\mathds{E}}\left(r + \gamma Q_{\theta'}(|\psi(s')\rangle, a') - Q_\theta(|\psi(s)\rangle, a)\right)^2\)
%         \STATE Optimize \(\omega\) w.r.t \(l(\omega) = - \underset{M}{\mathds{E}}\text{log}\left(G_\omega (a, |\psi(s)\rangle)\right)\)
%         \IF{target network update}
%             \STATE \(\theta' \leftarrow \theta\)
%         \ENDIF
%     \ENDWHILE
% \end{algorithmic}
% \end{algorithm*}

\medbreak

\begin{figure}[ht]
        \centering
        \includegraphics[width=0.99\textwidth]{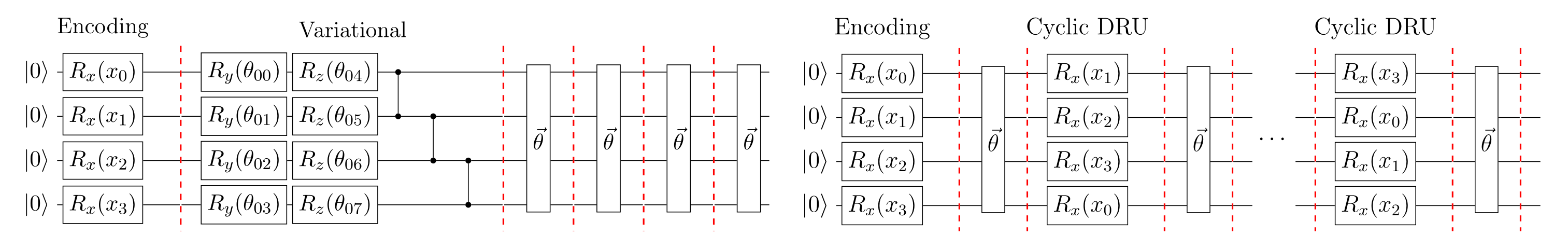}
        \caption{\textit{Left:} \gls{vqc} that is used as function approximator in the discrete \gls{bcqq} algorithm. \textit{Right:} \gls{vqc} with cyclic \gls{dru}. \textit{Note}: Each $\vec{\theta}$ block represents the repetition of the variational layer ansatz with different trainable parameters. Both taken from \cite{Periyasamy_2023}.}
        \label{fig:Periyasamy_2023}
    \end{figure}

\noindent\textit{Model Architecture.}
The \gls{vqc} used as the function approximator for the imitator and $Q$-network is shown in \cref{fig:Periyasamy_2023}. Each entry of the four-dimensional state vector returned by the \texttt{CartPole} environment is encoded using a single qubit Rx gate on an individual qubit. The variational block comprises five layers containing four parameterized Ry, and four parameterized $R_z$ gates each. In addition to the parameterized rotational gates, each layer also includes two-qubit CZ entanglement gates with nearest-neighbor connectivity. The \texttt{CartPole} environment has two discrete actions. Therefore, the expectation value of the Pauli-$ZZ$ observable on qubits 1 and 2 and Pauli-$ZZ$ observable on qubits 3 and 4 are used to decode the $Q$-values from the \gls{vqc}. Furthermore, trainable classical weights are applied on both expectation values to increase the range of possible $Q$-values.
%A four-qubit quantum system was chosen as the target system, because the \texttt{CartPole} environment has a four-dimensional state space and each feature of the observation is encoded using a single qubit Rx gate on each qubit. The variational block comprises five layers containing four parameterized Ry, and four parameterized Rz gates each. In addition to the parameterized rotational gates, each layer also includes two-qubit CZ entanglement gates with nearest-neighbor connectivity in the circuit layout. The \texttt{CartPole} has a two-dimensional action space. Therefore, the expectation value of the Pauli-$ZZ$ observable on qubits 1 and 2 and Pauli-$ZZ$ observable on qubits 3 and 4 was used to decode the Q-values from the \gls{vqc}.

Periyasamy et al.~\cite{Periyasamy_2022} established that spreading encoding gates for the feature vector of a given data point throughout the quantum circuit results in an improved representation of the data when the expectation values are measured for observables containing all Pauli strings. Following this, the authors use a re-uploading scheme, which exposes each qubit to all the entries of the current input state vector. Contrary to the standard data re-uploading, where the encoding scheme is re-introduced after each variational layer as such, the encoding scheme is re-introduced with the input state vector shifted by one step in a round-robin fashion. The structure of a \gls{vqc} with this cyclic \gls{dru} is shown on the right of \cref{fig:Periyasamy_2023}.

\medbreak

\noindent\textit{Experimental Results and Discussion.} In order to evaluate the performance of \gls{bcqq}, the authors train policies on buffers with varying sizes, filled with randomly sampled environment interactions. As a classical benchmark the authors train neural networks instead of \glspl{vqc} on the same buffers. For this benchmark, they first use a fully connected neural network with a total number of 67270 parameters and second a smaller network with just 55 parameters. The number of parameters in the smaller network is much more comparable to the \gls{vqc}. The authors find that the \gls{bcqq} agent is able to learn an optimal policy, achieving the maximum reward of 500, from a buffer of just 100 random environment interactions. Interestingly, the classical agents fail to learn a  policy in this low data regime, suggesting a potential quantum advantage in terms of the sample efficiency. 

Moreover, the cumulative reward these models can achieve beyond 500 is tested, which shows that the \gls{vqc} with cyclic \gls{dru} out-performs the \gls{vqc} with standard \gls{dru}. All these experiments were performed using an early stopping criteria, where during training the current policy is evaluated in the actual environment to save computational resources. Strictly speaking, this makes the training not fully offline. In a second experiment however, the authors train the \gls{vqc} with cyclic \gls{dru} on a buffer filled with 100 interactions obtained from an optimal policy with noise. From this, the authors show that without early stopping the \gls{bcqq} agent can learn an optimal policy from this noisy buffer.
    
\medbreak
    
\noindent\textit{Remarks.} It remains to be shown that the observed sample efficiency scales to more complex environments. Furthermore, a more elaborate analysis of the effectiveness of cyclic \gls{dru} could give insights for future \gls{vqc} design.

\medbreak

\begin{table}[ht]
    \centering
    \begin{minipage}{\textwidth}
        \renewcommand\footnoterule{}
        \renewcommand{\thefootnote}{\alph{footnote}}
    \caption*{\textbf{Algorithmic Characteristics - Periyasamy et al.}~\cite{Periyasamy_2023}}
    \small
    \begin{tabular*}{\textwidth}{p{\dimexpr 0.19\textwidth-2\tabcolsep-\arrayrulewidth}|p{\dimexpr 0.14\textwidth-2\tabcolsep-\arrayrulewidth}|p{\dimexpr 0.15\textwidth-2\tabcolsep-\arrayrulewidth}|p{\dimexpr 0.1\textwidth-2\tabcolsep-\arrayrulewidth}|p{\dimexpr 0.1\textwidth-2\tabcolsep-\arrayrulewidth}|p{\dimexpr 0.1\textwidth-2\tabcolsep-\arrayrulewidth}|p{\dimexpr 0.22\textwidth-2\tabcolsep}}
        \toprule
        \centering\multirow{2}{*}{\textbf{Environment}} & \centering\textbf{Algorithm} & \centering\textbf{Quantum} & \centering\textbf{State} & \centering\textbf{Action} & \centering\multirow{2}{*}{\textbf{Qubits}} & \centering\textbf{Parameterized} \tabularnewline
        & \centering\textbf{Type} & \centering\textbf{Component} & \centering\textbf{Space} & \centering\textbf{Space} & & \centering\textbf{~Gates}\footnotemark[1] \tabularnewline
        \midrule
        \midrule
        %%%% ROW 1
        \centering\multirow{3}{*}{\begin{tabular}{@{}c@{}}\texttt{CartPole}\\(OpenAI Gym)\end{tabular}} &
        \centering\multirow{3}{*}{\gls{bcq}} &
        \centering\multirow{3}{*}{\begin{tabular}{@{}c@{}} Imitator, \\ $Q$-function\end{tabular}} &
        \centering\multirow{3}{*}{\begin{tabular}{@{}c@{}}\tiny{continuous}\\$4$-dim\end{tabular}} &
        \centering\multirow{3}{*}{\begin{tabular}{@{}c@{}}\tiny{discrete}\\$2$\end{tabular}} &
        \centering\multirow{3}{*}{$4$} &
        \textcolor{black}{\centering $4 \times 1$ (encoding)} \tabularnewline
        &
        &
        &
        &
        &
        & 
        \textcolor{black}{\centering $4 \times 15 \times 2$ (weights)} \tabularnewline
        &
        &
        &
        &
        &
        & 
        \textcolor{black}{\centering $N/A$ (classical)}\footnotemark[2] \tabularnewline
        \bottomrule
    \end{tabular*}
    \footnotetext[1]{~encoding gates: $qubits \times per\_qubit$; variational gates: $qubits \times layers \times per\_qubit\_per\_layer$;}
    \footnotetext[2]{~model incorporates classical weights after measurement, details are not stated;}
    \end{minipage}
\end{table}

%------------------------------------------------------------------------------------------------------%
%------------------------------------------------------------------------------------------------------%
%------------------------------------------------------------------------------------------------------%

%------------------------------------------------------------
\paragraph{\label{subsubsec:Cheng_2023a}Offline Quantum Reinforcement Learning in a Conservative Manner, Cheng et al.~(2023)}\mbox{}\\
%\cite{Cheng_2023a}
%------------------------------------------------------------

    \vspace{-1em}
    \noindent\textit{Summary.} This work by Cheng et al.~\cite{Cheng_2023a} introduces the offline \gls{qrl} algorithm, \gls{cq2l}. In contrast to online \gls{rl}, offline \gls{rl} is used in scenarios where the agent cannot interact with the environment during training and is hence trained purely data-driven from a set of previously collected data. The proposed algorithm is based on the classical \gls{cql} algorithm by Kumar et al.~\cite{Kumar_2020}. Experiments are conducted in the OpenAI \texttt{CartPole}, \texttt{Acrobot} and \texttt{MountainCar} environments.
    
\medbreak
    
\noindent\textit{Algorithm.} The objective of offline \gls{rl} is to learn a near-optimal policy from a fixed dataset $\mathcal{D}$ sampled with a behavior policy $\pi_b$, without further environment interactions. A major challenge in this setting is that the fundamental assumption that agents can sample data online is violated. This means that agents have to learn a policy or value function from out-of-distribution (OOD) data, which is nontrivial. This distributional shift makes it hard to evaluate and consequently improve current Q-value functions, leading to an extrapolation error \cite{Kostrikov_2021}. 

Under the online setting, agents obtain corrective feedback through environment interactions. However, for offline training, the extrapolation error means that agents could overestimate $Q$-values for unseen state-action pairs, which could lead to poor performance. Hence, \gls{cql} suppresses the overestimation problem in offline \gls{rl} by learning a conservative $Q$-value function. In particular, this is achieved via double $Q$-learning \cite{Van_2016} and a penalty term to update the Q-values in a conservative manner. The resulting conservative update target is obtained as
\begin{equation}
    \underset{Q}{\text{argmin}} \; \alpha \cdot \underset{s \sim \mathcal{D}}{\mathbb{E}} \left(\log \sum_{a \in A} \exp(Q(s, a; \theta_k^A)) - \underset{a \sim \pi_b}{\mathbb{E}} Q(s, a; \theta_k^A)) \right) + \underset{(s, a, r, s') \sim \mathcal{D}}{\mathbb{E}} \left( Y_k^{\text{DoubleQ}} - Q(s, a) \right)^2,
\end{equation}
with the double $Q$-learning target update
\begin{equation}
\label{eqn:doubleQ}
    Y_k^{\text{DoubleQ}} := r + \gamma \cdot Q(s', \underset{\overline{a} \in A}{argmax} \; Q(s', \overline{a}; \theta_k^A); \theta_k^B).
\end{equation}
Here, $\theta_k^A$ and $\theta_k^B$ denote two independent sets of parameters, which are updated similarly to the target network in the \gls{dqn} algorithm, by symmetrically exchanging the roles of $\theta_k^A$ and $\theta_k^B$ in \cref{eqn:doubleQ}. Having these independent parameters helps to compute unbiased $Q$-value estimates. \gls{cq2l} is then obtained by implementing the $Q$-value function via the variational \gls{vqdqn} proposed by Franz et al.~\cite{Franz_2022}.

\medbreak

\noindent\textit{Model Architecture.} \Glspl{vqc} with 5 layers to represent $Q$-value functions are used. For \texttt{CartPole}, \texttt{Acrobot} and \texttt{MountainCar} 4, 6 and 2 qubit systems are used, respectively.
According to the feasible actions in these environments, quantum observables $[Z_0Z_1, Z_2Z_3]$, $[Z_0, Z_1, Z_2]$, and $[Z_0, Z_0  Z_1, Z_1]$ are chosen, where $Z_i$ denotes the readout of a Pauli Z gate on the $i$th qubit. Input data are encoded with X rotation gates, while the variational part includes X, Y, and Z rotation gates. Moreover, qubits are entangled in a circular topology. The variational part, entanglement, and data encoding are repeated several times, which is then measured by Pauli Z gates to determine the $Q$-values.

\medbreak

\noindent\textit{Experimental Results and Discussion.} To evaluate the offline \gls{qrl} algorithm, the authors create offline data sampled by a \gls{dqn} agent with epsilon-greedy policy, interacting with the corresponding environment. The sampled data are recorded in a replay buffer with length $1 \times 10^6$ and then saved for offline \gls{qrl}. The logged data contain tuples of $(s_t, a_t, r_t, s_{t+1}, d)$, where $d$ indicates whether an episode terminates. For training, a single trajectory from the collected buffer is selected.

The authors compare the performance of \gls{cq2l} with the off-policy \gls{vqdqn} trained offline on the same data. These experiments show that \gls{cq2l} is able to solve all given environments and outperform offline \gls{vqdqn}. The latter indicates that it is not feasible to directly extend off-policy \gls{qrl} algorithms like \gls{vqdqn} to the offline setting. Furthermore, the authors find that \gls{cq2l} performs only marginally worse than online \gls{vqdqn} in \texttt{CartPole}. Interestingly, online \gls{vqdqn} fails to solve \texttt{Acrobot} and \texttt{MountainCar} and is clearly outperformed by \gls{cq2l}.

Finally, the performance is compared to classical \gls{cql}, where a fully connected neural network with a similar number
of parameters as the \gls{vqc} is used. The results indicate that \gls{cq2l} could achieve comparable performance to the classical one. Besides, no significant advantages in the sample efficiency or the parameter size are observed. The authors hypothesize that this may indicate that the current structure of \glspl{vqc} or the limited number of qubits is not sufficient to exhibit quantum advantages for \gls{qrl}.
    
\medbreak
    
\noindent\textit{Remarks.} The performance is compared to classical \gls{cql}, where a fully connected neural network with a similar number of parameters as the \gls{vqc} is used. The results indicate that \gls{cq2l} could achieve comparable performance to the classical one. Besides, no significant advantages in the sample efficiency or the parameter size are observed. The authors hypothesize that this may indicate that the current structure of \glspl{vqc} or the limited number of qubits is not sufficient to exhibit quantum advantages for \gls{qrl}. This result contradicts other observations in the literature, where at least for small system sizes some improvement w.r.t.\ parameter complexity was observed. However, we agree with the statement, that such performance improvements might strongly depend on the specific \gls{vqc} architecture.

\begin{table}[ht]
    \centering
    \begin{minipage}{\textwidth}
        \renewcommand\footnoterule{}
        \renewcommand{\thefootnote}{\alph{footnote}}
    \caption*{\textbf{Algorithmic Characteristics - Cheng et al.}~\cite{Cheng_2023a}}
    \small
    \begin{tabular*}{\textwidth}{p{\dimexpr 0.19\textwidth-2\tabcolsep-\arrayrulewidth}|p{\dimexpr 0.14\textwidth-2\tabcolsep-\arrayrulewidth}|p{\dimexpr 0.15\textwidth-2\tabcolsep-\arrayrulewidth}|p{\dimexpr 0.1\textwidth-2\tabcolsep-\arrayrulewidth}|p{\dimexpr 0.1\textwidth-2\tabcolsep-\arrayrulewidth}|p{\dimexpr 0.1\textwidth-2\tabcolsep-\arrayrulewidth}|p{\dimexpr 0.22\textwidth-2\tabcolsep}}
        \toprule
        \centering\multirow{2}{*}{\textbf{Environment}} & \centering\textbf{Algorithm} & \centering\textbf{Quantum} & \centering\textbf{State} & \centering\textbf{Action} & \centering\multirow{2}{*}{\textbf{Qubits}} & \centering\textbf{Parameterized} \tabularnewline
        & \centering\textbf{Type} & \centering\textbf{Component} & \centering\textbf{Space} & \centering\textbf{Space} & & \centering\textbf{~Gates}\footnotemark[1] \tabularnewline
        \midrule
        \midrule
        %%%% ROW 1
        \centering\multirow{3}{*}{\begin{tabular}{@{}c@{}}\texttt{CartPole}\\(OpenAI Gym)\end{tabular}} &
        \centering\multirow{3}{*}{\gls{cql}} &
        \centering\multirow{3}{*}{$Q$-function} &
        \centering\multirow{3}{*}{\begin{tabular}{@{}c@{}}\tiny{continuous}\\$4$-dim\end{tabular}} &
        \centering\multirow{3}{*}{\begin{tabular}{@{}c@{}}\tiny{discrete}\\$2$\end{tabular}} &
        \centering\multirow{3}{*}{$4$} &
        \textcolor{black}{\centering $4 \times 1$ (encoding)} \tabularnewline
        &
        &
        &
        &
        &
        & 
        \textcolor{black}{\centering $4 \times 15 \times 2$ (weights)} \tabularnewline
        &
        &
        &
        &
        &
        & 
        \textcolor{black}{\centering $N/A$ (classical)}\footnotemark[2] \tabularnewline
        \midrule
        %%%% ROW 2
        \centering\multirow{3}{*}{\begin{tabular}{@{}c@{}}\texttt{Acrobot}\\(OpenAI Gym)\end{tabular}} &
        \centering\multirow{3}{*}{\gls{cql}} &
        \centering\multirow{3}{*}{$Q$-function} &
        \centering\multirow{3}{*}{\begin{tabular}{@{}c@{}}\tiny{continuous}\\$6$-dim\end{tabular}} &
        \centering\multirow{3}{*}{\begin{tabular}{@{}c@{}}\tiny{discrete}\\$3$\end{tabular}} &
        \centering\multirow{3}{*}{$6$} &
        \textcolor{black}{\centering $4 \times 1$ (encoding)} \tabularnewline
        &
        &
        &
        &
        &
        & 
        \textcolor{black}{\centering $4 \times 15 \times 2$ (weights)} \tabularnewline
        &
        &
        &
        &
        &
        & 
        \textcolor{black}{\centering $N/A$ (classical)}\footnotemark[2] \tabularnewline
        \midrule
        %%%% ROW 3
        \centering\multirow{3}{*}{\begin{tabular}{@{}c@{}}\texttt{MountainCar}\\(OpenAI Gym)\end{tabular}} &
        \centering\multirow{3}{*}{\gls{cql}} &
        \centering\multirow{3}{*}{$Q$-function} &
        \centering\multirow{3}{*}{\begin{tabular}{@{}c@{}}\tiny{continuous}\\$2$-dim\end{tabular}} &
        \centering\multirow{3}{*}{\begin{tabular}{@{}c@{}}\tiny{discrete}\\$3$\end{tabular}} &
        \centering\multirow{3}{*}{$2$} &
        \textcolor{black}{\centering $4 \times 1$ (encoding)} \tabularnewline
        &
        &
        &
        &
        &
        & 
        \textcolor{black}{\centering $4 \times 15 \times 2$ (weights)} \tabularnewline
        &
        &
        &
        &
        &
        & 
        \textcolor{black}{\centering $N/A$ (classical)} \footnotemark[2] \tabularnewline
        
        \bottomrule
    \end{tabular*}
    \footnotetext[1]{~encoding gates: $qubits \times per\_qubit$; variational gates: $qubits \times layers \times per\_qubit\_per\_layer$;}
    \footnotetext[2]{~model incorporates classical weights after measurement, details are not stated;}
    \end{minipage}
\end{table}

\subsubsection{Algorithmic and Conceptual Extensions}
\label{subsec:VQC_based_Extensions}
This section describes extensions to the \gls{vqc}-based \gls{qrl} framework, that have relevance for multiple of the previously classified methods. This entails tools to deal with partially observable (quantum) environments discussed in Kimura et al.~\cite{Kimura_2021}. A big emphasis is put on the explicit design of model architectures. Work by Hsiao et al.~\cite{Hsiao_2022,Truong_2023} demonstrates that this is indeed an important topic, as otherwise everything could be easily emulated with classical architectures. Different approaches to this design task are discussed in Refs.~\cite{Chen_2023a,Chen_2023b,Dragan_2022,Kruse_2023,Sun_2023,Andres_2023,Park_2020}. Avoiding the typical gradient-based training routines, a evolutionary approach is proposed by Chen et al.~\cite{Chen_2022a} and also discussed in Refs.~\cite{Ding_2023,Koelle_2023}.

\begin{table}[t!]
    \centering
    \begin{tabular}{p{\dimexpr 0.15\textwidth-2\tabcolsep-\arrayrulewidth}|p{\dimexpr 0.2\textwidth-2\tabcolsep-\arrayrulewidth}|p{\dimexpr 0.65\textwidth-2\tabcolsep}}
        \toprule
        \textbf{Citation} & \textbf{First Author} & \textbf{Title} \\
        \midrule
        \midrule
        \arrayrulecolor{black}\midrule
        \cite{Kimura_2021} & T. Kimura & \hyperref[subsubsec:Kimura_2021]{Variational Quantum Circuit-Based Reinforcement Learning for POMDP and Experimental Implementation} \\
        \arrayrulecolor{black}\midrule
        \cite{Hsiao_2022} & J.-Y. Hsiao & \hyperref[subsubsec:Hsiao_2022]{Unentangled quantum reinforcement learning agents in the OpenAI Gym} \\
        \arrayrulecolor{black}\midrule
        \cite{Truong_2023} & N. Truong & \hyperref[subsubsec:Hsiao_2022]{Investigating Quantum Reinforcement Learning structure to the CartPole control task} \\
        \arrayrulecolor{black!30}\midrule
        \cite{Chen_2023a} & S. Y.-C. Chen & \hyperref[subsubsec:Architecture]{Quantum deep recurrent reinforcement learning} \\
        \arrayrulecolor{black!30}\midrule
        \cite{Chen_2023b} & S. Y.-C. Chen & \hyperref[subsubsec:Architecture]{Efficient quantum recurrent reinforcement learning via quantum reservoir computing} \\
        \arrayrulecolor{black!30}\midrule
        \cite{Dragan_2022} & T.-A. Dr\u{a}gan & \hyperref[subsubsec:Architecture]{Quantum Reinforcement Learning for Solving a Stochastic Frozen Lake Environment and the Impact of Quantum Architecture Choices} \\
        \arrayrulecolor{black!30}\midrule
        \cite{Kruse_2023} & G. Kruse & \hyperref[subsubsec:Architecture]{Variational Quantum Circuit Design for Quantum Reinforcement Learning on Continuous Environments} \\
        \arrayrulecolor{black!30}\midrule
        \cite{Sun_2023} & Y. Sun & \hyperref[subsubsec:Architecture]{Differentiable Quantum Architecture Search for Quantum Reinforcement Learning} \\
        \arrayrulecolor{black!30}\midrule
        \cite{Andres_2023} & E. Andr\'{e}s & \hyperref[subsubsec:Architecture]{Efficient Dimensionality Reduction Strategies for Quantum Reinforcement Learning} \\
        \arrayrulecolor{black}\midrule
        \cite{Chen_2022a} & S. Y.-C. Chen & \hyperref[subsubsec:Chen_2022a]{Variational quantum reinforcement learning via evolutionary optimization} \\
        \arrayrulecolor{black!30}\midrule
        \cite{Ding_2023} & L. Ding & \hyperref[subsubsec:Chen_2022a]{Multi-objective evolutionary search for parameterized quantum cirucits} \\
        \arrayrulecolor{black!30}\midrule
        \cite{Koelle_2023} & M. Kölle & \hyperref[subsubsec:Chen_2022a]{Multi-Agent Quantum Reinforcement Learning using Evolutionary Optimization} \\
        \arrayrulecolor{black}\bottomrule
    \end{tabular}
    \caption{Work considered for ``\gls{qrl} with \glspl{vqc} -- Algorithmic and Conceptual Extensions'' (\cref{subsec:VQC_based_Extensions})}
\end{table}

%------------------------------------------------------------
\paragraph{\label{subsubsec:Kimura_2021}Variational Quantum Circuit-Based Reinforcement Learning for POMDP and Experimental Implementation, Kimura et al.~(2021)}\mbox{}\\
% \cite{Kimura_2021}
%------------------------------------------------------------

    \vspace{-1em}
    \noindent\textit{Summary.} The paper by Kimura et al.~\cite{Kimura_2021} extends the concept of \gls{vqc}-based \gls{rl} to partially observable environments. The approach is inspired by classical model-free, complex-valued \gls{rl}~\cite{Hamagami_2006}. Additionally, a novel \gls{vqc} architecture (novel with regard to measurement procedure) is proposed. A detailed description of the gradient computation with backpropagation techniques is provided (it is not quite clear how this method generalizes to quantum hardware).
    
    \medbreak
    
    \noindent\textit{Partially Observable \gls{mdp}.} A \gls{pomdp} is described as a tuple $(S, A, T, R, \Omega, O)$ and is a generalization of a \gls{mdp}. The variable $S$ denotes a discrete state space, $A$ is a discrete set of actions, $T(s' | s,a)$ describes the state transition probabilities and $R(s,a)$ is a reward function. Extending the fully-observable case, $\Omega$ is a discrete set of observations and $O(o | s,a)$ is an observation probability matrix with $o \in \Omega$.
    
    One caveat of partially observable environments is the \noindent\textit{perceptual aliasing problem}. This refers to the property, that the agent cannot distinguish two different states due to the limited observation ability. An example of such an environment is the partially observable maze used in Kimura et al.~\cite{Kimura_2021}. Similar to most gridworld environments, the task is to navigate from the start state to the goal state on the shortest path possible. However, the observations provided to the agent are ambiguous as several cells return the same state indicator.
    
    \medbreak
    
    \noindent\textit{Solving \glspl{pomdp} with Complex Valued \gls{rl}.} One way to bypass this state ambiguity is to introduce a belief distribution over possible states. Unfortunately, this is computationally expensive. An alternative approach is complex-valued \gls{rl}~\cite{Hamagami_2006,Mochida_2017}. It incorporates time series information into the action-value function, which represented as complex numbers. More concretely, the complex $\dot{Q}$-function ($\dot{x}$ denotes complex values) encodes the history of the agent, i.e.\ the previously visited states. The cumulative reward value is expressed by the absolute value of $\dot{Q}$-function, while the path length of the propagated reward is represented by the phase of the $\dot{Q}$-function on the complex plane. Therefore, $\dot{Q}$-function-Learning keeps continuity w.r.t.\ the described internal reference value. This helps distinguish states which are affected by the perceptual aliasing problem. Formally, this is achieved by updating the complex values in the opposite phase direction. The complex-valued $\dot{Q}$-function can be represented with tabular methods~\cite{Hamagami_2006}, or with complex-valued \glspl{nn}~\cite{Mochida_2017}.
    
    The update mechanism represents a generalized $Q$-learning approach, i.e.\ the objective is to optimize the loss function $L_{\theta} = \frac{1}{2} \cdot | \dot{Q}(o_{t-k},a_{t-k}) - (r_{t+1}+\gamma \cdot \dot{Q}_{max}(t)) \dot{u_t}(k) |$. Here, $\dot{u_t}(k)=\dot{\beta}^{k+1}$ is a complex hyperparameter and $k$ is the trace length. The \gls{nn} is replaced with a \gls{vqc} as action-value function approximator in the following.
    
    \medbreak
    
    \noindent\textit{\gls{vqc} Architecture and Gradient Computation.} The paper deviates in several design choices from the standard method. Most importantly, the $\dot{Q}$-values for the different actions are not extracted from the same circuit (e.g.\ measurement on different qubits corresponding to different actions). Instead, the actions are encoded into the \gls{vqc} with a feature map similar to the one used for state encoding. Consequently, different circuits have to be evaluated for each action. This encoding can happen either directly in the feature map, or alternatively into the decoding unitary. However, the three-part structure of the circuit is preserved.

    \medbreak
    
    \begin{figure}[ht]
        \centering
        \includegraphics[width=0.7\textwidth]{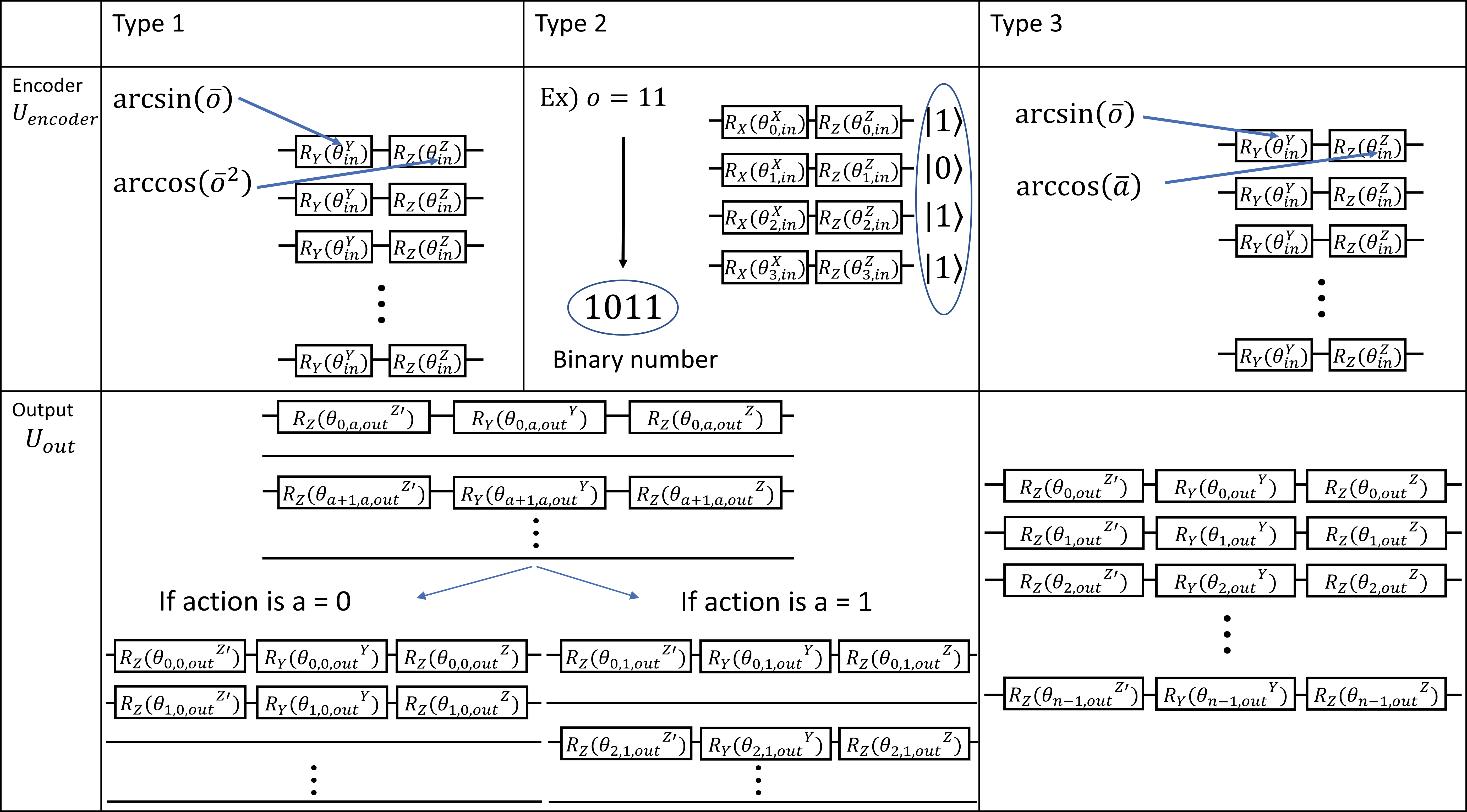}
        \caption{Encoding and decoding architectures proposed by and taken from \cite{Kimura_2021};}
        \label{fig:PartiallyObservable_arc}
    \end{figure}
    
    The encoding unitary $U_{encoder}$ consists of simple parameterized $1$-qubit rotations, where three different concrete encodings are considered as shown in \cref{fig:PartiallyObservable_arc}. The \texttt{Type 1} feature map encodes the observations directly with an $\arcsin$ function. \texttt{Type 2} uses a computational encoding for the observations, basically equivalent to the one proposed by Chen et al.~\cite{Chen_2020}. \texttt{Type 3} also uses an $\arcsin$ transform, but directly encodes the action information into the feature map.

    The variational part repeats several layers of parameterized $1$-qubit rotations, followed by a circular entanglement structure. In the experimental part, the authors consider different circuit depths.
    
    The output of the circuit is evaluated with the Hadamard Test, which measures the prepared state against an output unitary $U_{out}$. This introduces an overhead since a controlled version of the unitary $U_{out}$ needs to be implemented (details in \cref{fig:PartiallyObservable_arc}). Additionally, an ancilla qubit, and three $1$-qubit gates are required. The output unitary itself consists also of learnable $1$-qubit rotations. If encoding unitaries of \texttt{Type 1} or \texttt{Type 2} are used, it additionally encodes the action information. The real and imaginary output of the Hadamard test are used to construct the complex-valued $\dot{Q}$-function.
    
    The evaluation of the circuits is straightforward on quantum hardware. However, this does not apply to evaluating gradients w.r.t.\ the parameters, which is necessary for training. The paper gives a detailed derivation on how to compute the gradients via simulation on classical hardware. The idea is inspired by classical backpropagation and somewhat looks like the \emph{adjoint method}~\cite{Luo_2020}. This makes it infeasible, at least in the given form, for actual quantum hardware.
    
    \medbreak
    
    \noindent\textit{Experimental Results and Discussion.} The paper compares the training results (on the described maze environment) for the three types of quantum agents to different classical agents. The classical tabular approach outperforms all other methods, as the underlying algorithm guarantees an optimal solution. The authors argue, that there seems to be some intrinsic advantage of the \texttt{Type 2} quantum circuits, as these perform better then the other approximate algorithms.
    
    \medbreak
    
    \noindent\textit{Remarks.} We think there needs to be some further investigation regarding the applicability of the algorithm to actual quantum hardware. Currently, we propose to consider the approach as \gls{qirl}. We agree, that \gls{qc} offers great potential for complex-valued \gls{rl}, as \gls{qc} itself deals with complex numbers. However, there are still open questions regarding the most promising way to exploit this connection. A quantum version of a \gls{pomdp} is discussed in Ref.~\cite{Barry_2014}, which might provide for an interesting extension of this paper.
    
    % \medbreak
    
    % \input{Tables/Kimura_2021}
    
%------------------------------------------------------------------------------------------------------%
%------------------------------------------------------------------------------------------------------%
%------------------------------------------------------------------------------------------------------%

%------------------------------------------------------------
\paragraph{\label{subsubsec:Hsiao_2022}Unentangled quantum reinforcement learning agents in the OpenAI Gym, Hsiao et al.~(2022)}\mbox{}\\
% \cite{Hsiao_2022}
%------------------------------------------------------------

    \vspace{-1em}
    \noindent\textit{Summary.} The paper by Hsiao et al.~\cite{Hsiao_2022} uses an hybrid \gls{ppo} algorithm, with a combination of \gls{vqc} and \gls{nn} as policy function approximator. The quantum circuit architecture is untypical, as it only uses $1$-qubit rotations. Consequently, no entanglement is created, and all qubits can be considered as independent systems. Still, the resulting \gls{rl} agent is able to learn good policies on some standard environments (\texttt{CartPole}, \texttt{Acrobot}, and \texttt{LunarLander}). The learned parameters are ported to quantum hardware and tested with sophisticating results.
    
    \medbreak
    
    \noindent\textit{Underlying \gls{rl} Algorithm and Model Architecture.} The classical \gls{rl} algorithm is \gls{ppo}, i.e.\ an policy-based approach. It follows the typical hybrid setup, as the \gls{vqc} is used as function approximator, and parameter updates are computed on classical hardware. To enhance the expressivity of the model, a classical \gls{nn} is appended. It uses the measured expectation values as inputs. The outputs of the network are post-processed using a softmax function.
    
\begin{figure}[ht]
        \centering
        \includegraphics[width=0.6\textwidth]{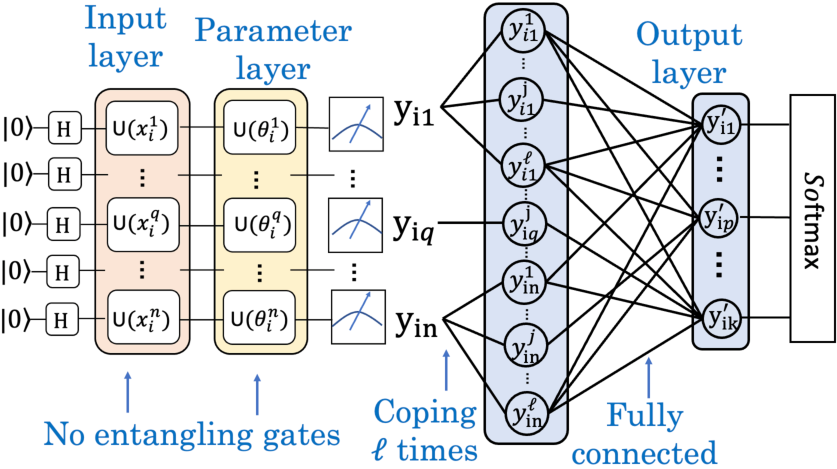}
        \caption{Hybrid quantum-classical model proposed by and taken from Hsiao et al.~\cite{Hsiao_2022};}
        \label{fig:UnentangledAgents_model}
    \end{figure}

    The structure of the hybrid model is displayed in \cref{fig:UnentangledAgents_model}. The feature map consists of $1$-qubit rotations, which is a common choice in the literature. The variational (`parameter' in \cref{fig:UnentangledAgents_model}) layer incorporates $1$-qubit parameterized rotations. It is important to highlight that the circuit does not contain any multi-qubit gates. Consequently, no entanglement between the qubits is created. As efficient classical simulation of the circuit is possible, the approach should be counted towards \gls{qirl}. Despite this, the authors demonstrate, that a good \gls{rl} training performance can be achieved with this model.
    
    \medbreak
    
    \noindent\textit{Experimental Results.} The described hybrid agent is trained on three tasks from the OpenAI Gym, i.e.\ \texttt{CartPole}, \texttt{Acrobot}, and \texttt{LunarLander}. The quantum agents outperforms several classical architectures. As this is achieved with much fewer parameters, the authors claim that the approach points towards potential advantage.
    
    The results on \texttt{LunarLander} are remarkable in that regard, that it might be the most complex environment solved with \gls{vqc}-based \gls{rl} thus far. While the classical simulability prohibits any intrinsic quantum advantage, the models still are able to achieve a good performance. This gives rise to the questions, whether one can draw inspiration from quantum mechanics for purely classical approaches.
    
    \medbreak
    
    \noindent\textit{Testing on Quantum Hardware.} Once the models are trained, they are tested with the learned parameters on \texttt{IBMQ} hardware (with up to $8$ qubits, depending on the environment). The models are able to replicate the learned near-optimal behavior.
    
    \medbreak
    
    \noindent\textit{Remarks.} As without entanglement the \glspl{vqc} can be simulated classically, we agree with the authors that the proposed algorithm should be considered as a \gls{qirl} approach. As the proposed model incorporates also a classical network, it is nor clear, what part of the learning is conducted with the \gls{vqc}. The simple circuit structure might also explain, that the results for testing on hardware are stable. Usually, a big portion of the noise is caused by two-qubit gates, which are not present in the used \gls{vqc}. A partial re-implementation of this work can also be found in Ref.~\cite{Truong_2023}.
    
    % \medbreak
    
    % \input{Tables/Hsiao_2022}

%------------------------------------------------------------
\paragraph{\label{subsubsec:Architecture}Compendium of Architecture Discussions}\mbox{}\\
%------------------------------------------------------------
    
    \vspace{-1em}
    \noindent As demonstrated by the previously discussed work of Hsiao et al.~\cite{Hsiao_2022}, it is important to put careful consideration into the design of the employed quantum model architecture. In the following, we briefly summarize several works that make contributions in that direction. The idea of incorporating the information of multiple timesteps via recurrent networks is discussed by S. Y.-C. Chen~\cite{Chen_2023a} and extended in Ref.~\cite{Chen_2023b}. Several explicit \gls{vqc} architectures are compared and analyzed by  Dr\v{a}gan et al.~\cite{Dragan_2022}. An automated approach for architecture generation is proposed in Sun et al.~\cite{Sun_2023}. Different encoding techniques are discussed by Andr\'{e}s et al.~\cite{Andres_2023}. Drawing a connection to a different context, the work by Park et al.~\cite{Park_2020} proposes to vary the architecture itself, by dynamically in- and excluding two-qubit gates.

    \medbreak

    \noindent\textit{Recurrent Quantum Neural Networks.} The work by S. Y.-C. Chen~\cite{Chen_2023a} proposes the use of \glspl{qrnn} in the $Q$-learning setting (see~\cref{subsec:VQC_based_ValueFunction}), specifically \gls{qlstm}~\cite{Chen_2022b}. This enables the agent to also incorporate information from previous timesteps into the decision process. In is experimentally demonstrated on the \texttt{CartPole} environment, that the \gls{qrnn} is at least least competitive -- if not superior -- to purely classical models of similar size. It is also discussed that the method might be well suited for partially observable environments, establishing a connection to~\cite{Kimura_2021}. A continuation of this line of research in Ref.~\cite{Chen_2023b} proposes a more efficient training routine for \gls{qrnn}, based on reservoir computing~\cite{Lukovsevivcius_2009} and the \gls{qa3c} approach discussed in Ref.~\cite{Chen_2023c}.

    \medbreak

    \noindent\textit{Explicit Architecture Comparison.} A study by Dr\v{a}gan et al.~\cite{Dragan_2022} compares various circuit architectures for a modified version of the \texttt{FrozenLake} environment. The underlying algorithm is a quantum version of \gls{ppo} (see~\cref{subsec:VQC_based_Policy}) and the \glspl{vqc} are combined with classical \glspl{nn} to a hybrid model. The results suggest that the performance is strongly dependent on the choice of \gls{vqc} architecture. Measures like expressibility~\cite{Sim_2019}, entanglement capability~\cite{Sim_2019}, and effective dimension~\cite{Abbas_2021} provide an a priori indicator for the potential suitability of the architecture. However, there seems to be no clear correlation between the concrete value of these measures and the \gls{rl} performance.

    \medbreak

    \noindent\textit{Continuous Environments and Encoding.} The work by Kruse et al.~\cite{Kruse_2023} extends the actor-critic paradigm (discussed e.g. in Ref.~\cite{Dragan_2022}) to continuous action spaces. The authors demonstrate that the quantum agent is able to learn in the environments \texttt{Pendulum-v1} and \texttt{LunarLander-v2}. It is conjectured, that applying an \texttt{arctan} function to data points -- as often done in literature -- is indeed counter-productive for the overall performance. Moreover, a \emph{stacked} encoding is proposed, which uses angle encoding on multiple qubits for a single data point. This allows to avoid pre-processing with a classical neural network, ensuring potential performance improvements can really be attributed to the quantum agent. On both benchmarks a reduction in parameter complexity compared to classical agents is reported. However, this only holds true for certain design choices, which again highlights the importance of architecture selection.

    \medbreak

    \noindent\textit{Automatic Generation of Architectures.} Sun et al.~\cite{Sun_2023} propose an automated tool for the generation of \gls{qrl}-suitable circuit architectures. The method is based on \gls{dqas}~\cite{Zhang_2022}, i.e. the architecture itself is trained using gradient-based methods. The approach is studied within the framework of quantum $Q$-learning (see \cref{subsec:VQC_based_ValueFunction}) on the \texttt{FrozenLake} environment. Using \gls{dqas}, the authors are able to identify a \gls{vqc} architecture that seems to be very well-suited for the given problem and outperforms some typically used problem-agnostic circuit designs.

    \medbreak

    \noindent\textit{Encoding Considerations.} The work by Andr\'{e}s et al.~\cite{Andres_2023} compares different strategies for encoding data into the \gls{vqc}, all within the context of quantum $Q$-learning (see~\cref{subsec:VQC_based_ValueFunction}). Evaluations are conducted on three environments within the energy-efficiency and management context. The authors compare three different architecture layouts: (1) classical data is pre-processed and reduced in dimensionality using a \gls{nn} and encoded via rotational parameters; (2) similar, but data re-uploading~\cite{Perez_2020} is employed; (3) classical data is normalized and encoded via amplitude encoding~\cite{Schuld_2018}, output is post-processed with a \gls{nn}; The authors claim superior performance compared to classical models of similar size, especially using amplitude encoding. However, it has to be noted, that the experiments were quite small-scale. The combination with \glspl{nn} complicates statements on the actual contribution of the quantum part. It also has to be noted, that amplitude encoding might not be \gls{nisq}-compatible in the general case.
    
%------------------------------------------------------------------------------------------------------%
%------------------------------------------------------------------------------------------------------%
%------------------------------------------------------------------------------------------------------%

%------------------------------------------------------------
\paragraph{\label{subsubsec:Chen_2022a}Variational quantum reinforcement learning via evolutionary optimization, Chen et al.~(2022)}\mbox{}\\
% \cite{Chen_2022a}
%------------------------------------------------------------

    \vspace{-1em}
    \noindent\textit{Summary.} The main focus of the paper by Chen er al.~\cite{Chen_2022a} is the investigation of gradient-free evolutionary optimization for $Q$-learning with \glspl{vqc}. This routine is tested in two different scenarios, for each of which also a state encoding scheme is proposed. More concretely, amplitude encoding is applied to the \texttt{CartPole} environment. For the gridworld environment \texttt{MiniGrid} with larger state space ($147$ dimensional), the paper proposes a hybrid model with an encoding mechanism based on \gls{tn} techniques.
    
    \medbreak
    
    \noindent\textit{Amplitude Encoding.} The observation space of the \texttt{CartPole} environment is $4$-dimensional. The state values are continuous. This allows the use of amplitude encoding, i.e.\ two qubits can be used to encode the (re-scaled) values into the four amplitudes of the system. The authors follow the method described in Schuld and Petruccione~\cite{Schuld_2018}. This works fine for small systems, but requires not \gls{nisq}-compatible operators for bigger instances.
    
    \medbreak
    
    \noindent\textit{\gls{tn}-based Encoding.} The \texttt{MiniGrid} environment is similar to \texttt{FrozenLake}, as the goal in both environments is to navigate from a start to a goal state on the shortest way possible. The paper uses simple environment configurations, with state spaces of size $5 \times 5$, $6 \times 6$, and $8 \times 8$. The observation space is of dimensionality $7 \times 7 \times 3$. The agent has to decide between $6$ actions, of which only $4$ are relevant in the simplified scenario. The reward is defined as $1 - 0.9 \cdot \mathrm{number\_steps}/\mathrm{max\_number\_steps}$. Apart from the larger observation space, we assume this environment to be about the same complexity as \texttt{FrozenLake}.
    
\begin{figure}[ht]
        \centering
            \subfloat[\centering \Gls{tn} for performing dimensionality reduction;]{\includegraphics[width=0.36\textwidth]{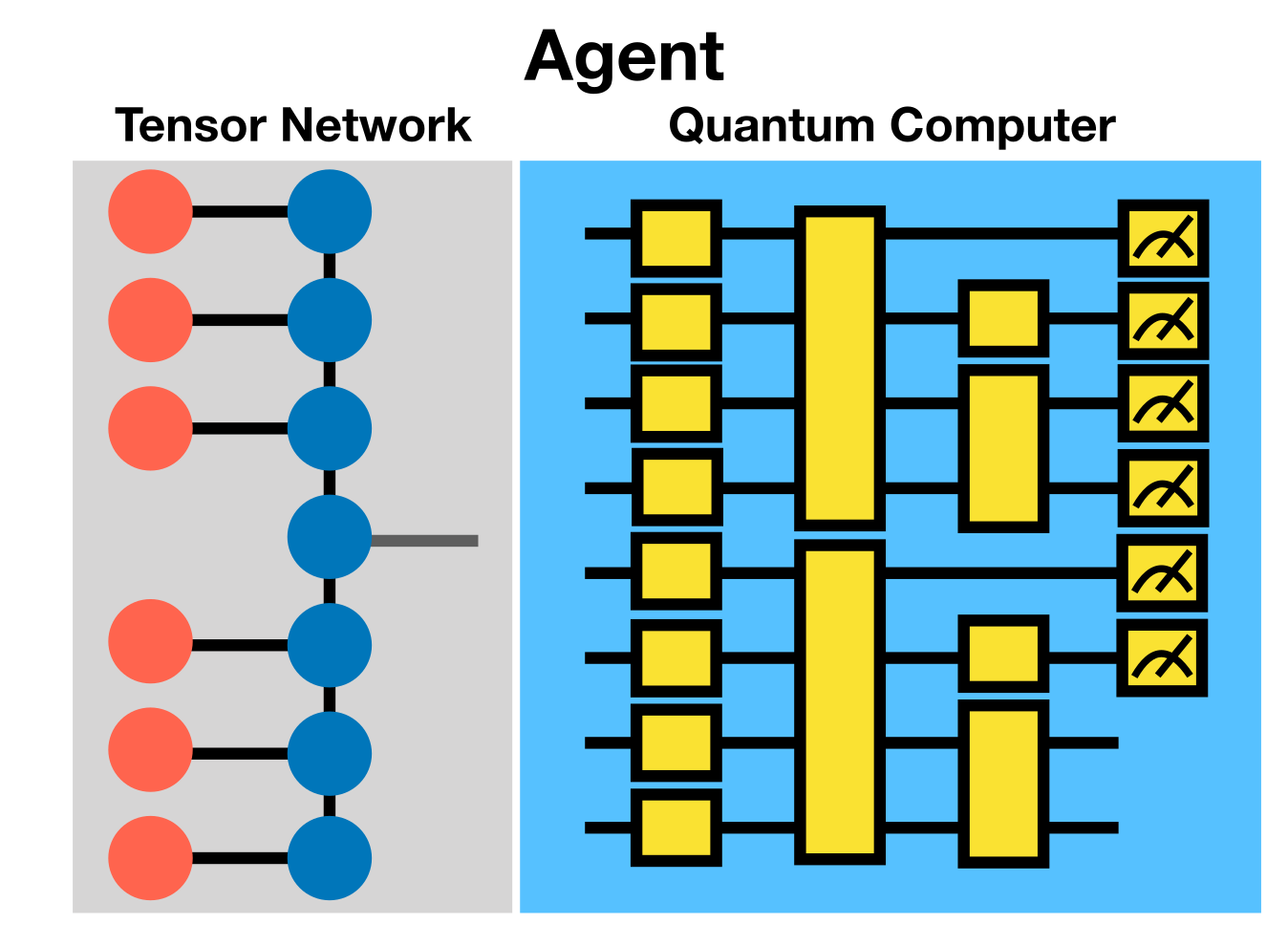} \label{fig:chen_tensor}}
        \qquad
            \subfloat[\centering \gls{vqc} with feature map, several variational layers, and $1$-qubit measurements;]{\includegraphics[width=0.54\textwidth]{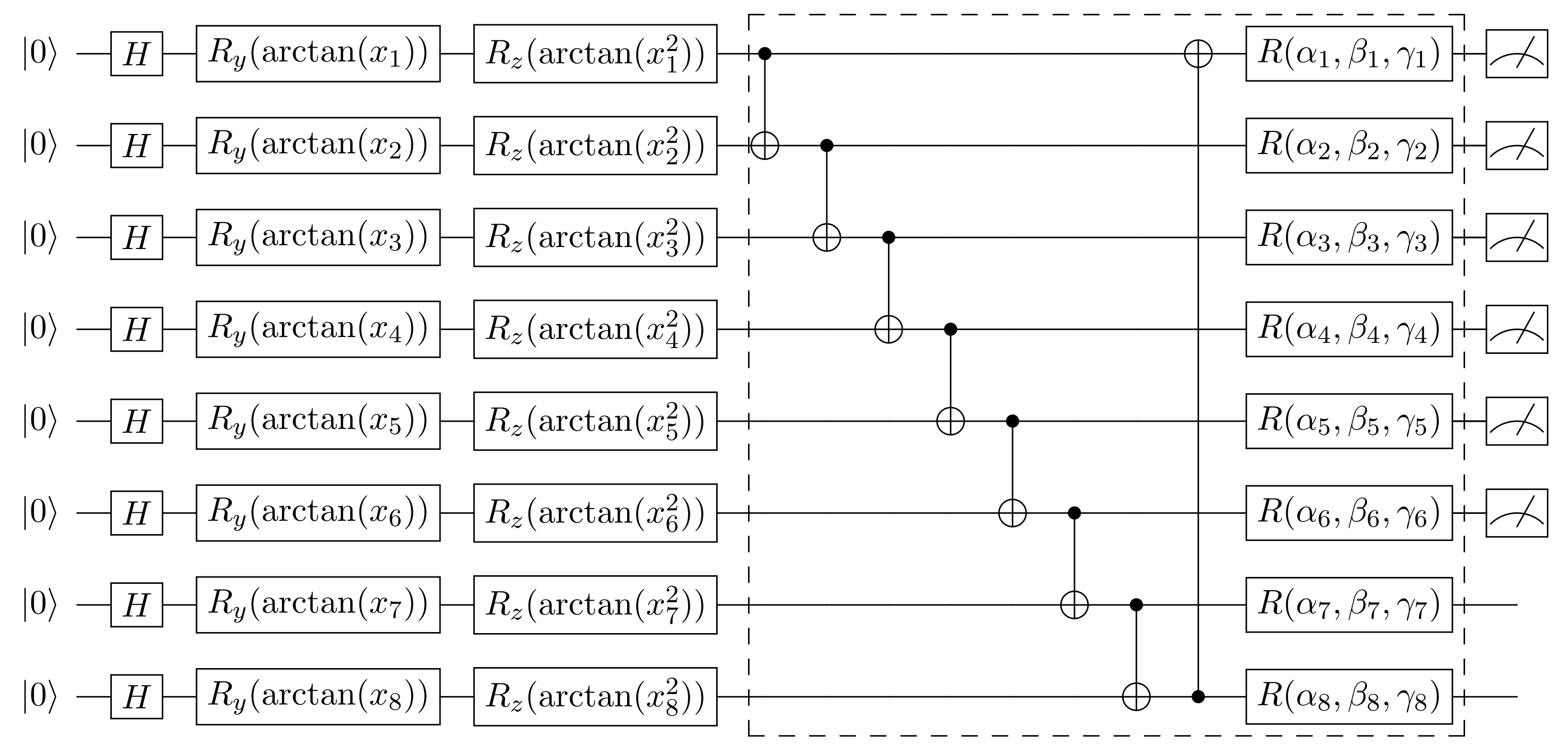} \label{fig:chen_vqc}}
        \caption{Components of the architecture proposed by and taken from Chen et al.~\cite{Chen_2022a}.}
    \end{figure}

    The paper addresses the problem of encoding the $147$-dimensional state into a quantum feature map with just $8$ variational parameters. Other work uses e.g.\ \glspl{cnn} to reduce the dimensionality of the feature space~\cite{Lockwood_2021}. As the encoding networks have to be pre-trained, it is not quite clear, what part of the work is really done by the \gls{vqc}. The authors suggest to use a hybrid encoding scheme based on \glspl{tn}, similar to Chen et al.~\cite{Chen_2021}. The proposed \gls{tn} technique encodes the observation $[v_1, \cdots, v_{147} ]^t$ into the product state $[1-v_1, v_1]^t \otimes [1-v_2, v_2]^t \otimes \cdots \otimes [1-v_N, v_N]^t$, where the individual elements are normalized. Those encoded states represented by the red nodes in \cref{fig:chen_tensor}. The trainable part of the \gls{mps} outputs an $8$-dimensional compressed feature vector. This is represented by the $147+1$ blue nodes and the open leg (i.e.\ outgoing edge) in \cref{fig:chen_tensor}. The bond dimension is a hyperparameter of the \gls{mps}, which correlates with the number of trainable parameters~\cite{Perez_2006}.
    
    \medbreak
    
    \noindent\textit{\gls{vqc} Architecture.} The model follows the typical three-part architecture, i.e.\ first the feature map, then the variational part, and finally some measurements. For the \texttt{CartPole} environment, a simple $2$-qubit circuit with amplitude encoding and $4$ variational layers is used. Both qubits are measured in the Pauli-$Z$ basis and the action corresponding to the higher expectation value is selected. For the \texttt{MiniGrid} environment, the $8$-qubit circuit with just one repetition of the variational layer is used. The encoding is done with the \gls{tn}-compressed state, i.e. the output from the \gls{tn} is encoded into the circuit as shown in \cref{fig:chen_vqc}. As the environment has $6$ actions, the top $6$ qubits are measured, and the action corresponding to the highest expectation value is executed.
    
    \medbreak
    
    \noindent\textit{\gls{rl} with Evolutionary Optimization.} The underlying algorithm is a $Q$-learning \gls{rl} approach. The updates of the \gls{qnn} representing the action-value function are conducted via evolutionary optimization. This implies, that no gradients have to be computed. Usually, this is one major bottleneck of \gls{vqc}-based \gls{rl}, which might be circumvented by this approach.
    
    The paper uses a simplistic instance of an evolutionary algorithm, where mutation, but no recombination operations are employed. An initial population of $M$ individuals is generated, which are used to simulate some episodes on the environment. The best $T$ agents (the ones producing the highest reward averaged over several runs) are selected as parents for the next generation. Random Gaussian noise is applied to this parents (mutation), until $M-1$ children are generated. Additionally, the best individual from the previous generation is kept, i.e.\ again $M$ individuals. This procedure is repeated until a certain convergence criteria is met, e.g.\ a high enough reward.
    
    \medbreak
    
    \noindent\textit{Experimental Findings and Discussion.} The paper applies the two different encoding methods, combined with the evolutionary optimization idea, to the respective environments. All experiments are conducted as noiseless simulations. On the \texttt{CartPole} environment, the $2$-qubit architecture achieves an near-optimal performance with only $26$ parameters, which is significantly less than in most state-of-the-art \glspl{nn}. The authors claim, that with their method the number of parameters can be reduced to $\mathcal{O}(\mathrm{polylog}(n))$. In contrast, classical \gls{ml} requires $\mathcal{O}(\mathrm{poly}(n))$ parameters.
    
    The experiments on the \texttt{MiniGrid} environment employ the described hybrid \gls{tn}-based architecture. Results are compared to an encoding based on a simple \gls{nn}, presumably similar to Lockwood and Si~\cite{Lockwood_2021}. All approaches achieve a near-optimal performance. Overall, the \gls{tn}-model (with large enough bond dimension) slightly outperforms the classical approach. The authors consider this as a proof-of-principle for effectiveness of the \gls{mps} encoding for \gls{rl} learning.
    
    \medbreak
    
    \noindent\textit{Remarks.} The amplitude encoding is currently not feasible for more complex problems, due to the lack of an \gls{nisq}-compatible state-preparation routine. The evolutionary optimization approach could circumvent some of the problems typically associated with gradient based techniques. Experiments on larger-scale environments might be an interesting direction for future work, to investigate how the evolutionary algorithm deals with more complex optimization landscapes. We suggest to incorporate some recombination procedures into the evolutionary algorithm, to enhance its performance.
    
    \medbreak

    \noindent\textit{Multi-Objective Formulation.} Related work by Ding and Spector~\cite{Ding_2023} proposes a version of evolutionary search for the automated generation of \gls{qrl} architectures (see also the discussions on \gls{vqc} architecture above in Ref.~\cite{Hsiao_2022} and related work). The training itself is done with a \gls{qpg} approach~\cite{Jerbi_2021a} (see~\cref{subsec:VQC_based_Policy}) and nested with evolutionary architecture search~\cite{Ding_2022}. This procedure is conducted w.r.t.\ several objectives, including enforcing a as-small-as-possible model size and several noise-related considerations. The approach is validated on the three benchmark environments \texttt{CartPole}, \texttt{MountainCar}, and \texttt{Acrobot}. The results demonstrate improved training behavior -- with smaller model size -- compared to previous work~\cite{Jerbi_2021a}. The authors also further analyze the learned architectures for recurring patterns. However, it is acknowledge that larger-scale experiments are necessary to identify a general guideline for architecture selection.

    \medbreak

    \noindent\textit{Multi-Agent Scenario.} The work by Kölle et al.~\cite{Koelle_2023} extends the framework of Ref.~\cite{Chen_2022a} to the multi-agent setting (see~\cref{subsec:VQC_based_CombinationApproximations}). The authors compare different evolutionary strategies, including mutation-only and two different setups with additional recombination steps. The evaluation is conducted on the \texttt{CoinGame} environment and yields results that are competitive with classical approaches -- using significantly fewer parameters. It has to be noted, that the experiments are too small-scale to make reliable statements about the scaling behaviour of this approach. While evolutionary optimization is certainly an interesting consideration compared to gradient-based techniques, the stated advantage regarding reduced proneness to barren plateaus is not sufficiently documented and should therefore be viewed with some scepticism.

    % \medbreak

    % \input{Tables/Chen_2022a}
    
%------------------------------------------------------------------------------------------------------%
%------------------------------------------------------------------------------------------------------%
%------------------------------------------------------------------------------------------------------%

\subsubsection{Application-Focused Work}
\label{subsec:VQC_based_Application}
This section summarizes work that discusses \gls{vqc}-based \gls{qrl} techniques for specific applications. On the one hand, this is a very important area of research, in order to identify practically relevant \gls{qrl} one day. On the other hand, it has to be noted, that all current work is limited to relatively small problem setups. This can be justified by current hardware restrictions -- but also casts some doubt on the scalability of the stated results. Nonetheless, an overview of the considered ideas might be beneficial for further research:

Applications related to robotics and similar control tasks are discussed in Refs.~\cite{Acuto_2022,Heimann_2022,Cobussen_2023,Bar_2022,Sinha_2023,Hickmann_2023,Kim_2023}. Planning tasks of different form are the focus of Refs.~\cite{Correll_2023,Sanches_2022,Andres_2022,Liu_2023,Kumar_2023,Rainjonneau_2023,Shahid_2023,Rezazadeh_2022}. Collaborative environments are addressed with multi-agent methods in Refs.~\cite{Yan_2022,Park_2023b,Narottama_2023,Park_2023a,Park_2023c,Yun_2023b,Ansere_2023}. The field of finances is discussed in Refs.~\cite{Cherrat_2023b,Yang_2023}. A back-to-the-roots work considers \gls{qrl} for board games in Ref.~\cite{Chao_2023}. Last but not least, the task of designing \gls{vqc} architectures is addressed in Ref.~\cite{Chen_2023d}.

\begin{table}[ht!]
    \centering
    \begin{tabular}{p{\dimexpr 0.15\textwidth-2\tabcolsep-\arrayrulewidth}|p{\dimexpr 0.2\textwidth-2\tabcolsep-\arrayrulewidth}|p{\dimexpr 0.65\textwidth-2\tabcolsep}}
        \toprule
        \textbf{Citation} & \textbf{First Author} & \textbf{Title} \\
        \midrule
        \midrule
        \cite{Acuto_2022} & A. Acuto & \hyperref[subsubsec:Application_robotics]{Variational Quantum Soft Actor-Critic for Robotic Arm Control} \\
        \arrayrulecolor{black!30}\midrule
        \cite{Heimann_2022} & D. Heimann & \hyperref[subsubsec:Application_robotics]{Quantum Deep Reinforcement Learning for Robot Navigation Tasks} \\
        \arrayrulecolor{black!30}\midrule
        \cite{Cobussen_2023} & J. Cobussen & \hyperref[subsubsec:Application_robotics]{Quantum Reinforcement Learning for Sensor-Assisted Robot Navigation Tasks} \\
        \arrayrulecolor{black!30}\midrule
        \cite{Bar_2022} & N. F. Bar & \hyperref[subsubsec:Application_robotics]{An Approach Based on Quantum Reinforcement Learning for Navigation Problems} \\
        \arrayrulecolor{black!30}\midrule
        \cite{Sinha_2023} & A. Sinha & \hyperref[subsubsec:Application_robotics]{Nav-Q: Quantum Deep Reinforcement Learning for Collision-Free Navigation of Self-Driving Cars} \\
        \arrayrulecolor{black!30}\midrule
        \cite{Hickmann_2023} & M. L. Hickmann & \hyperref[subsubsec:Application_robotics]{Potential analysis of a Quantum RL controller in the context of autonomous driving} \\
        \arrayrulecolor{black!30}\midrule
        \cite{Kim_2023} & G. S. Kim & \hyperref[subsubsec:Application_robotics]{Realizing Stabilized Landing for Computation-Limited Reusable Rockets: A Quantum Reinforcement Learning Approach} \\
        \arrayrulecolor{black}\midrule
        \cite{Correll_2023} & R. Correll & \hyperref[subsubsec:Application_planning]{Quantum Neural Networks for a Supply Chain Logistics Application} \\
        \arrayrulecolor{black!30}\midrule
        \cite{Sanches_2022} & F. Sanches & \hyperref[subsubsec:Application_planning]{Short quantum circuits in reinforcement learning policies for the vehicle routing problem} \\
        \arrayrulecolor{black!30}\midrule
        \cite{Andres_2022} & E. Andr\'{e}s & \hyperref[subsubsec:Application_planning]{On the Use of Quantum Reinforcement Learning in Energy-Efficiency Scenarios} \\
        \arrayrulecolor{black!30}\midrule
        \cite{Liu_2023} & D. Liu & \hyperref[subsubsec:Application_planning]{Multi-agent quantum-inspired deep reinforcement learning for real-time distributed generation control of 100\% renewable energy systems} \\
        \arrayrulecolor{black!30}\midrule
        \cite{Kumar_2023} & M. Kumar & \hyperref[subsubsec:Application_planning]{Blockchain Based Optimized Energy Trading for E-Mobility Using Quantum Reinforcement Learning} \\
        \arrayrulecolor{black!30}\midrule
        \cite{Rainjonneau_2023} & S. Rainjonneau & \hyperref[subsubsec:Application_planning]{Quantum Algorithms applied to Satellite Mission Planning for Earth Observation} \\
        \arrayrulecolor{black!30}\midrule
        \cite{Shahid_2023} & M. Shahid & \hyperref[subsubsec:Application_planning]{Introducing Quantum Variational Circuit for Efficient Management of Common Pool Resources} \\
        \arrayrulecolor{black!30}\midrule
        \cite{Rezazadeh_2022} & F. Rezazadeh & \hyperref[subsubsec:Application_planning]{Towards Quantum-Enabled 6G Slicing} \\
        \arrayrulecolor{black}\bottomrule
    \end{tabular}
    \caption{[Part 1] Work considered for ``\gls{qrl} with \glspl{vqc} -- Application-Focused Work'' (\cref{subsec:VQC_based_Application})}
\end{table}

\begin{table}[ht!]
    \centering
    \begin{tabular}{p{\dimexpr 0.15\textwidth-2\tabcolsep-\arrayrulewidth}|p{\dimexpr 0.2\textwidth-2\tabcolsep-\arrayrulewidth}|p{\dimexpr 0.65\textwidth-2\tabcolsep}}
        \toprule
        \textbf{Citation} & \textbf{First Author} & \textbf{Title} \\
        \midrule
        \midrule
        \cite{Yan_2022} & R. Yan & \hyperref[subsubsec:Application_collaborative]{A Multiagent Quantum Deep Reinforcement Learning Method for Distributed Frequency Control of Islanded Microgrids} \\
        \arrayrulecolor{black!30}\midrule
        \cite{Park_2023b} & S. Park & \hyperref[subsubsec:Application_collaborative]{Quantum Multi-Agent Actor-Critic Networks for Cooperative Mobile Access in Multi-UAV System} \\
        \arrayrulecolor{black!30}\midrule
        \cite{Narottama_2023} & B. Narottama & \hyperref[subsubsec:Application_collaborative]{Layerwise Quantum Deep Reinforcement Learning for Joint Optimization of UAV Trajectory and Resource Allocation} \\
        \arrayrulecolor{black!30}\midrule
        \cite{Park_2023a} & S. Park & \hyperref[subsubsec:Application_collaborative]{Quantum Multi-Agent Reinforcement Learning for Autonomous Mobility Cooperation} \\
        \arrayrulecolor{black!30}\midrule
        \cite{Park_2023c} & S. Park & \hyperref[subsubsec:Application_collaborative]{Quantum Reinforcement Learning for Large-Scale Multi-Agent Decision-Making in Autonomous Aerial Networks} \\
        \arrayrulecolor{black!30}\midrule
        \cite{Yun_2023b} & W. J. Yun & \hyperref[subsubsec:Application_collaborative]{Quantum Multi-Agent Actor-Critic Neural Networks for Internet-Connected Multi-Robot Coordination in Smart Factory Management} \\
        \arrayrulecolor{black!30}\midrule
        \cite{Ansere_2023} & J. A. Ansere & \hyperref[subsubsec:Application_collaborative]{Quantum Deep Reinforcement Learning for Dynamic Resource Allocation in Mobile Edge Computing-based IoT Systems} \\
        \arrayrulecolor{black}\midrule
        \cite{Cherrat_2023b} & E. A. Cherrat & \hyperref[subsubsec:Application_finances]{Quantum Deep Hedging} \\
        \arrayrulecolor{black!30}\midrule
        \cite{Yang_2023} & J. Yang & \hyperref[subsubsec:Application_finances]{Apply Deep Reinforcement Learning with Quantum Computing on the Pricing of American Options} \\
        \arrayrulecolor{black}\midrule
        \cite{Chao_2023} & J. Chao & \hyperref[subsubsec:Application_games]{Quantum Enhancements for AlphaZero} \\
        \midrule
        \cite{Chen_2023d} & S. Y.-C. Chen & \hyperref[subsubsec:Application_design]{Quantum Reinforcement Learning for Quantum Architecture Search} \\
        \midrule
        \bottomrule
    \end{tabular}
    \caption{[Part 2] Work considered for ``\gls{qrl} with \glspl{vqc} -- Application-Focused Work'' (\cref{subsec:VQC_based_Application})}
\end{table}

%------------------------------------------------------------------------------------------------------%
%------------------------------------------------------------------------------------------------------%
%------------------------------------------------------------------------------------------------------%

\paragraph{\label{subsubsec:Application_robotics}\gls{qrl} for Robotics and other Control Tasks}\mbox{}\\

\vspace{-1em}
\noindent The work by Acuto et al.~\cite{Acuto_2022} applies the quantum \gls{sac} approach proposed in Ref.~\cite{Lan_2021} to the control of an robotic arm. The environment is implemented as an extension of the \texttt{Acrobot-v1} environment. On this small-scale setup the hybrid quantum model demonstrates reduced parameter complexity compared to classical methods.

\medbreak

\noindent A robot navigation scenario is discussed by Heimann et al.~\cite{Heimann_2022} in a simulated environment. The quantum Q-learning (see~\cref{subsec:VQC_based_ValueFunction}) approach demonstrates parameter reduction compared to classical approaches. The setup is extended to a more complex environment by J. Cobussen~\cite{Cobussen_2023}.

\medbreak
\noindent A similar robot navigation task is considered in Bar et al.~\cite{Bar_2022}, which employs the Q-learning method proposed in Ref.~\cite{Chen_2020}. The authors report a reduction in the number of parameters, which however also yields a decreased success rate for the considered scenarios.

\medbreak
\noindent Collision-free navigation of self-driving cars is considered in Sinha et al.~\cite{Sinha_2023}. The authors employ an actor-critic quantum A2C approach, which is similar to the \gls{qa3c} introduced by Ref.~\cite{Chen_2023c}. On a small $4$-qubit toy environment the proposed approach shows improved training stability compared to classical A2C. A similar problem is considered with tools from quantum $Q$-learning (see~\cref{subsec:VQC_based_ValueFunction}) by Hickmann et al.~\cite{Hickmann_2023}.

\medbreak

\noindent The task of steering reusable rockets is considered in Kim et al.~\cite{Kim_2023}. The unspecified \gls{qrl} method demonstrates reduced memory requirements (by requiring fewer parameters) on an $8$-qubit toy environment.

%------------------------------------------------------------------------------------------------------%
%------------------------------------------------------------------------------------------------------%
%------------------------------------------------------------------------------------------------------%

\paragraph{\label{subsubsec:Application_planning}\gls{qrl} for Planning Tasks}\mbox{}\\

\vspace{-1em}
\noindent The \gls{vrp} is considered by Correll et al.~\cite{Correll_2023} via an quantum-enhanced attention mechanism. Several parts of a classical encoder-decoder model with attention mechanism~\cite{Kool_2018} are replaced with medium-scale \glspl{vqc} (up to $10$ qubits). With using quantum methods to implement orthogonal \glspl{nn}~\cite{Kerenidis_2021}, a potential speed-up during inference is reported. Experimental on a simple instance of the \gls{tsp} are conducted to support this claim. A simpler approach for the same task is considered in Sanches et al.~\cite{Sanches_2022}, where only the attention heads are replaced with $4$-qubit \glspl{vqc}.

\medbreak

\noindent The work by Andr\'{e}s et al.~\cite{Andres_2022} considers different planing tasks related to energy-efficiency scenarios.	The authors employ quantum actor-critic methods (see~\cref{subsec:VQC_based_Policy}) to address these tasks. The authors report a slower convergence compared to classical methods, however therefore a reduced parameter complexity. Similar scenarios within the energy context are also discussed by Liu et al.~\cite{Liu_2023} and Kumar et al.~\cite{Kumar_2023}.

\medbreak

\noindent The task of satellite mission planning is formulated as a scheduling problem and addressed by Rainjonneau et al.~\cite{Rainjonneau_2023}. The authors apply two different quantum-enhanced methods within this context: (1) policy approximation (see~\cref{subsec:VQC_based_Policy}) with \glspl{vqc}; (2) replacing several components of AlphaZero with quantum components, similar as to discussed in Ref.~\cite{Chao_2023}; The experiments with $4$-qubit circuits demonstrate a clear improvement compared to straightforward greedy methods.

\medbreak 

\noindent The problem of distributing common pool resources is discussed by Shahid and Hassan~\cite{Shahid_2023}. Quantum-enhanced $Q$-learning (see~\cref{subsec:VQC_based_ValueFunction}) is applied to an $8$-qubit toy environment, and superior training performance compared to classical models of similar size is reported.

\medbreak

\noindent A task from mobile communication (6G slicing) is considered in Rezazadeh et al.~\cite{Rezazadeh_2022}. The authors employ the \gls{vqc}-based $Q$-learning approach proposed in Ref.~\cite{Chen_2020} and claim improvements w.r.t. parameter complexity and the potential for distributed computing.

%------------------------------------------------------------------------------------------------------%
%------------------------------------------------------------------------------------------------------%
%------------------------------------------------------------------------------------------------------%

\paragraph{\label{subsubsec:Application_collaborative}\gls{qrl} in Collaborative Scenarios}\mbox{}\\

\vspace{-1em}
\noindent Different tasks that are based on the collaboration of multiple entities are discussed in a series of work by Yan et al.~\cite{Yan_2022}, Park et al.~\cite{Park_2023b,Park_2023a,Park_2023c}, Yun et al.~\cite{Yun_2023b}, Narottama et al.~\cite{Narottama_2023}, and Ansere et al.~\cite{Ansere_2023}. The foundation is the multi-agent approach \gls{qmarl} proposed in Ref.~\cite{Yun_2022} with smaller extensions. On respective toy environments, the approaches demonstrate faster convergence and reduced parameter complexity compared to classical implementations.

%------------------------------------------------------------------------------------------------------%
%------------------------------------------------------------------------------------------------------%
%------------------------------------------------------------------------------------------------------%

\paragraph{\label{subsubsec:Application_finances}\gls{qrl} for Finances}\mbox{}\\

\vspace{-1em}
\noindent The work by Cherrat et al.~\cite{Cherrat_2023b} addresses the task of \emph{deep hedging} with distributional actor-critic methods. Classical methods are modified with quantum-enhanced orthogonal \glspl{nn}~\cite{Kerenidis_2021}, which promises speed-ups during inference. This is supported by medium-scale hardware test on up to $16$ qubits -- which makes this one of the largest-scale demonstrations of \gls{vqc}-based \gls{qrl}.

\medbreak

\noindent Another work within the context of finances, conducted by J. Yang~\cite{Yang_2023}, proposes the use of quantum Q-learning (see~\cref{subsec:VQC_based_ValueFunction}) to speed up calculations.

%------------------------------------------------------------------------------------------------------%
%------------------------------------------------------------------------------------------------------%
%------------------------------------------------------------------------------------------------------%

\paragraph{\label{subsubsec:Application_games}\gls{qrl} for Games}\mbox{}\\

\vspace{-1em}
\noindent The work by Chao et al.~\cite{Chao_2023} thinks back to the origins of classical \gls{rl} and consider es the board game \texttt{Orthello}, which basically is a simplified version of \texttt{Go}. To solve this toy environment, the authors modify two components of AlphaZero~\cite{Silver_2018}: (1) replacing function approximators with \glspl{vqc}; (2) using tensor network methods for feature extraction; For simulations on up to $12$ qubits, the methods show performance compared to classical approaches.

%------------------------------------------------------------------------------------------------------%
%------------------------------------------------------------------------------------------------------%
%------------------------------------------------------------------------------------------------------%

\paragraph{\label{subsubsec:Application_design}\gls{qrl} for Architecture Design}\mbox{}\\

\vspace{-1em}
\noindent S. Y.-C. Chen~\cite{Chen_2023d} addresses the task of quantum circuit design. The author uses the actor-critic method \gls{qa3c}~\cite{Chen_2023c} to generate circuits that prepare $2$-qubit Bell states and GHZ states on up to $3$ qubits.

\subsection{Projective Simulation for Quantum Reinforcement Learning}
\label{subsec:projective_simulaton}
%------------------------------------------------------------

\begin{table}[ht!]
    \centering
    \begin{tabular}{p{\dimexpr 0.15\textwidth-2\tabcolsep-\arrayrulewidth}|p{\dimexpr 0.2\textwidth-2\tabcolsep-\arrayrulewidth}|p{\dimexpr 0.65\textwidth-2\tabcolsep}}
        \toprule
        \textbf{Citation} & \textbf{First Author} & \textbf{Title} \\
        \midrule
        \midrule
        \cite{Briegel_2012} & H. J. Briegel & \hyperref[subsubsec:Briegel_2012]{Projective simulation for artificial intelligence} \\
        \arrayrulecolor{black!30}\midrule
        \cite{Melnikov_2017} & A. A. Melnikov & \hyperref[subsubsec:Briegel_2012]{Projective simulation with generalization} \\
        \midrule
        \cite{Boyajian_2020} & W. L. Boyajian & \hyperref[subsubsec:Briegel_2012]{On the convergence of projective-simulation--based reinforcement learning in Markov decision processes} \\
        \midrule
        \cite{Paparo_2014} & G. D. Paparo & \hyperref[subsubsec:Briegel_2012]{Quantum Speedup for Active Learning Agents} \\
        \midrule
        \cite{Teixeira_2021a} & M. Teixeira & \hyperref[subsubsec:Briegel_2012]{Quantum Reinforcement Learning Applied to Games} \\
        \midrule
        \cite{Teixeira_2021b} & M. Teixeira & \hyperref[subsubsec:Briegel_2012]{Quantum Reinforcement Learning Applied to Board Games} \\
        \midrule
        \cite{Dunjko_2015a} & V. Dunjko & \hyperref[subsubsec:Briegel_2012]{Quantum-enhanced deliberation of learning agents using trapped ions} \\
        \midrule
        \cite{Sriarunothai_2018} & T. Sriarunothai & \hyperref[subsubsec:Briegel_2012]{Speeding-up the decision making of a learning agent using an ion trap quantum processor} \\
        \midrule
        \cite{Flamini_2023} & F. Flamini & \hyperref[subsubsec:Briegel_2012]{Towards interpretable quantum machine learning via single-photon quantum walks} \\
        \arrayrulecolor{black}\bottomrule
    \end{tabular}
    \caption{Work considered for ``Projective Simulation for \gls{qrl}'' (\cref{subsec:projective_simulaton})}
\end{table}

%------------------------------------------------------------
\paragraph{\label{subsubsec:Briegel_2012}Projective simulation for artificial intelligence, Briegel et al.~(2012) and related work}\mbox{}\\
%------------------------------------------------------------

\vspace{-1em}
\noindent\textit{Summary.} \textit{Projective simulation for artificial intelligence} by Briegel et al.~\cite{Briegel_2012} is the first in a series of articles, which propose a learning scheme for creative behavior. This is understood in the sense that the agent can deal with unseen experiences by relating to other conceivable situations. The method is developed for classical agents. There is only a brief final paragraph, outlining a quantum-mechanical implementation. Since subsequent papers ‘quantize’ the original idea heavily, a brief summary is in order:
The approach is based on a random walk on a previous-experience network (memory), simulating an agent pondering its next action. More specifically, previous experiences compose a network of clips, which is dynamically modified by new experiences. It is important to note that clips, in contrast to actual experiences, are e.g.~remembered observations, states or actions. To select the next action, an observation of the agent activates a clip, followed by a random walk through the network (\glsentrylong{ps}). This is repeated until an action is `excited' and coupled out from the network and the action is selected. It is worthwhile noting that the term \textit{projective} as used here is not related to its use in quantum physics, such as in \textit{projective measurement}.
\medbreak
\noindent
\textit{Action Selection.} The process of action selection is slightly more sophisticated than described above. If a percept $s$ is observed, a random walk through the network starts from the corresponding percept clip. After some deliberation time the random walk reaches an action clip, which is only out-coupled and taken in reality if the percept-action pair $(s,a)$ was rewarded in the past (i.e.\ tagged positively). If not, a new simulation is started. This process repeats until an action clip with positive tag, or a predefined reflection time is reached; in the latter case the action is out-coupled irrespective of the tag.

\medbreak
\noindent
\textit{Learning Procedure.} The actual learning process can be summarized as follows:
\begin{enumerate}
\item If a transition $(s,a)$ is rewarded, increase the network weight of the direct transition $s \rightarrow a$. (Note that the agent might have chosen $s \rightarrow a$ after many steps of \gls{ps}; by reinforcing the direct transition, it might be exploited directly next time);
\item Increase the weights of the indirect transition (all weights of the network that led to the transition $s \rightarrow a$ in the random walk through the network). Thus, the agent discovers useful actions after deliberation of fictitious clips;
\item Introduce damping of all weights to let the agent forget, in order to be able to adapt to new situations (as appearing for example in a time-dependent environment);
\item If a new situation is discovered, a corresponding clip is added to the network and directed edges from all the other clips to the new one are added;
\item Additional extensions can be implemented, such as modifications of clips and creation of completely fictitious compositions of episodes;
\end{enumerate}

This line of research has been continued in Ref.~\cite{Melnikov_2017} (generalization) and \cite{Boyajian_2020} (convergence). In the last paragraph of Ref.~\cite{Briegel_2012} a quantum version of the algorithm is briefly discussed. The idea is to replace the random walk on the network by a quantum walk. A number of subsequent papers investigate the quantum approach more rigorously:

In order to define a quantum walk algorithm as done in Ref.~\cite{Paparo_2014}, the \gls{ps} approach is viewed slightly different. The given clip network with the percept set $S$ is separated in $|S|$ disjoint networks. Thus one obtains a directed weighted graph (a Markov chain) for each percept with action clips as absorber states. Each of the actions is initially flagged (corresponding to the emotion tags of the initial projected simulation proposal). If an actual out-coupled action did not lead to a reward, this particular flag is removed. Now the action selection proceeds in the following way: If the agent observes a percept $s$, a random walk starts through the graph (deliberation) until an action is reached, which is out-coupled only if the action is flagged (reflection). Thus, action selection corresponds to sampling from the conditional probability distribution over the flagged action space. Given the transition matrix $P$ of the Markov chain, subsequent applications of $P$ to the initial state (probability one for the percept clip) realizes the approximate stationary distribution (subsequently referred to as diffusion). Sampling from this distribution and disregarding un-flagged actions produces the correct samples. As $p_s$ is the probability to sample a flagged action from the equilibrium distribution obtained by diffusion, one needs to repeat the sampling process $\mathcal{O}(1/p_s)$ times until a flagged action is sampled. The quantum random walk search algorithm is closely related to Grover's algorithm. By elevating the transition matrix to a diffusion operator and introducing an oracle that marks flagged actions, the quantum algorithm only needs $\mathcal{O}(1/\sqrt{p_s})$ oracle calls. Consequently, a quadratic speed-up for the deliberation process can be achieved. Therefore, this quantum algorithm speeds up the agent's internal computation time for action selection. This technique is extended and applied to board games in Refs.~\cite{Teixeira_2021a,Teixeira_2021b}.

\medbreak
\noindent
\textit{Experimental Implementation.} In Ref.~\cite{Dunjko_2015a} the authors investigate the implementation of the algorithm proposed by Ref.~\cite{Paparo_2014} on an ion trap quantum computer. Results are also backed up by numerical simulations. The actual proof-of-principle experiment with two qubits is discussed in Ref.~\cite{Sriarunothai_2018}, where signatures of the quadratic speed up are observed. Ref.~\cite{Flamini_2023} proposes a quantum-optics based implementation of the projective simulation paradigm. Here, the random walk through the clip network is promoted to a quantum walk of a single photon through an optical interferometer. Outcoupling of an action then corresponds to an occupation number measurement of output modes.

%------------------------------------------------------------
\subsection{Boltzmann Machines for Quantum Reinforcement Learning}
\label{subsec:boltzman_machines}
%------------------------------------------------------------

\begin{table}[ht]
    \centering
    \begin{tabular}{p{\dimexpr 0.15\textwidth-2\tabcolsep-\arrayrulewidth}|p{\dimexpr 0.2\textwidth-2\tabcolsep-\arrayrulewidth}|p{\dimexpr 0.65\textwidth-2\tabcolsep}}
        \toprule
        \textbf{Citation} & \textbf{First Author} & \textbf{Title} \\
        \midrule
        \midrule
        \cite{Jerbi_2021b} & S. Jerbi & \hyperref[subsubsec:Jerbi_2021b]{Quantum Enhancements for Deep Reinforcement Learning in Large Spaces} \\
        \arrayrulecolor{black!30}\midrule
        \cite{Crawford_2018} & D. Crawford & \hyperref[subsubsec:Jerbi_2021b]{Reinforcement Learning Using Quantum Boltzmann Machines} \\
        \arrayrulecolor{black!30}\midrule
        \cite{Schenk_2022} & M. Schenk & \hyperref[subsubsec:Jerbi_2021b]{Hybrid actor-critic algorithm for quantum reinforcement learning at CERN beam lines} \\
        \arrayrulecolor{black!30}\midrule
        \cite{Levit_2017} & A. Levit & \hyperref[subsubsec:Jerbi_2021b]{Free energy-based reinforcement learning using a quantum processor} \\
        \arrayrulecolor{black}\bottomrule
    \end{tabular}
    \caption{Work considered for ``Boltzmann Machines for \gls{qrl}'' (\cref{subsec:boltzman_machines})}
\end{table}

%------------------------------------------------------------
\paragraph{\label{subsubsec:Jerbi_2021b}Quantum Enhancements for Deep Reinforcement Learning in Large Spaces, Jerbi et al.~(2021) and related work}\mbox{}\\
% \cite{Jerbi_2021b}
%------------------------------------------------------------

\begin{figure}[ht]
    \centering
    \includegraphics[width=0.7\textwidth]{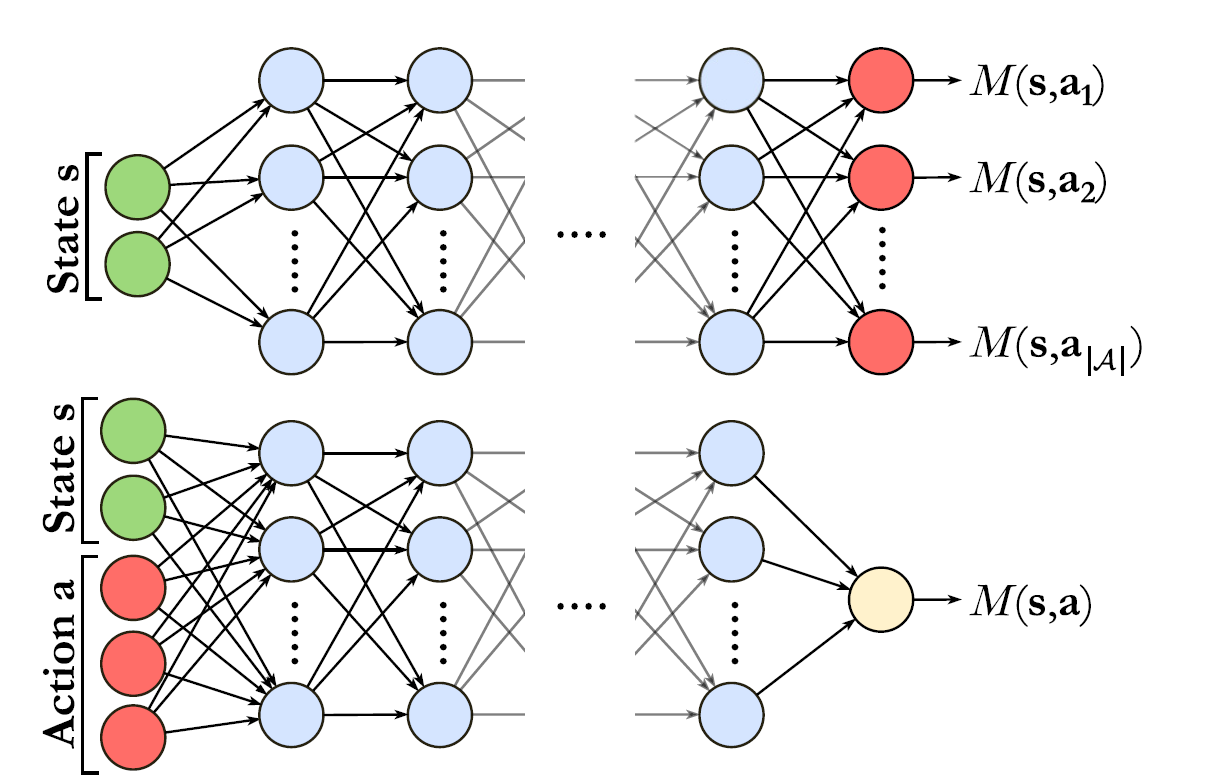}
    \caption{Difference between architectures in $Q$-learning (upper part of the figure) and energy-based models (lower part of the figure) as shown in Jerbi et al.~\cite{Jerbi_2021b}}
    \label{EnergybasedModel}
\end{figure}

\vspace{-1em}
\noindent\textit{Summary}. The work presented in Ref.~\cite{Jerbi_2021b} investigates an alternative \gls{nn} architecture to those often used for learning the $Q$-function (or more generally the merit function) in \gls{rl} tasks. The authors argue that these alternative models perform advantageously in large action spaces. This is due to their capability to represent multimodal functions better than standard network architectures, while using a similar number of parameters. It is further found that these alternative architectures are closely related to energy-based models, some of which admit quantum representations. In turn, this allows quantum evaluations, enabling a provable quantum speed-up for fault-tolerant quantum computing.
\medbreak
\noindent
\textit{Motivation.} The standard architecture for $Q$-learning with \glspl{nn} is depicted in \cref{EnergybasedModel} (upper part). The representation of a state is fed into a \gls{nn}, which outputs the values of the so-called merit function (the $Q$-value in case of $Q$-learning) for each possible action (given the state). The policy can be derived from this function by with softmax post-processing. The effective-temperature parameter is decreased over time to reduce exploration and enhance exploitation.

The authors argue that this \gls{nn} architecture is not suited for large action spaces. It has to output a high dimensional function, i.e.\ the merit functions for all actions simultaneously for a given state. The authors argue that this network is unable to approximate a multimodal merit function in case of complex state-action correlations. Instead, the authors discuss the \gls{nn} structure shown in the lower part of \cref{EnergybasedModel}. Here, the state and action is fed to the \gls{nn}, which outputs the corresponding merit function. Action selection is done by sampling from the probability distribution, given by a softmax function on the values of the merit function. Therefore, sampling requires $|A|$ forward passes, where $A$ is the action set, making action selection a computationally expensive  task for large action spaces.

Experiments are conducted on a generalized \texttt{GridWorld} environment with a large set of actions. The associated complex transition function gives rise to one optimal and many sub-optimal policies. The authors find that the \gls{nn} architecture shown in the lower part of \cref{EnergybasedModel} indeed performs better, but at the cost of the expensive sampling described before.
\medbreak
\noindent
\textit{Energy-based Models.} The potential for quantum speed up comes from the observation that the second architecture in \cref{EnergybasedModel} is equivalent to a certain kind of energy-based model. Energy-based function approximators are used for generative modeling of probability distributions based on the Boltzmann-Gibbs distribution with respect to an energy functional. Boltzmann machines are one instance of such energy-based models where the energy functional is given by a spin-spin interaction model. However, Boltzmann machines are hard to train which led to the development of restricted Boltzmann machines where a special interaction structure with a hidden layer enables more efficient training. In Ref.~\cite{Jerbi_2021b} the authors observe that the lower architecture in \cref{EnergybasedModel} is equivalent to a generalized form of restricted Boltzmann machines.

\medbreak
\noindent
\textit{Quantum Speed-Up.}
Inspired by this insight, the authors next investigate quantum energy-based models. Here, the classical spin-spin interaction energy is promoted to a spin Hamiltonian, known as quantum Boltzmann machines and restricted quantum Boltzmann machines. Some of these models allow efficient training, while the hardness of sampling remains. To speed up sampling in the classical and quantum setting, the following quantum subroutines for a \gls{rl} algorithm are discussed:

(1) Quantum Gibbs sampling: The Gibbs-Boltzmann distribution is prepared as a qsample, from which expectations values can be sampled with quadratic speed-up, compared to classical Monte-Carlo sampling methods. (2) Gibbs-state preparation by Hamilton simulation: Using Hamilton-simulation techniques, an approximation to the Gibbs qsample can be prepared, leading to quadratic speed-up compared to exact sampling (calculating all energies and explicitly normalizing the probability distribution). (3) Quantum simulated annealing: This method uses a quantum method for the approximate Monte-Carlo sampling of the Gibbs state itself by leveraging quantum random walks on graphs.

All methods discussed so far need oracularized access to the Hamiltonian and it is unlikely that they could be realized on current hardware. A realization on near-term hardware might be achieved by (4) Variational Gibbs-state preparation: Here, a variational circuit can be employed to approximate a Gibbs qsample, using the free energy as an objective. Any quantum speed up, however, for this method is heuristic and has not been made rigorous so far.

\medbreak
\noindent
\textit{Remarks.} Related work~\cite{Crawford_2018,Schenk_2022,Levit_2017} proposes models based on quantum Boltzmann machines for quantum annealing hardware. Since this literature survey focuses on algorithms proposed for gate-based \gls{qc}, we do not include a detailed summary here.

%------------------------------------------------------------
\subsection{Quantum Policy and Value Iteration}
\label{subsec:QPI}
%------------------------------------------------------------
 
So far, we have considered \gls{qrl} algorithms that employ \gls{qc} for function approximation or propose quantum approaches to alternative learning frameworks such as \gls{ps}. We now turn to proposals that replace subroutines of existing \gls{rl} frameworks by quantum algorithms such as amplitude estimation, quantum maximum finding and, respectively, quantum matrix inversion. As a result, the proposed \gls{qrl} algorithms guarantee improved sample or computational complexity.
As these methods need oracular access to the environment, they should be categorized as post-\gls{nisq} algorithms. 

\medbreak

\begin{table}[ht!]
    \centering
    \begin{tabular}{p{\dimexpr 0.15\textwidth-2\tabcolsep-\arrayrulewidth}|p{\dimexpr 0.2\textwidth-2\tabcolsep-\arrayrulewidth}|p{\dimexpr 0.65\textwidth-2\tabcolsep}}
        \toprule
        \textbf{Citation} & \textbf{First Author} & \textbf{Title} \\
        \midrule
        \midrule
        \cite{Wang_2021a} & D. Wang & \hyperref[subsubsec:Wang_2021a]{Quantum algorithms for reinforcement learning with a generative model} \\
        \midrule
        \cite{Ganguly_2023a} & B. Ganguly & \hyperref[subsubsec:Ganguly_2023a]{Quantum Computing Provides Exponential Regret Improvement in Episodic Reinforcement Learning} \\
        \arrayrulecolor{black!30}\midrule
        \cite{Zhong_2023} & H. Zhong & \hyperref[subsubsec:Ganguly_2023a]{Provably Efficient Exploration in Quantum Reinforcement Learning with Logarithmic Worst-Case Regret} \\
        \arrayrulecolor{black!30}\midrule
        \cite{Ganguly_2023b} & B. Ganguly & \hyperref[subsubsec:Ganguly_2023a]{Quantum Acceleration of Infinite Horizon Average-Reward Reinforcement Learning} \\
        \arrayrulecolor{black}\midrule
        \cite{Cherrat_2023a} & E. A. Cherrat & \hyperref[subsubsec:Cherrat_2023a]{Quantum Reinforcement Learning via Policy Iteration} \\
        \midrule
        \cite{Wiedemann_2022} & S. Wiedemann & \hyperref[subsubsec:Wiedemann_2022]{Quantum Policy Iteration via Amplitude Estimation and Grover Search - Towards Quantum Advantage for Reinforcement Learning} \\
        \bottomrule
    \end{tabular}
    \caption{Work considered for ``Quantum Policy and Value Iteration'' (\cref{subsec:QPI})}
\end{table}

%------------------------------------------------------------

%------------------------------------------------------------------------------------------------------%
%------------------------------------------------------------------------------------------------------%
%------------------------------------------------------------------------------------------------------%

%------------------------------------------------------------
\paragraph{\label{subsubsec:Wang_2021a}Quantum algorithms for reinforcement learning with a generative model, Wang et al.~(2021)}\mbox{}\\
% \cite{Wang_2021a}
%------------------------------------------------------------

\vspace{-1em}
\noindent\textit{Summary.} The work in Ref.~\cite{Wang_2021a} proposes two algorithms for \gls{rl} with a generative model and rigorously derives bounds for their sample complexity.

\medbreak
\noindent
\textit{Classical Generative Models.} Classically, the term generative model describes a simulator, which queried with a state-action pair $(s,a)$, produces a sample $s^\prime \sim P(\cdot|s,a)$. Thus, by repeated sampling for each state-action pair, one can estimate the transition matrix of the underlying \gls{mdp}. This allows to subsequently obtain an approximation of the optimal policy by means of value iteration. Over the years there has been tremendous effort devoted to improving sample efficiency (defined as the number of times the simulator has to be queried). This performance metric is meaningful if one assumes that every query of the simulator is costly. The best classical algorithm \cite{Li_2020b} requires a total number of $\mathcal{O}(|S| |A| \Gamma^3/\epsilon^2)$ samples, where $|S|$ and $|A|$ are the number of states and actions, $\Gamma=1/(1-\gamma)$ is the effective horizon of the \gls{mdp}, and $\epsilon$ is the deviation of the optimal value function from the approximation. The sample complexity is linear in the product $|S||A|$, since the transition matrix has to be estimated for each $(s,a)$. The factor $1/\epsilon^2$ originates from Hoeffding's inequality (indeed bounding the deviation of a sample average from its real value by $\epsilon$, requires $\mathcal{O}(1/\epsilon^2)$ samples). The origin of the third power of $\Gamma$, in contrast, is less intuitive. Note that the sample complexity of the classical algorithm is also a lower bound (in the classical case) and therefore optimal.

\medbreak
\noindent
\textit{Incorporating Quantum Subroutines.} As shown in Ref.~\cite{Wang_2021a}, the classical sample complexity can be reduced by replacing the classical mean-estimation subroutine in Ref.~\cite{Li_2020b} by a quantum routine based on the quantum mean-estimation algorithm \cite{Brassard_2002}. Even though the optimal classical algorithm is more sophisticated as outlined above and so is its quantization, the following discussion captures the essential features. The quantum subroutine requires the generative model in oracle form and can then be used to estimate the expectation value $\mathbb{E}(V)=\sum_{s^\prime} P(s^\prime|s,a)V(s^\prime)$ (which appears in the Bellman equation) individually for every pair $(s,a)$ in time $\mathcal{O}(1/\epsilon)$. This quadratic speed-up originates from Grover's algorithm, on which the quantum-mean estimation algorithm is based upon. As a consequence, the quantum-policy iteration algorithm achieves the sample complexity $\mathcal{O}(|S||A| \Gamma^{1.5}/\epsilon)$ with an quadratic improvement in $\Gamma$ and $\epsilon$.

The dependence on the size of the action space can be further reduced by using quantum maximum finding \cite{Montanaro_2015} to calculate the maximum over actions in the Bellman optimality equation. However, using this quantum routine, one can not fully exploit the power of the classical optimal algorithm. Hence, while the dependence on $|A|$ is reduced quadratically and the improvement in $\epsilon$ is kept, the improvement in $\Gamma$ is lost. As a result, the algorithm based on both quantum-mean estimation and quantum maximum finding achieves a sample complexity $\mathcal{O}(|S| \sqrt{|A|} \Gamma^3/\epsilon)$.

Finally, the lower bound $\mathcal{O}(|S||A|\Gamma^{1.5}/\epsilon)$ is derived and possible improvements of the algorithm to reach this limit are discussed.

%------------------------------------------------------------
\paragraph{\label{subsubsec:Ganguly_2023a}Quantum computing provides exponential regret improvement in episodic reinforcement learning, Ganguly et al.~(2023)}\mbox{}\\
%\cite{Ganguly_2023a}
%------------------------------------------------------------

\vspace{-1em}
\noindent\textit{Summary.} 
In Ref.~\cite{Ganguly_2023a} and independently in Ref.~\cite{Zhong_2023} the authors consider the problem of an agent operating in a finite-horizon episodic tabular \gls{mdp} and investigate if quantum computing can alleviate the exploration-exploitation trade-off. This problem has been considered for the case of bandits \cite{Wan_2023, Lumbreras_2022, Li_2022} but is here generalized to the full multi-state \gls{rl} problem. In the online setting, the agent only has access to the next state and reward given its current state and chosen action. In contrast, previous work \cite{Wang_2021a} assumed access to a generative model, which can be queried with arbitrary state-action pairs producing samples of the next state and reward. This setting does not consider the exploration-exploitation trade-off that arises from online interaction with the environment. Here, the agent must learn to discover high-reward states by a suitable exploration strategy. The performance of the agent in this problem can be measured by the regret, which is defined as the cumulative difference between the optimal value function and its approximation after $K$ episodes. The goal is to design an algorithm with the weakest scaling of the regret in $K$, indicating a more effective trade-off between exploration and exploitation. The classical UCB-VI algorithm achieves the lower bound $\Omega(\sqrt{K})$ \cite{Jaksch_2010, Azar_2017} of the regret. The proposed quantum algorithm in Ref.~\cite{Ganguly_2023a} builds upon this classical algorithm by replacing the mean estimation routine with a quantum algorithm. Given a state-action pair, the quantum algorithm assumes a `transition oracle' which generates a quantum superposition over all possible next states with amplitudes given by the square root of the respective transition probabilities. A similar oracle is used for generating rewards. The algorithm utilizes the quantum multivariate mean estimation algorithm \cite{Hamoudi_2021}, which reduces the number of samples required to satisfy a given error bound for mean estimation quadratically. The result is a decrease of the regret of the quantum algorithm from $\mathcal{O}(\sqrt{K})$ to $\mathcal{O}(1)$ up to logarithmic factors. This is an exponential improvement over classical results. In a follow-up work by the same authors \cite{Ganguly_2023b}, the results were extended to infinite horizon problems, where an exponential reduction in regret from $\mathcal{O}(\sqrt{T})$ to $\mathcal{O}(1)$ ($T$ being the total number of time steps) is achieved. Additionally, Ref.~\cite{Zhong_2023} consideres linear function approximation and demonstrates that the exponential improvement is maintained.

\paragraph{\label{subsubsec:Cherrat_2023a}Quantum Reinforcement Learning via Policy Iteration, Cherrat et al.~(2023)}
% \cite{Cherrat_2022}
%------------------------------------------------------------

\vspace{-1em}
\noindent\textit{Summary.}
Ref.~\cite{Cherrat_2023a} proposes a quantum algorithm for an iterative scheme of $Q$-value evaluation and policy improvement. The algorithm evaluates the $Q$-value on a quantum computer, with the state vector representing the $Q$-values, being extracted by measurements. The policy afterwards is improved on a classical device. The algorithm can achieve quantum advantage in certain situations.

To set up the general framework, the authors first formulate the Bellman equation for $Q$-value evaluation as a matrix equation \cite{Lagoudakis_2003}
\begin{equation*}
Q=R+\gamma P \Pi Q\,.
\end{equation*}
Denoting the size of the action and state space as $|A|$ and $|S|$, the $|A||S|$ dimensional vectors $Q$ and $R$ represent the $Q$-values and the reward vector, respectively; the environment transition function is the $|A||S|\times|S|$ dimensional matrix $P$; the policy is represented by an $|S| \times|A||S|$-dimensional matrix $\Pi$; $\gamma$ denotes the usual discounting factor; The authors propose to compute $(1\!\!1-\gamma P \Pi)^{-1}R$ on a quantum device.

\medbreak
\noindent
\textit{Quantum Subroutine: Block Encodings and Linear Algebra.} To perform this task, Ref.~\cite{Cherrat_2023a} relies on so-called block encodings of matrices \cite{Gilyen_2019b}. This powerful framework gives rise to various quantum algorithms for encoding general complex (not necessarily rectangular) matrices in the leading principal block of a larger unitary matrix. Once the data has been loaded, the framework further provides linear-algebra routines such as matrix multiplication, addition \cite{Gilyen_2019b} and inversion \cite{Childs_2017}. The encoding algorithms need quantum access to the data, i.e.~via oracles. Therefore, the methods can be attributed to the post-\gls{nisq} algorithms category. A well-known data-loading scheme is the sparse-input model, viable for sparse matrices. The authors of Ref.~\cite{Cherrat_2023a} apply a more general scheme, the so-called $\mu_p(A)$ \cite{Chakraborty_2019} block encoding of a matrix $A$. Here, the quality (i.e.\ the probability to obtain the correct output of the algorithm, e.g.\ after matrix-vector multiplication and a subsequent measurement) of the encoding depends on the maximum of the column and row norms of the matrix. The aforementioned norm is a function of $p$ and can be chosen freely to optimize the encoding quality. Based on this formalism, the authors show that policy evaluation requires time
\begin{equation}
\label{Cherrat_eq_1}
 \mathcal{O}(\mu_P \Gamma \mathrm{polylog}(|S||A|\Gamma/\epsilon))\,.
\end{equation}
In \cref{Cherrat_eq_1}, the parameter $\Gamma=(1-\gamma)^{-1}$, $\epsilon$ denotes the accuracy of the matrix inversion subroutine. The term $\mu_P$ describes the quality of the encoding of the environment-transition matrix, which depends on the structure of the environment. In the worst case it scales as $\sqrt{|S||A|}$. Due to the sparsity of the transition function of many environments, a better scaling is often expected. As discussed below, for the \texttt{frozen-lake} environment one even finds $\mu_P=\mathcal{O}(1)$. The complexity in \cref{Cherrat_eq_1} assumes an efficient loading routine for the matrices. To achieve efficient loading also for the policy matrix, a QRAM data structure for the policy needs to be constructed. This needs to happen in time $\mathcal{O}(|S||A|)$ for each policy-evaluation step. Afterwards, the matrix can be loaded efficiently for each cycle of the measurement protocol.

\medbreak
\noindent
\textit{Classical Subroutine: Policy Improvement.}
The policy improvement step on a classical device requires reading out the $Q$-vector from the quantum computer after matrix inversion. Naively, one would expect that the measurement process introduces exponential overhead. However, since convergence results for the Bellman equations are based on the maximum norm ($L_\infty$ norm), the authors employ $L_\infty$-norm state tomography \cite{Kerenidis_2020}. This is efficient, i.e.\ requires $\mathcal{O}(1/\epsilon^2)$ shots, where $\epsilon$ now is the target accuracy for the optimal $Q$-values (under $L_\infty$-norm). Consequently, the overall time complexity (neglecting logarithmic terms) of the algorithm is
\begin{equation}
\label{Cherrat_eq_2}
\mathcal{O}(|S||A| +\mu_P\Gamma/\epsilon^2)\,.
\end{equation}
In \cref{Cherrat_eq_2} the factor $1/\epsilon^2$ appears in the second term since the matrix inversion subroutine is called for each of the $1/\epsilon^2$ shots. The first term is the classical complexity of calculating the $argmax$ function for policy improvement and construction of the policy oracle prior to each evaluation step.
\medbreak
\noindent
\textit{Example Environments.}
The authors consider the \texttt{FrozenLake} and the \texttt{InvertedPendulum} environments as examples. We will briefly discuss the insights from the former here: The simple form of the environment allows choosing $\mu_P=1/2$, which thus is independent of the size of the action and state space. Note that the gate complexity is still of the order of $|S||A|$. It only becomes efficient for special structured instances of the environment such as all `holes' on the diagonal of the grid.
\medbreak
\noindent
\textit{Quantum advantage.}
The leading term in \cref{Cherrat_eq_2} is linear in $|S|$ and $|A|$, showing a speed-up with respect to classical linear-system of equations solvers. These exhibit complexity $\mathcal{O}((|S||A|)^\omega)$, with $\omega >1$, and vanilla $Q$-value iteration with complexity $\mathcal{O}(|S|^2|A|)$. Even though a more detailed characterization of possible quantum advantage is not provided in Ref.~\cite{Cherrat_2023a}, it is clear that the speed up can be at most polynomial.
\medbreak
\noindent
\textit{Least-Squares Policy Iteration.}
Finally, the authors generalize the method to least-squares policy iteration \cite{Lagoudakis_2003}, where the $Q$-vector is approximated by a set of basis functions. For details refer to Refs.~\cite{Lagoudakis_2003, Cherrat_2023a}.

%------------------------------------------------------------------------------------------------------%
%------------------------------------------------------------------------------------------------------%
%------------------------------------------------------------------------------------------------------%

%------------------------------------------------------------
\paragraph{\label{subsubsec:Wiedemann_2022}Quantum Policy Iteration via Amplitude Estimation and Grover Search - Towards Quantum Advantage for Reinforcement Learning, Wiedemann et al.~(2022)}\mbox{}\\
% \cite{Wiedemann_2022}
%------------------------------------------------------------

\vspace{-1em}
\noindent\textit{Summary.} In the \gls{qrl} scheme proposed in Refs.~\cite{Wiedemann_2022,Wiedemann_2021}, a policy is evaluated by constructing a superposition of all possible trajectories of an \gls{mdp} with fixed-horizon and with finite action and state space. Making use of amplitude estimation \cite{Brassard_2002}, the number of calls to a state-transition oracle for estimation of the value function (up to some fixed additive error) can be quadratically reduced. A second algorithm finds the optimal policy in the policy space quadratically faster compared to direct policy search by means of Grover's algorithm.

\begin{figure}[ht]
    \centering
    \includegraphics[width=0.4\textwidth]{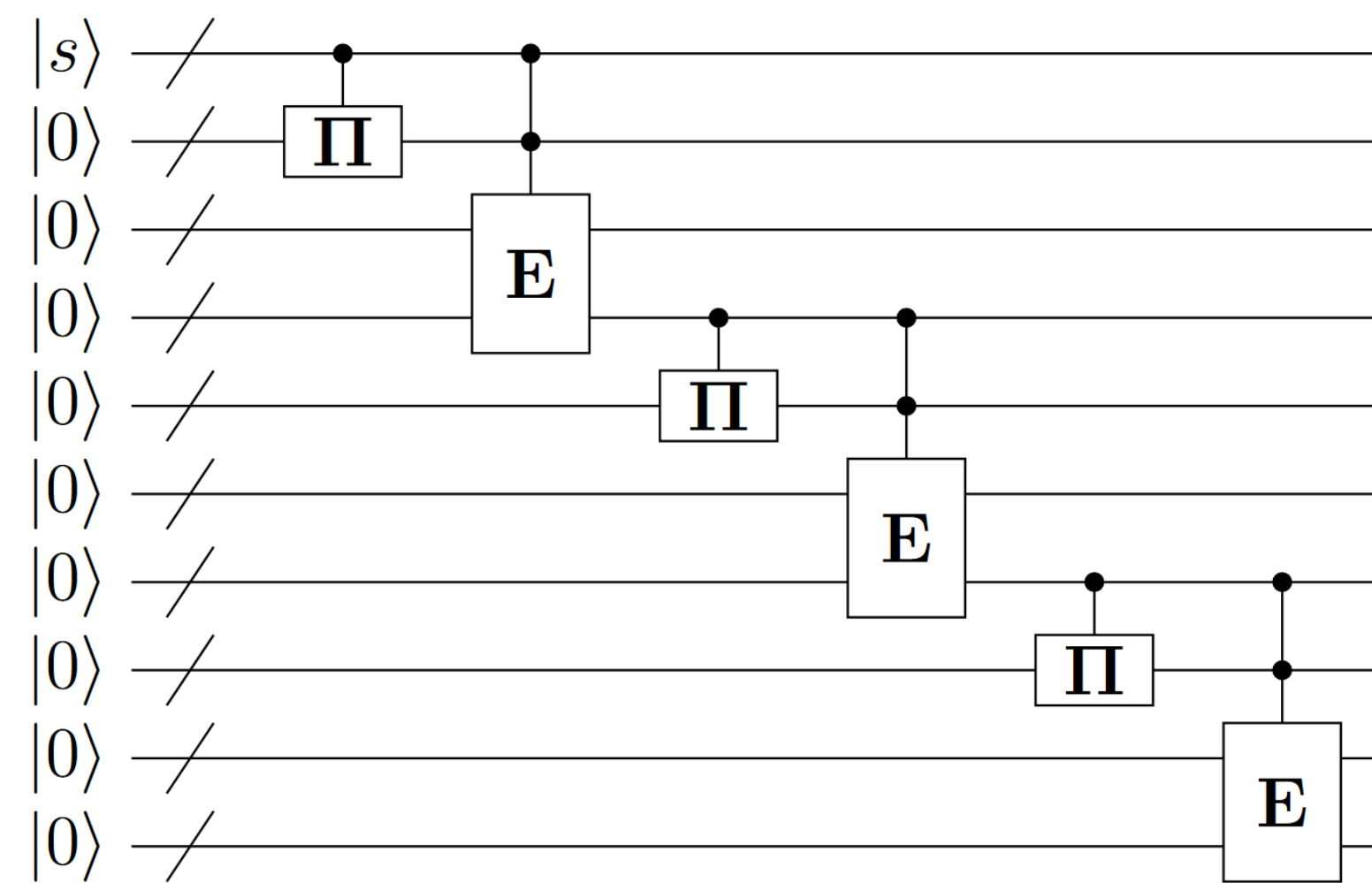}
    \caption{Sequence of policy and environment operator application to an initial state $s$. This constructs a superposition of all possible trajectories for a fixed horizon \gls{mdp} as shown in Wiedemann et al.~\cite{Wiedemann_2021}.}
    \label{fig:Wiedemann1}
\end{figure}

\medbreak
\noindent
\textit{First Algorithm.} The first algorithm assumes access to a policy oracle $\Pi$ and an environment oracle $E$ which act on an initial state $|s\rangle$ as

\begin{equation*}
\Pi (|s\rangle|0\rangle_\mathcal{A})=\sum_a \sqrt{\pi(a|s)}|s\rangle|a\rangle
\end{equation*}

\begin{equation*}
E (|s\rangle|a\rangle|0\rangle_\mathcal{R}|0\rangle_\mathcal{S})=\sum_{r,s^\prime}\sqrt{p(r,s^\prime|s,a)}|s\rangle|a\rangle|r\rangle|s^\prime\rangle\,.
\end{equation*}
Applying these operators sequentially on partially fresh registers as shown in \cref{fig:Wiedemann1} results in a superposition of all possible trajectories
\begin{equation*}
\ket{t}=|s_0\rangle|a_0\rangle|r_1\rangle|s_1\rangle\,.\,.\,.\,|r_H\rangle|s_H\rangle
\end{equation*}
where $H$ is the horizon of the MDP, such that the quantum state reads
\begin{equation*}
|\psi^\pi\rangle=\sum_t \sqrt{p_t}|t\rangle|G_t\rangle\,.
\end{equation*}
Here, $p_t$ is the probability of trajectory $t$. An additional unitary operator has been applied that calculates the return $G_t$ of trajectory $t$ and encodes the value into an additional register entangled with the corresponding trajectory. The superscript $\pi$ on $|\psi^\pi\rangle$ denotes that the state corresponds to the superposition of trajectories for a given policy $\pi$.

The next step of the algorithm attaches an ancilla qubit. With bit-by-bit rotations of the state the digital encoding of $G_t$ is transformed into amplitude encoding (assuming here for simplicity $G_t \in [0,1]$). A simple calculation reveals that the probability of finding the ancilla qubit in state $|1\rangle$ is given by the average return, that is the value function of the initial state $s$. With this insight in mind, the authors propose amplitude estimation \cite{Brassard_2002}. This involves the phase-estimation algorithm, to extract the value function. While classically sampling from the superposition of trajectories would require $\mathcal{O}(1/\epsilon^2)$ preparations of the state, the quantum algorithm achieves the same error with $O(1/\epsilon)$, resulting in a quadratic speed-up. Hereby, $\epsilon$ denotes the fixed additive error to which the value function is to be determined.

\medbreak
\noindent
\textit{Second Algorithm.} The second algorithm shown in Ref.~\cite{Wiedemann_2022} is a quantum version of direct policy search. The authors propose to create a superposition
\begin{equation*}
\frac{1}{\sqrt{|\text{P}|}}\sum_\pi|\pi\rangle|\psi^\pi\rangle
\end{equation*}
where $|\pi\rangle$ is a digital representation of the policy, $|\psi^\pi\rangle$ the superposition of all trajectories corresponding to policy $\pi$ as before, and $|P|$ the size of the policy space. Quantum minimum finding \cite{Durr_1996} can now be applied to find the optimal policy (the one with maximal expected return starting from initial state $s$), requiring only $\mathcal{O}(\sqrt{|P|})$ preparations of the state. This is opposed by $O(|P|)$ in classical direct policy search. Note, however, that the space of all policies scales as $O(|A|^{|S|})$, where $|A|$ and $|S|$ are the sizes of action and state space, respectively. Consequently, the quantum algorithm scales exponentially worse compared to policy iteration where the Bellman optimality equation is iterated with polynomial complexity in $|S|$ and $|A|$. The method proposed in Refs.~\cite{Wiedemann_2022,Wiedemann_2021} therefore should be seen as a quantum version of direct policy search.

%------------------------------------------------------------
\subsection{Quantum Reinforcement Learning with Oracularized Environments}
\label{subsec:Oracles}
%------------------------------------------------------------

In this final section we summarize work that proposes fully quantum-mechanical approaches to \gls{qrl}. In the articles we survey below, the environment is a quantum system or oracle that can be queried by superpositions of states and actions. Interactions with a quantum-mechanical agent create superpositions of trajectories as input for subroutines like Grover search, quantum-maximum finding, and amplitude estimation. 
Provable quantum advantage renders some of these proposals interesting candidates for the post-\gls{nisq} era.

\medbreak

\begin{table}[ht!]
    \centering
    \begin{tabular}{p{\dimexpr 0.15\textwidth-2\tabcolsep-\arrayrulewidth}|p{\dimexpr 0.2\textwidth-2\tabcolsep-\arrayrulewidth}|p{\dimexpr 0.65\textwidth-2\tabcolsep}}
        \toprule
        \textbf{Citation} & \textbf{First Author} & \textbf{Title} \\
        \midrule
        \midrule
        \cite{Dunjko_2016} & V. Dunjko & \hyperref[subsubsec:Dunjko_2016]{Quantum-Enhanced Machine Learning} \\
        \arrayrulecolor{black!30}\midrule
        \cite{Dunjko_2015b} & V. Dunjko & \hyperref[subsubsec:Dunjko_2016]{Framework for learning agents in quantum environments} \\
        \midrule
        \cite{Dunjko_2017} & V. Dunjko & \hyperref[subsubsec:Dunjko_2016]{Advances in quantum reinforcement learning} \\
        \midrule
        \cite{Hamann_2021} & A. Hamann & \hyperref[subsubsec:Dunjko_2016]{Quantum-accessible reinforcement learning beyond strictly epochal environments} \\
        \midrule
        \cite{Wang_2021b} & D. Wang & \hyperref[subsubsec:Dunjko_2016]{Quantum exploration algorithms for multi-armed bandits} \\
        \midrule
        \cite{Wan_2023} & Z. Wan & \hyperref[subsubsec:Dunjko_2016]{Quantum Multi-Armed Bandits and Stochastic Linear Bandits Enjoy Logarithmic Regrets} \\
        \midrule
        \cite{Saggio_2021a} & V. Saggio & \hyperref[subsubsec:Dunjko_2016]{Experimental quantum speed-up in reinforcement learning agents} \\
        \midrule
        \cite{Hamann_2022} & A. Hamann & \hyperref[subsubsec:Dunjko_2016]{Performance analysis of a hybrid agent for quantum-accessible reinforcement learning} \\
        \arrayrulecolor{black}\midrule
        \cite{Cornelissen_2018} & A. Cornelissen & \hyperref[subsubsec:Cornelissen_2018]{Quantum gradient estimation and its application to quantum reinforcement learning} \\
        \arrayrulecolor{black}\bottomrule
    \end{tabular}
    \caption{Work considered for ``\gls{qrl} with Oracularized Environments'' (\cref{subsec:Oracles})}
\end{table}

%------------------------------------------------------------
\paragraph{\label{subsubsec:Dunjko_2016}Quantum-Enhanced Machine Learning, Dunjko et al.~(2016) and related work}\mbox{}\\

\vspace{-1em}
\noindent\textit{Summary.} In Ref.~\cite{Dunjko_2016} and in a more detailed preprint \cite{Dunjko_2015b} a general framework of an agent-environment interaction where both entities are quantum-mechanical systems is developed. To query the environment by a superposition of action states (intuitively the agent learns in parallel), clearly the environment must be modeled by some form of an oracle. As it turns out, this oracularization is much more involved than one might naively think. The focus of the work is therefore:
\begin{itemize}
\item Formalizing a quantum mechanical version of agent-environment interaction
\item Investigation of the classical limit
\item Properties of the general quantum mechanical set-up
\item Treatment of special oracularizable environments
\item Identification of quantum advantage for these environments
\end{itemize}
\medbreak
\noindent
\textit{General Setup.} The interaction between agent and environment is modeled as shown in Fig.~2a in Ref.~\cite{Dunjko_2015b}.
The register $R_A$ processes the computations of the agent, while the register $R_E$ represents the environment. The communication register stores one action and one state. The interaction is described by \gls{cptp} maps or, if we wish, unitary maps on a larger system. The first map $M_1^E$ outputs the initial state and stores it into the communication register. The map $M_1^A$ (modeling the agent) reads this state and, after some processing on $R_A$, outputs an action state which is added to the communication register. Now this action processed by $M_2^E$, which outputs a new state. Consecutively, the previous state in the communication register is overwritten, and so on. The particular form of the states of $R_C$ (if in superpositions of action or not) will be discussed later.

While $R_C$ only contains a state-action pair, the agent's register stores all previous states and actions (because the next action proposed by the learning algorithm depends on all actions and states encountered before, note here the distinction between algorithm and policy). The same is true for the (in general non-Markovian) environment.

\begin{figure}[ht]
\centering
\includegraphics[width=0.9\textwidth]{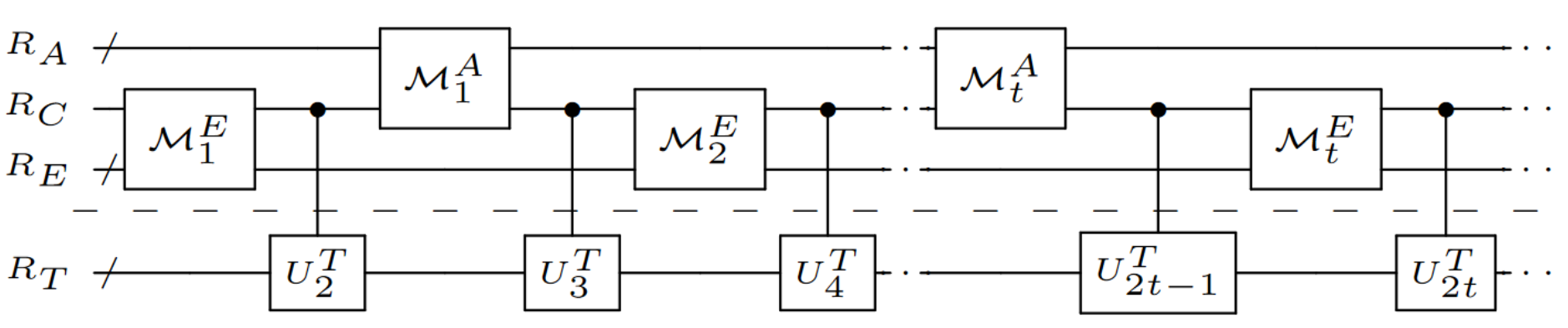}
\caption{Adding a tester as proposed in Dunjko et al.~\cite{Dunjko_2016}.}
\label{Atester}
\end{figure}

Next, as shown in \cref{Atester}, a tester register $R_T$ is introduced, which is designed to `observe' the elapsed history (all encountered states and actions during a learning sequence). This copying from $R_C$ to $R_T$ is modeled by controlled unitaries (so they do not modify $R_C$). Each of them act on a fresh part of the register $R_T$.

The term copying the register here means that a superposition of computational basis states is concatenated with a second register, on which then each basis state is copied to. This produces in general a highly entangled state, which cannot be factorized into the initial state on the first register and a copy on the second (note the no-cloning theorem only rules out a transformation producing this factorized copy for a general initial state). The most general form of the tester interaction treated in this work allows additional unitary transformations, such that the \textit{copying} can be described in the form of controlled unitaries. A tester interaction that merely copies the states will be referred to as \textit{classical}.

After training, the register $R_T$ contains the sequence of actions and states, the so-called history. Any metric measuring performance of learning can be phrased as a function of the history probabilities. Therefore, it can be formulated as the expectation value of an observable on $R_T$.

%
%------------------------------------------------------------
%------------------------------------------------------------
%

\medbreak
\noindent
\textit{Classical Limit.} For recovering the classical learning set-up, the notion of classical interaction is defined by restricting the form of the maps, such that the state in $R_A-R_C-R_E$ remains separable (note that no entanglement between the registers does not prohibit entangled agent or environment states, thus quantum mechanical environments and agents equipped with a quantum computer are not excluded). Additionally, the tester interaction is supposed to be classical (in the sense as defined above). For this setup it is shown that for every scenario with separable register state there exists a classical environment and a classical agent that produce the same history. Consequently, no quantum improvements are possible. Hence, there can be no improvement in the figure of merit, even when the agent has access to a quantum computer.
\medbreak
\noindent
\textit{General Quantum-Mechanical Set-Up.} What happens when we allow general maps and general states on the registers? The authors prove that the state on $R_T$ is still an incoherent mixture, and therefore no quantum advantage can be expected. The reason for this result lies in the memory of agent and possibly the environment:
The agent in general has to remember all previous encountered states and actions, because the learning algorithm run by the agent is a function of that particular elapsed history. The quantum state therefore is a superposition of histories entangled with a state, which describes the agent that has seen this particular history. The states of this agent are orthogonal, since a different agent state translates into a different bit state of the memory. Thus, when tracing out these degrees of freedoms, the resulting reduced density matrix on $R_T$ is an incoherent mixture and no quantum advantage can be achieved in the figure of merit. (Side remark: This does not exclude a quantum advantage in terms of computational complexity in the internal processing of the agent. The result is about exploiting the `quantumness' of the environment-agent interaction)

We note that one has to be careful with the interpretation of density matrices. One might be inclined to think that an incoherent mixture of history states weighted by their probability in some sense corresponds to traversing all of the histories simultaneously but note that the correct expectation value with respect to this density matrix is only obtained in the limit of infinitely many runs corresponding to sampling trajectories one after another.
\medbreak
\noindent
\textit{Oracularization of Environments.} The next part of the work focuses on a special class of environments and learning setting without memory, which overcome the decoherence problem. These oracularized environments are of the following form:
\begin{itemize}
\item episodic with fixed horizon $\rightarrow$ fixed sequence of interactions
\item deterministic $\rightarrow$ action sequence fully determines the history, states can be disregarded
\item binary rewards issued at final state $\rightarrow$ allows use of Grover search
\end{itemize}
\medbreak
\noindent
\textit{Quantum Advantage.} With these assumptions a proper oracle can be constructed, that can be queried with a superposition of actions. This allows to use it as a phase flip oracle, as in the Deutsch-Jozsa or Grover algorithm. The time required for finding a rewarded-action sequence is therefore quadratically reduced. Consequently, this setting is meaningful for learning tasks, where the reward is very sparse. That is, the agent cannot learn until it has first seen a reward. After this initial exploration phase, the agent can now be further trained in simulation. Finally, some of the assumptions are relaxed. The authors also show, how stochastic oracles can be constructed.
\medbreak
\noindent
\textit{Further Work.} There is further work that builds upon the results of Refs.~\cite{Dunjko_2015b,Dunjko_2016}. In Ref.~\cite{Dunjko_2017}, the algorithm is applied to the optimization of parameters describing the properties of the agent (hyperparameter). It also discusses the notion of register hijacking, where the agent has access to hidden memory registers of the environment. This assumption allows the oracularization of more general environments, which is also discussed in Ref.~\cite{Dunjko_2018}. This class is further generalized in Ref.~\cite{Hamann_2021} beyond episodic environments. A closer investigation of amplitude amplification techniques for the special case of multi-armed bandits environments is conducted in Refs.~\cite{Wang_2021b,Wan_2023}. In Ref.~\cite{Saggio_2021a}, the learning setting is implemented experimentally for a two-qubit system and an experimental quantum advantage is observed. Finally, the performance of an agent in this setting is investigated in Ref.~\cite{Hamann_2022}.

%------------------------------------------------------------
%------------------------------------------------------------
%------------------------------------------------------------

\paragraph{\label{subsubsec:Cornelissen_2018}Quantum gradient estimation and its application to quantum reinforcement learning, Cornelissen (2018)}\mbox{}\\
% \cite{Cornelissen_2018}
%------------------------------------------------------------

\vspace{-1em}
\noindent\textit{Summary.}
The master's thesis~\cite{Cornelissen_2018} considers model-based \gls{rl} and develops quantum algorithms for policy evaluation and policy optimization. For the former method a quadratic improvement in sample complexity is found.

\medbreak
\noindent
\textit{Quantum Policy Evaluation.} A quantum algorithm for quantum policy evaluation is presented in Sec.~6.2 of the thesis and will be summarized in the following: The algorithm is executed on a register that is capable to store $T$ states and actions of a $T$-step \gls{mdp}. To generate a sequence, a transition-probability oracle and a policy oracle are defined. They generate a superposition of all possible action-state sequences of the Markov problem, weighted by the square root of the corresponding probabilities. Note that the state is normalized as the probabilities sum up to one. Next, a reward oracle is defined which, when acting on a state-action pair, multiplies the state with a phase factor. The phase is the discounted reward for this state action pair. The discount factors are introduced by making use of fractional phase oracles. This is discussed in detail in Sec.~4 and 5 of the thesis, which are based on Refs.~\cite{Gilyen_2019a, Gilyen_2019b}. The fractional reward oracle is applied to every state-action pair in the register, resulting in the product of phase factors containing the individual discounted rewards. Thus, when merging the exponentials to one exponential, the full quantum state is a superposition of all state-action sequences, weighted by the square root of the individual probability and a phase factor containing the corresponding return. Next, it is shown how the phase factor can be encoded in the amplitude by a controlled operation on an ancilla qubit. Consequently, the probability of measuring the ancilla in, say, state $\ket{0}$ is given by the expectation value of the return, that is the value function. It can be measured using quantum-amplitude estimation, which works based on the phase estimation algorithm. The amplitude-estimation algorithm is a Grover-type algorithm. Hence, it is not surprising that the quadratic speed up compared to classical Monte-Carlo sampling results from this algorithmic step.

\medbreak
\noindent
\textit{Quantum Policy Optimization.} In Sec.~6.4 of the thesis a policy optimization algorithm is developed. This method can be seen as a quantum analogue of policy gradient. First of all, the policy needs to be parameterized. This is done by introducing the parameters $x_{sa}$ such that $\pi(a|s)=x_{sa}$ for all $a$ but one arbitrarily chosen $a^*$ and $\pi(a^*|s)=1-\sum_a x_{sa}$ otherwise. By that definition, the policy is properly normalized and all $x_{sa} \in [0,1]$. Consequently, the expected return is a high-dimensional polynomial in the parameters $x_{sa}$. For taking the derivative of this objective, Jordan's quantum gradient algorithm~\cite{Jordan_2005} in it's advanced form~\cite{Gilyen_2019a} is employed. This leads to a finite-difference approximation of the gradients, written in a phase factor, which can be read out after applying phase estimation. Following Ref.~\cite{Gilyen_2019b}, significant amount of work is devoted to transform the probability oracle for the policy and the transition matrix described above into a phase oracle. Once the superposition of state-action sequences is prepared, an oracle call multiplies each state in the superposition by the corresponding discounted reward. Consecutively, the gradient estimation algorithm is applied and the gradients can be read out. This step is followed by adapting the policy through gradient ascent. It is concluded in the thesis that this policy optimization algorithm does not necessarily lead to quantum speed-up. However, as the author argues, it is conceivable that improvement of the algorithm might lead to a quantum speed-up.

%------------------------------------------------------------
\section{Outlook}
\label{sec:Outlook}
We have given a rather detailed account of the various instances \gls{qrl} that have appeared throughout the literature. We observed, that the dichotomy found at the hardware level, i.e., currently available \gls{nisq} devices vs.\ fault-tolerant and error-corrected \glspl{qpu}, manifests also at the algorithmic level.

With \gls{nisq} devices in mind, \glspl{vqc} have been suggested as function approximators. These replace their classical counterparts in \gls{rl} algorithms with function approximation in policy space, value space, or both. Here, one typically replaces a classical learning heuristic by a learning heuristic with a quantum component. Any sort of potential quantum advantage, however, is not immediately apparent. We eventually can obtain theoretical insight into the properties of \glspl{vqc} viewed as \gls{ml} models and function approximators. However, a direct comparison to their classical cousins, such as neural networks, is anything but easy and might strongly depend on the chosen metric. How can we meaningfully deploy an agent trained with \gls{vqc}-components? What are the requirements for quantum advantage in such a heuristic setting? What does non-simulability of quantum circuits imply for e.g.\ generalization bounds of \glspl{vqc} as \gls{ml} models? Can we scale \glspl{vqc} while maintaining their desirable properties? What is the intrinsic inductive bias of \glspl{vqc} viewed as \gls{ml} models? What are the implications for \gls{rl} and its application domains? All these questions are currently being investigated in the research community, and we are looking forward to new results. 

While quantum algorithms for fault-tolerant and error-corrected \glspl{qpu} have been put forward, we are still far from being able to deploy these algorithms for meaningful problem sizes. Given the necessary advancements of hardware platforms, it will be exciting to see whether these types of quantum algorithms will become competitive with classical learning approaches in practice.

We hope that our survey on the \gls{qrl} literature and the various types of \gls{qrl} algorithms will help guide newcomers to the field and will serve as a valuable reference for researchers.
%------------------------------------------------------------

\section*{Acknowledgments}

We acknowledge collaboration and exchange with M. Franz, L. Wolf, M. Schönberger and W. Mauerer as well as M. J. Hartmann on the topic of quantum reinforcement learning. We further acknowledge exchange and discussion with W. Hauptmann, D. Hein, S. Udluft, V. Tresp, Y. Ma, A. Auer, M. Weber, B. Bisgin, L. Bleiziffer, C. Mendl, S. Wiedemann, S. Wölk, J. M. Lorenz, M. Monnet, T.-A. Dragan, G. Kruse and G. Kontes. We would like to thank M. Leib for feedback on an early version of the manuscript. This work was supported by the German Federal Ministry of Education and Research (BMBF), funding program “quantum technologies – from basic research to market”, grant number 13N15645.

\newpage

\printglossary[type=\acronymtype, title=List of Abbreviations, toctitle=List of Abbreviations]

\newpage

%------------------------------------------------------------
\printbibliography
%------------------------------------------------------------

\end{document}